\documentclass[a4paper,fleqn,usenatbib]{mnras}
\usepackage{longtable}
\usepackage{lscape}
\usepackage{subfig}
\usepackage{graphicx}

\usepackage[T1]{fontenc}
\usepackage{ae,aecompl}

\usepackage{graphicx}	
\usepackage{amsmath}	
\usepackage{amssymb}	

\title[New L subdwarfs and population properties]{Primeval very low-mass stars and brown dwarfs -- IV.  New L subdwarfs, {\sl Gaia} astrometry, population properties, and a blue brown dwarf binary}

\author[Z. H. Zhang et al.]{Z. H. Zhang,$^{1,2,3}$\thanks{E-mail:
zenghuazhang@hotmail.com}\thanks{PSL Fellow}
M. C. G\'{a}lvez-Ortiz,$^{4}$  D. J. Pinfield,$^{5}$ A. J. Burgasser,$^{6}$ 
\newauthor 
  N. Lodieu,$^{2,3}$ H. R. A. Jones,$^{5}$ E. L. Mart{\'i}n,$^{4}$ B. Burningham,$^{5}$ D. Homeier,$^{7}$
\newauthor
 F. Allard,$^{8}$ M. R. Zapatero Osorio,$^{4}$  L. C. Smith,$^{9,5}$  R. L. Smart,$^{10}$ 
\newauthor
B. L{\'o}pez Mart{\'i},$^{11}$  F. Marocco$^{12,5}$\thanks{NASA Postdoctoral Program Fellow} and R. Rebolo$^{2,3}$  \\
$^{1}$GEPI, Observatoire de Paris, Universit{\'e} PSL, CNRS, 5 Place Jules Janssen, 92190 Meudon, France \\
$^{2}$Instituto de Astrof{\'i}sica de Canarias, E-38205 La Laguna, Tenerife, Spain \\
$^{3}$Universidad de La Laguna, Dept. Astrof{\'i}sica, E-38206 La Laguna, Tenerife, Spain \\
$^{4}$Centro de Astrobiolog\'{i}a (CSIC-INTA), Ctra. Ajalvir km 4, E-28850 Torrej\'{o}n de Ardoz, Madrid, Spain \\
$^{5}$Centre for Astrophysics Research, Science and Technology Research Institute, University of Hertfordshire, Hatfield AL10 9AB, UK \\
$^{6}$Center for Astrophysics and Space Science, University of California San Diego, La Jolla, CA 92093, USA \\
$^{7}$Zentrum f{\"u}r Astronomie der Universit{\"a}t Heidelberg, Landessternwarte, K{\"o}nigstuhl 12, D-69117 Heidelberg, Germany  \\
$^{8}$Univ Lyon, ENS de Lyon, Univ Lyon 1, CNRS, Centre de Recherche Astrophysique de Lyon, UMR5574, F-69007, Lyon, France \\
$^{9}$Institute of Astronomy, University of Cambridge, Madingley Road, Cambridge, CB3 0HA, UK \\
$^{10}$Istituto Nazionale di Astrofisica, Osservatorio Astronomico di Torino, Strada Osservatrio 20, I-10025 Pino Torinese, Italy \\
$^{11}$Saint Louis University - Madrid Campus, Avenida del Valle 34, E-28003 Madrid, Spain \\
$^{12}$Jet Propulsion Laboratory, California Institute of Technology, 4800 Oak Grove Dr., Pasadena, CA 91109, USA \\
}

\date{Accepted 2018 July 27. Received 2018 June 28; in original form 2018 March 28}

\pubyear{2018}

\begin{document}
\label{firstpage}
\pagerange{\pageref{firstpage}--\pageref{lastpage}}
\maketitle

\begin{abstract}
We present 27 new L subdwarfs and classify five of them as esdL and 22 as sdL. Our L subdwarf candidates were selected with the UKIRT Infrared Deep Sky Survey and Sloan Digital Sky Survey. Spectroscopic follow-up was carried out primarily with the OSIRIS spectrograph on the Gran Telescopio Canarias. Some of these new objects were  followed up with the X-shooter instrument on the Very Large Telescope. We studied the photometric properties of the population of known L subdwarfs using colour--spectral type diagrams and colour--colour diagrams, by comparison with L dwarfs and main-sequence stars, and identified new colour spaces for L subdwarf selection/study in current and future surveys. We further discussed the brown dwarf transition-zone and the observational stellar/substellar boundary. We found that about one-third of 66 known L subdwarfs are substellar objects, with two-thirds being very low-mass stars.  We also present the Hertzsprung--Russell diagrams, spectral type--absolute magnitude corrections, and tangential velocities of 20 known L subdwarfs observed by the {\sl Gaia} astrometry satellite. One of our L subdwarf candidates, ULAS J233227.03+123452.0, is a mildly metal-poor spectroscopic binary brown dwarf: a $\sim$L6p dwarf and a $\sim$T4p dwarf. This binary is likely a thick disc member according to its kinematics. 

\end{abstract}




\section{Introduction}
Very low-mass stars (VLMS) and brown dwarfs (BDs) are classified using M, L, T, and Y types according to spectral morphology that is dominated by temperature-dependent chemistry and thermal properties \citep{bess91,kir91,jone94,tsuj96,kir99a,kir12,mart99,bur02,bur03b,cus11}. The spectral types of field VLMS extend from M to early-type L \citep{diet14,dupu17}. Most massive BDs have spectral types between late-type M and mid-type L depending on their age \citep{zha17b}. For 0.1--1 Gyr massive BDs are late-type M, evolving to later type as they cool \citep[L type for $\sim$1--10 Gyr age, and T type for ages $>$10 Gyr; e.g., fig. 8 of][]{burr01}.

The classification of VLMS and BDs with subsolar metallicity is complicated by the population's wide metallicity range, which leads to significant diversity in the strength of spectral features. And available sample size also places some limitations on the scope of classification \citep{giz97,kir05,lep07,jao08,dhi12,bur07,kir14,zha17a}. However, previous work has established a set of subdwarf classes that are indicative of metallicity. \citet{giz97} refined M classification by setting up three subclasses; dwarf (dM), subdwarf (sdM) and extreme subdwarf (esdM). An additional intermediate sub-class d/sd for late-type M and L subdwarfs was also suggested by \citet[][]{bur07}, aimed at intermediate metallicity to further encompasses the Galactic thick disc population. \citet{lep07} revised the M dwarf classification into four subclasses; dM, sdM, esdM, and ultra-subdwarf (usdM). 

\citet{kir05} proposed a three-parameter classification strategy for late-type M, L, and T dwarfs to indicate their metallicity, temperature/clouds, and gravity. For example, a metal-poor halo object 2MASS J05325346+8246465 \citep{bur03} was classified esdL7 and an intermediate metal-poor object SDSS J141624.12+134827.4 \citep[SD1416;][]{bowl10,burn10,sch10} was classified sdL7 in \citet{kir10}. And more recently \citet{zha17a} classifies L subdwarfs using three subclasses: sdL, esdL, and usdL, based on the relative strength of subsolar metallicity sensitive spectral features across the optical and near-infrared (NIR). The metallicity ranges of the usdL, esdL, and sdL subclasses are approximately [Fe/H] $\la -1.7$;  $-1.7 \la$ [Fe/H] $\la -1.0$; and $-1.0 \la$ [Fe/H] $\la -0.3$, respectively \citep[see][]{zha17a}.

The L subdwarf population is composed of metal-deficient low-mass stars and high-mass BDs with effective temperature ($T_{\rm eff}$) in the range of $\sim$ 1300--2700 K. They have strong FeH absorption bands, weak or absent VO and CO bands, and enhanced collision-induced H$_2$ absorption \citep[CIA H$_2$;][]{bat52,sau12}, as summarized in \citet{zha17a}. Sample identification has been enabled by modern large-scale optical and NIR surveys: the Two Micron All Sky Survey \citep[2MASS;][]{skr06}, the Sloan Digital Sky Survey \citep[SDSS;][]{yor00}, the UKIRT Infrared Deep Sky Survey \citep[UKIDSS;][]{law07}, and the {\sl Wide-field Infrared Survey Explorer} \citep[{\sl WISE};][]{wri10}, with a total of 39 L subdwarfs previously reported in the literature \citep{bowl10,bur03,bur04a,bur04b,bur06b,burn10,sch10,cus09,giz06,kir10,kir14,kir16,lep02,luhm14,lod10,lod12,lod17,sch04,sch04b,schn16,siv09,smit18,zha17a,zha17b,zha18}.

In general, subdwarfs with sdM--sdL subclass are kinematically associated with the Galactic thick disc, while those with esdM--esdL and usdM--usdL subclasses have kinematics of the Galactic halo \citep{bur08b,cus09,zha13,zha17a}. By comparison to the population of known L dwarfs, L subdwarfs in the solar neighbourhood are rare. This is partly because they are predominantly thick disc and halo objects \citep[cf.][have shown that the fractions of thick disc and halo stars in the solar neighbourhood are 7 and 0.6 per cent]{redd06}, but also because L subdwarfs occupy a narrower mass range than L dwarfs \citep{zha17b}. \citet{lod17} report a UKIDSS/SDSS surface density (for late M and early L dwarfs) of around 0.04 per deg$^2$, pointing to potential for a much larger (hundreds strong) detectable population across surveyed sky. Inroads into this population are important for the detailed study of observed diversity and population make-up amongst the L subdwarfs.

This is the fourth paper in a series titled {\sl Primeval very low-mass stars and brown dwarfs}. The first paper reported the discovery of six new L subdwarfs, defined a new L subdwarf classification scheme, and studied the atmospheric properties of ultra-cool subdwarfs based on 22 late-type M and L subdwarfs \citep[][hereafter Paper I]{zha17a}. In the second paper, we presented the most metal-poor substellar object, and a procedure to distinguish massive halo BDs from the least-massive stars. We also found that mid-type L to early-type T subdwarfs of the Galactic halo are located in a substellar subdwarf gap, known also as the halo BD transition-zone. This zone covers a narrow mass range but spans a wide $T_{\rm eff}$ range due to unsteady nuclear fusion \citep[][hereafter Paper II]{zha17b}. In the third paper, we presented the discovery of three new halo transitional BDs \citep[][hereafter Paper III]{zha18}. Here, we present the discovery of 27 new L subdwarfs and their population properties, as well as a spectroscopic blue BD binary. Observations are presented in Section \ref{sobs}. Classification is carried out in Section \ref{scla}. Section \ref{spop} assesses L subdwarf population properties. Section \ref{sgaia} presents {\sl Gaia} observations of L subdwarfs. Section \ref{ssum} sums up and presents our conclusions.

\begin{table*}
 \centering
  \caption[]{Optical and NIR photometry of L subdwarfs} 
\label{tsdlm}
  \begin{tabular}{l l  c c c c c c c }
\hline
    Name  & SpT &  SDSS \emph{i} & SDSS  \emph{z} &  \emph{Y} (MKO) &  \emph{J} (MKO) &  \emph{H} (MKO) &  \emph{K} (MKO)   & Ref$^g$ \\
\hline
SDSSS J010448.47+153501.9 & usdL1.5 & 20.37$\pm$0.05 & 19.29$\pm$0.06 & 18.48$\pm$0.05 & 17.93$\pm$0.05 & 18.06$\pm$0.11 & 18.08$\pm$0.17 &  17,24 \\
SSSPM J10130734$-$1356204$^a$ & usdL0 & 17.19$\pm$0.01 & 16.09$\pm$0.01 & 15.16$\pm$0.01 & 14.56$\pm$0.01 & 14.39$\pm$0.01 & 14.30$\pm$0.01 &  19, 5 \\
SDSS J125637.13$-$022452.4 & usdL3 & 19.40$\pm$0.02 & 17.71$\pm$0.02 & 16.77$\pm$0.01 & 16.08$\pm$0.01 & 16.05$\pm$0.01 & 16.08$\pm$0.01 &  21, 6 \\
ULAS J135058.85+081506.8 & usdL3 & 21.22$\pm$0.08 & 19.47$\pm$0.06 & 18.66$\pm$0.05 & 17.93$\pm$0.04 & 18.07$\pm$0.10 & 17.95$\pm$0.15 &  15,11 \\
2MASS J16262034+3925190 & usdL4 & 17.90$\pm$0.01 & 16.16$\pm$0.01 & --- & 14.37$\pm$0.11 & 14.52$\pm$0.10 & 14.46$\pm$0.09 &  3 \\
WISEA J213409.15+713236.1 & usdL0.5 & --- & --- & --- & 16.12$\pm$0.15 & 16.30$\pm$0.28 & $\geq$16.64 &  12 \\
ULAS J230711.01+014447.1 & usdL4.5 & 22.51$\pm$0.24 & 19.81$\pm$0.09 & 18.99$\pm$0.08 & 18.15$\pm$0.06 & 18.34$\pm$0.12 & 18.17$\pm$0.18 &  25 \\
\hline
WISEA J001450.17$-$083823.4 & esdL0 & 17.43$\pm$0.01 & 16.05$\pm$0.01 & --- & 14.42$\pm$0.11 & 14.00$\pm$0.11 & 13.74$\pm$0.10 &  11,14 \\
WISEA J020201.25$-$313645.2$^b$ & esdL0.5 & 18.44$\pm$0.01 & 16.99$\pm$0.01 & 15.89$\pm$0.01 & 15.15$\pm$0.01 & 14.96$\pm$0.01 & 14.80$\pm$0.01 &  11 \\
ULAS J020858.62+020657.0 & esdL3 & 21.55$\pm$0.08 & 19.83$\pm$0.07 & 18.76$\pm$0.05 & 18.00$\pm$0.04 & 17.88$\pm$0.13 & 17.62$\pm$0.16 &  25 \\
WISEA J030601.66$-$033059.0 & esdL1 & --- & --- & --- & 14.39$\pm$0.11 & 14.11$\pm$0.11 & 13.96$\pm$0.11 &  11,14 \\
ULAS J033351.10+001405.8 & esdL0 & 19.24$\pm$0.01 & 17.87$\pm$0.02 & 16.81$\pm$0.01 & 16.11$\pm$0.01 & 15.77$\pm$0.01 & 15.50$\pm$0.02 &  16 \\
WISEA J043535.82+211508.9 & esdL1 & 18.59$\pm$0.01 & 16.96$\pm$0.01 & --- & 14.96$\pm$0.11 & 14.73$\pm$0.12 & 14.57$\pm$0.12 & 11,14  \\
2MASS J05325346+8246465 & esdL7 & 20.36$\pm$0.05 & 17.59$\pm$0.02 & --- & 15.08$\pm$0.11 & 14.96$\pm$0.10 & 14.92$\pm$0.09 &  1 \\
2MASS J06164006$-$6407194 & esdL6 & --- & --- & --- & 16.34$\pm$0.16 & 16.34$\pm$0.25 & $\geq$16.39 &  8 \\
ULAS J111429.54+072809.5 & esdL0 & 20.62$\pm$0.05 & 19.26$\pm$0.06 & 18.29$\pm$0.03 & 17.59$\pm$0.03 & 17.26$\pm$0.04 & 17.12$\pm$0.07 & 26  \\
ULAS J124425.75+102439.3 & esdL0.5 & 19.49$\pm$0.02 & 18.01$\pm$0.02 & 16.98$\pm$0.01 & 16.26$\pm$0.01 & 16.00$\pm$0.01 & 15.77$\pm$0.02 &  17 \\
SDSS J124410.11+273625.8 & esdL0.5 & 20.40$\pm$0.05 & 19.12$\pm$0.06 & 18.28$\pm$0.04 & 17.58$\pm$0.03 & 17.32$\pm$0.05 & 17.12$\pm$0.06 & 16  \\
ULAS J135216.31+312327.0 & esdL0.5 & 20.01$\pm$0.04 & 18.66$\pm$0.05 & 17.69$\pm$0.02 & 16.93$\pm$0.02 & 16.66$\pm$0.03 & 16.41$\pm$0.04 &  26 \\
SDSS J141405.74$-$014202.7 & esdL0 & 19.85$\pm$0.03 & 18.45$\pm$0.03 & 17.50$\pm$0.03 & 16.81$\pm$0.02 & 16.45$\pm$0.03 & 16.14$\pm$0.03 &  17 \\
SSSPM J144420.67-201922.2$^c$ & esdL1 & --- & --- & 13.23$\pm$0.01 & 12.45$\pm$0.01 & 12.19$\pm$0.11 & 11.93$\pm$0.01 & 18,12  \\
ULAS J145234.65+043738.4 & esdL0.5 & 20.50$\pm$0.06 & 18.92$\pm$0.05 & 18.11$\pm$0.03 & 17.28$\pm$0.03 & 16.79$\pm$0.04 & 16.62$\pm$0.06 &  26 \\
ULAS J151913.03$-$000030.0 & esdL4 & 21.42$\pm$0.09 & 19.26$\pm$0.06 & 18.19$\pm$0.03 & 17.21$\pm$0.02 & 17.07$\pm$0.03 & 16.97$\pm$0.04 &  23 \\
2MASS J16403197+1231068 & esdL0 & 19.01$\pm$0.02 & 17.56$\pm$0.02 & --- & 15.90$\pm$0.13 & 15.65$\pm$0.15 & 15.49$\pm$0.18 & 2, 9  \\
WISEA J204027.30+695924.1 & esdL0.5 & --- & --- & --- & 13.68$\pm$0.13 & 13.36$\pm$0.12 & 13.09$\pm$0.11 &  11,14 \\
ULAS J223302.03+062030.8 & esdL0.5 & 21.12$\pm$0.06 & 19.68$\pm$0.06 & 18.60$\pm$0.04 & 17.90$\pm$0.05 & 17.67$\pm$0.08 & 17.44$\pm$0.11 &  26 \\
ULAS J231924.35+052524.5 & esdL1 & 20.41$\pm$0.05 & 18.99$\pm$0.05 & 18.22$\pm$0.03 & 17.33$\pm$0.02 & 17.08$\pm$0.04 & 16.93$\pm$0.06 &  26 \\
\hline
2MASS J00412179+3547133 & sdL0.5 & 20.28$\pm$0.03 & 18.33$\pm$0.02 & --- & 15.89$\pm$0.13 & 15.78$\pm$0.18 & 15.14$\pm$0.15 & 2  \\
WISEA J005757.65+201304.0 & sdL7 & 21.19$\pm$0.11 & 18.97$\pm$0.06 & --- & 16.26$\pm$0.14 & 15.52$\pm$0.14 & 14.99$\pm$0.16 & 11,14  \\
WISEA J011639.05$-$165420.5$^d$ & sdL0 & 19.49$\pm$0.02 & 17.66$\pm$0.02 & --- & 15.76$\pm$0.12 & 15.38$\pm$0.14 & 14.94$\pm$0.16 &  20 \\
ULAS J011824.89+034130.4 & sdL0 & 21.93$\pm$0.20 & 20.47$\pm$0.21 & 19.10$\pm$0.07 & 18.18$\pm$0.05 & 17.73$\pm$0.07 & 17.56$\pm$0.11 & 26  \\
WISEA J013012.66-104732.4 & sdL0 & 19.33$\pm$0.03 & 17.50$\pm$0.02 & --- & 15.58$\pm$0.12 & 15.18$\pm$0.14 & 14.80$\pm$0.15 &  20 \\
ULAS J021258.08+064115.9 & sdL1 & 21.11$\pm$0.08 & 19.38$\pm$0.08 & 18.20$\pm$0.03 & 17.43$\pm$0.03 & 17.06$\pm$0.03 & 16.78$\pm$0.05 & 26  \\
ULAS J021642.96+004005.7 & sdL4 & 22.10$\pm$0.16 & 19.99$\pm$0.10 & 18.41$\pm$0.05 & 17.30$\pm$0.03 & 16.96$\pm$0.04 & 16.51$\pm$0.04 & 23 \\
ULAS J023803.12+054526.1 & sdL0 & 20.48$\pm$0.04 & 18.57$\pm$0.03 & 17.26$\pm$0.02 & 16.43$\pm$0.01 & 16.02$\pm$0.02 & 15.59$\pm$0.02 &  26 \\
2MASS J06453153$-$6646120 & sdL8 & --- & --- & --- & 15.54$\pm$0.13 & 14.77$\pm$0.12 & 14.40$\pm$0.13 & 10  \\
ULAS J075335.23+200622.4 & sdL0 & 19.76$\pm$0.02 & 17.90$\pm$0.02 & 16.68$\pm$0.01 & 15.87$\pm$0.01 & 15.48$\pm$0.01 & 15.09$\pm$0.01 & 26  \\
ULAS J082206.61+044101.8 & sdL0 & 20.33$\pm$0.04 & 18.51$\pm$0.03 & 17.12$\pm$0.02 & 16.29$\pm$0.01 & 15.96$\pm$0.02 & 15.53$\pm$0.02 & 26  \\
WISEA J101329.72$-$724619.2$^e$ & sdL2? & --- & --- & --- & 15.89$\pm$0.01 & 15.49$\pm$0.15 & 14.94$\pm$0.02 &  12 \\
2MASS J11582077+0435014 & sdL7 & 21.08$\pm$0.05 & 18.18$\pm$0.02 & 16.61$\pm$0.01 & 15.43$\pm$0.01 & 14.88$\pm$0.01 & 14.37$\pm$0.01 &  10 \\
ULAS J123142.99+015045.4 & sdL0 & 21.38$\pm$0.11 & 19.45$\pm$0.09 & 18.46$\pm$0.04 & 17.54$\pm$0.03 & 17.21$\pm$0.04 & 16.78$\pm$0.05 & 26  \\
ULAS J124104.75$-$000531.4 & sdL0 & 22.41$\pm$0.27 & 20.42$\pm$0.17 & 19.12$\pm$0.08 & 18.46$\pm$0.10 & 18.08$\pm$0.12 & 18.04$\pm$0.20 & 26  \\
ULAS J124947.04+095019.8 & sdL1 & 20.40$\pm$0.04 & 18.66$\pm$0.04 & 17.62$\pm$0.02 & 16.83$\pm$0.02 & 16.40$\pm$0.03 & 16.12$\pm$0.04 &  23 \\
ULAS J125226.62+092920.1 & sdL0 & 20.78$\pm$0.05 & 19.03$\pm$0.05 & 17.69$\pm$0.02 & 16.87$\pm$0.02 & 16.47$\pm$0.02 & 16.05$\pm$0.03 &  26 \\
VVV J12564163$-$6202039 & sdL3$^{f}$ & 19.70$\pm$0.08 & --- & 17.10$\pm$0.02 & 16.10$\pm$0.01 & 15.89$\pm$0.02 & 15.72$\pm$0.03 & 22  \\
ULAS J130710.22+151103.4 & sdL8 & --- & --- & 19.31$\pm$0.07 & 18.14$\pm$0.04 & 17.53$\pm$0.07 & 17.24$\pm$0.09 &  26 \\
ULAS J133348.27+273505.5 & sdL1 & 20.52$\pm$0.05 & 18.76$\pm$0.04 & 17.47$\pm$0.02 & 16.62$\pm$0.01 & 16.27$\pm$0.02 & 15.98$\pm$0.02 & 23  \\
ULAS J133836.97$-$022910.7 & sdL7 & 22.53$\pm$0.28 & 20.10$\pm$0.15 & 18.56$\pm$0.06 & 17.37$\pm$0.03 & 16.81$\pm$0.04 & 16.37$\pm$0.05 & 23  \\
ULAS J134206.86+053724.9 & sdL0.5 & 21.95$\pm$0.16 & 19.62$\pm$0.08 & 18.38$\pm$0.04 & 17.43$\pm$0.03 & 17.01$\pm$0.03 & 16.56$\pm$0.04 & 26  \\
ULAS J134423.98+280603.8 & sdL4 & 22.67$\pm$0.28 & 19.85$\pm$0.10 & 18.40$\pm$0.03 & 17.19$\pm$0.02 & 16.72$\pm$0.03 & 16.13$\pm$0.03 & 26  \\
ULAS J134749.79+333601.7 & sdL0 & 19.87$\pm$0.03 & 18.07$\pm$0.02 & 16.66$\pm$0.01 & 15.85$\pm$0.01 & 15.46$\pm$0.01 & 15.27$\pm$0.02 & 23  \\
ULAS J134852.93+101611.8 & sdL0 & 20.76$\pm$0.05 & 18.99$\pm$0.04 & 17.54$\pm$0.02 & 16.69$\pm$0.01 & 16.32$\pm$0.01 & 15.87$\pm$0.02 &  26 \\
ULAS J135359.58+011856.7 & sdL0 & 21.49$\pm$0.14 & 19.37$\pm$0.08 & 18.26$\pm$0.04 & 17.36$\pm$0.03 & 16.89$\pm$0.03 & 16.52$\pm$0.04 & 26  \\
WISEA J135501.90$-$825838.9$^e$ & sdL5? & --- & --- & --- & 16.32$\pm$0.01 & 15.37$\pm$0.17 & 14.90$\pm$0.01 &  12 \\
ULAS J141203.85+121609.9 & sdL5 & 21.29$\pm$0.09 & 19.14$\pm$0.05 & 17.54$\pm$0.02 & 16.33$\pm$0.01 & 15.85$\pm$0.01 & 15.43$\pm$0.02 & 26  \\
SDSS J141624.12+134827.4 & sdL7 & 18.38$\pm$0.01 & 15.91$\pm$0.01 & 14.26$\pm$0.01 & 12.99$\pm$0.01 & 12.47$\pm$0.01 & 12.05$\pm$0.01 &  7, 10 \\
ULAS J141832.35+025323.0 & sdL0 & 20.11$\pm$0.04 & 18.27$\pm$0.03 & 16.86$\pm$0.01 & 16.00$\pm$0.01 & 15.61$\pm$0.01 & 15.19$\pm$0.01 &  26 \\
ULAS J144151.55+043738.5 & sdL4 & --- & --- & 18.45$\pm$0.04 & 17.23$\pm$0.03 & 16.95$\pm$0.04 & 16.34$\pm$0.04 &  26 \\
ULAS J151649.84+083607.1 & sdL5 & 22.65$\pm$0.24 & 20.25$\pm$0.10 & 18.74$\pm$0.04 & 17.35$\pm$0.02 & 16.71$\pm$0.03 & 16.26$\pm$0.03 & 26  \\
ULAS J154638.34$-$011213.0 & sdL3 & 22.10$\pm$0.14 & 19.98$\pm$0.11 & 18.56$\pm$0.05 & 17.51$\pm$0.04 & 17.21$\pm$0.05 & 16.95$\pm$0.08 & 26  \\
2MASS J17561080+2815238 & sdL1 & --- & --- & --- & 14.66$\pm$0.11 & 14.19$\pm$0.11 & 13.79$\pm$0.10 &  10 \\
LSR J182611.3+301419.1 & sdL0 & --- & --- & --- & 11.61$\pm$0.11 & 11.22$\pm$0.10 & 10.78$\pm$0.10 &  13 \\
\hline
\end{tabular}
\end{table*}

\addtocounter{table}{-1}
\begin{table*}
 \centering
  \caption[]{Continued.} 
\label{tsdlm}
  \begin{tabular}{c l  c c c c c c c }
\hline
    Name  & SpT & SDSS \emph{i} & SDSS \emph{z} &  \emph{Y} (MKO) &  \emph{J} (MKO) &  \emph{H} (MKO) &  \emph{K} (MKO)   & Ref$^g$ \\
\hline
ULAS J223440.80+001002.6 & sdL1 & 22.05$\pm$0.14 & 20.16$\pm$0.12 & 18.97$\pm$0.07 & 17.63$\pm$0.04 & 17.15$\pm$0.05 & 16.90$\pm$0.07 &  26 \\
ULAS J225902.14+115602.1 & sdL0 & 21.30$\pm$0.16 & 19.23$\pm$0.11 & 17.97$\pm$0.02 & 17.05$\pm$0.02 & 16.63$\pm$0.03 & 16.21$\pm$0.03 & 26  \\
ULAS J230256.53+121310.2 & sdL0 & 21.74$\pm$0.15 & 19.46$\pm$0.08 & 18.30$\pm$0.04 & 17.49$\pm$0.03 & 17.05$\pm$0.06 & 16.73$\pm$0.05 &  26 \\
ULAS J230443.30+093423.9 & sdL0 & 21.65$\pm$0.16 & 19.27$\pm$0.08 & 18.21$\pm$0.03 & 17.23$\pm$0.02 & 16.74$\pm$0.03 & 16.27$\pm$0.03 &  26 \\
\hline
\end{tabular}
\begin{list}{}{}
\item[Notes.] $^a$ATLAS $i,z$ and VHS $YJHK$ photometry. $^b$ ATLAS $i,z$ and VIKING $YJH$ photometry. $^c$VHS $YJK$ photometry. $^d$ATLAS $i,z$ photometry. $^e$VHS $JK$ photometry. 
The other NIR photometry of objects with $Y$ band detection are  from  UKIDSS. $J$, $H$, and $K$ photometry of the other objects without UKIDSS $Y$ band detection are converted from 2MASS with equations (\ref{ej}--\ref{ek}). $^{f}$VVV J12564163$-$6202039 is an sdL3 subdwarf according to a new optical-NIR spectrum we obtained with the X-shooter. 
$^{g}$Reference: 1. \citet{bur03}; 2. \citet{bur04a}; 3. \citet{bur04b}; 4. \citet{bur06b}; 5. \citet{bur07}; 6. \citet{bur09}; 7. \citet{bowl10,burn10,sch10}; 8. \citet{cus09}; 9. \citet{giz06};  10. \citet{kir10}; 11. \citet{kir14}; 12. \citet{kir16}; 13. \citet{lep02}; 14. \citet{luhm14};  15. \citet{lod10}; 16. \citet{lod12}; 17. \citet{lod17}; 18. \citet{sch04}; 19. \citet{sch04b}; 20. \citet{schn16}; 21. \citet{siv09}; 22. \citet{smit18}; 23. Paper I; 24. Paper II; 25. Paper III; 26. This work.
\end{list}
\end{table*}

\begin{figure}
\begin{center}
   \includegraphics[width=\columnwidth]{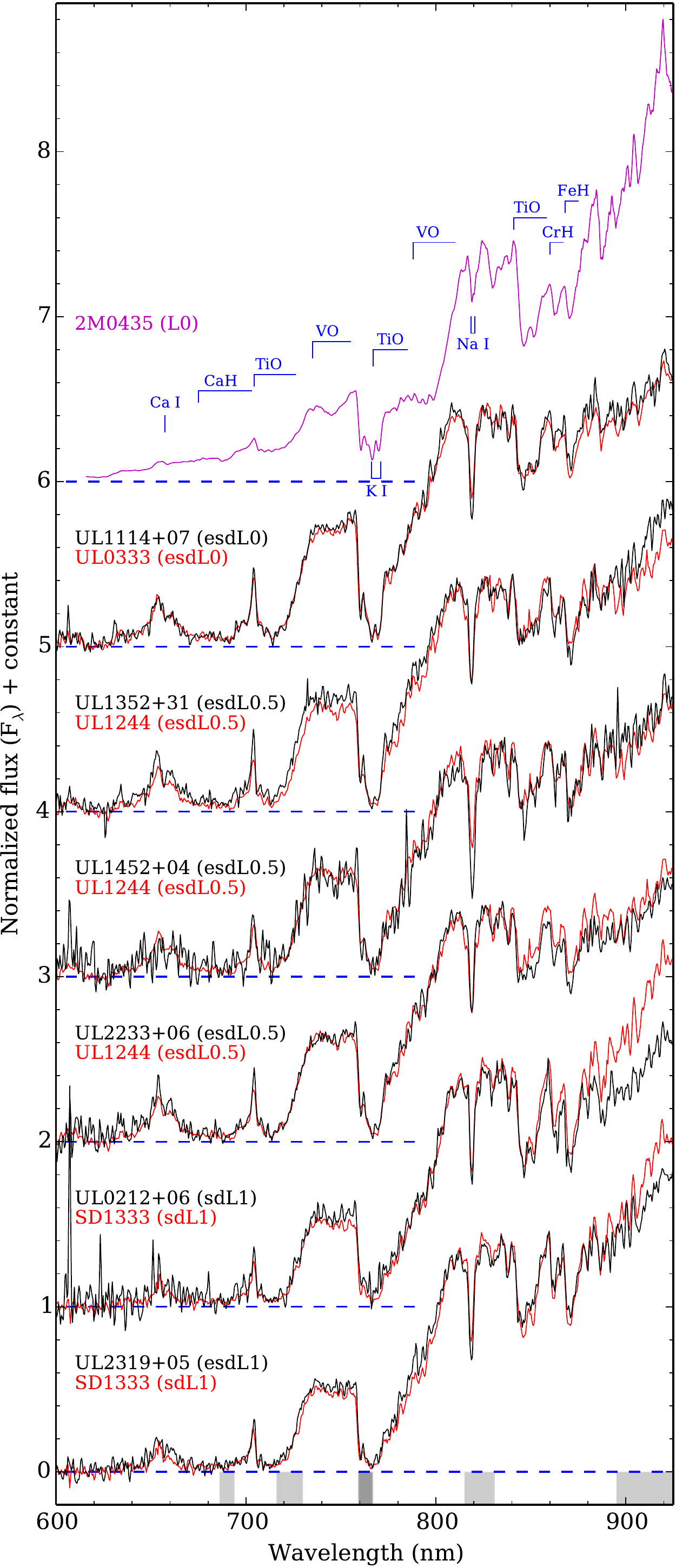}
\caption[]{Optical spectra of 6 new L subdwarfs (black) showing strong optical metal-poor features in comparison to the L0 dwarf (magenta) 2MASP J0345432+254023 \citep[2M0435;][]{kir99a}, and three known L subdwarfs (red): UL0333, UL1244 \citep{lod12}, and SD1333 (Paper I). Spectra are normalized near 800 nm. Spectra of UL0333, UL1244, and SD1333 were shifted by $-150$ to $-300$ km s$^{-1}$ to align their Na I lines with those of our objects (except for UL0212+06). Telluric absorption is indicated with shaded grey bands but not corrected. Lighter and thicker shaded bands indicate regions with weaker and stronger telluric effects.}
\label{f6sdl}
\end{center}
\end{figure}

\begin{figure*}
\begin{center}
   \includegraphics[width=\textwidth]{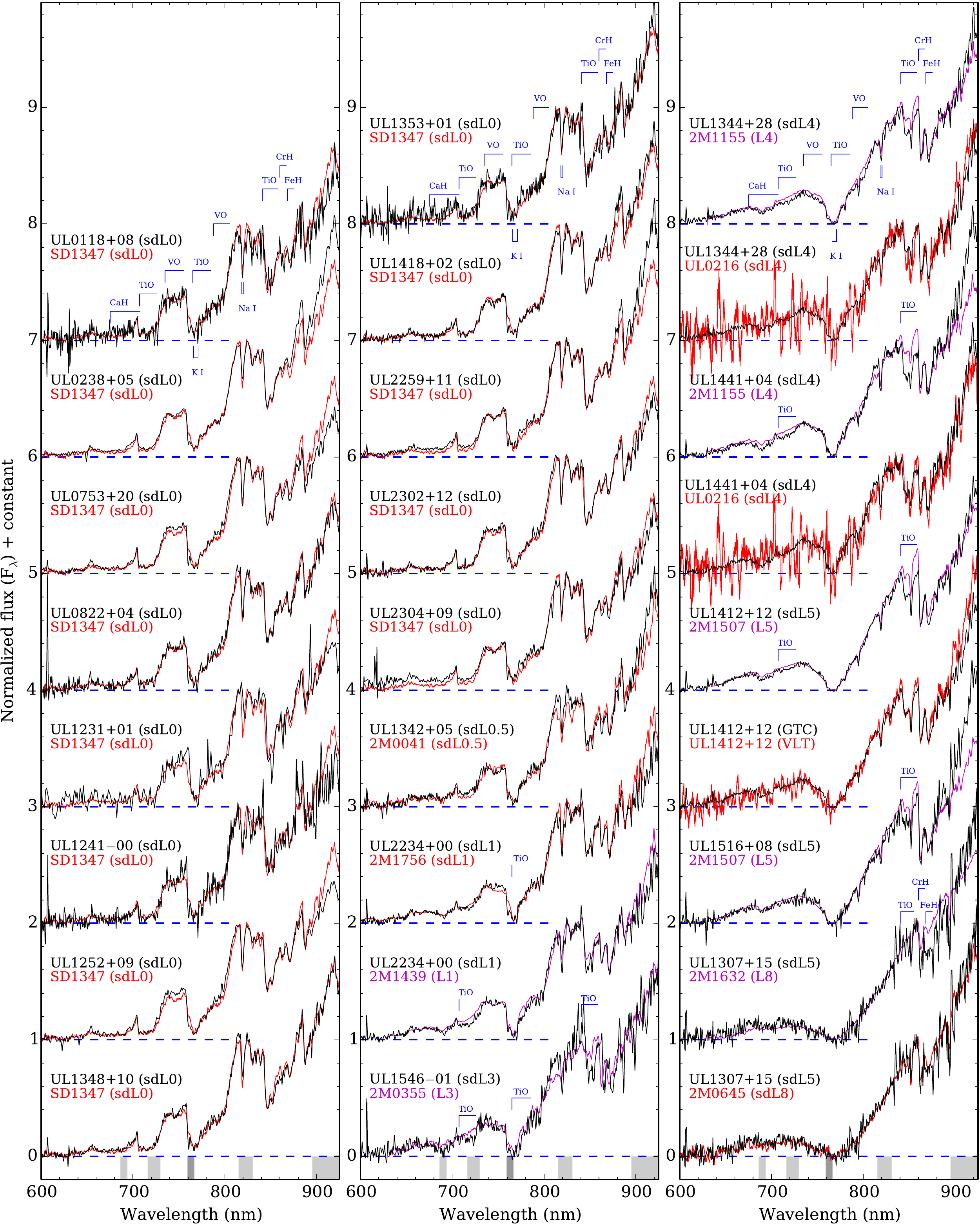}
\caption[]{Optical spectra of 21 new L subdwarfs (black) compared to known L subdwarfs (red) and dwarfs (magenta). Spectra are normalized near 840 nm. Spectra of SD1347, 2M0041, and UL0216 are from Paper I. Spectra of 2M1439, 2M0355, 2M1155, and 2M1632 are from \citet{kir99a}. The spectrum of 2M1756 is from \citet{kir10} and was corrected for telluric absorption. The spectrum of 2M1507 is from \citet{rei00}. 2M0645 was observed by our X-shooter follow-up programme.}
\label{f20sdl}
\end{center}
\end{figure*}

\begin{figure}
\begin{center}
  \includegraphics[width=\columnwidth]{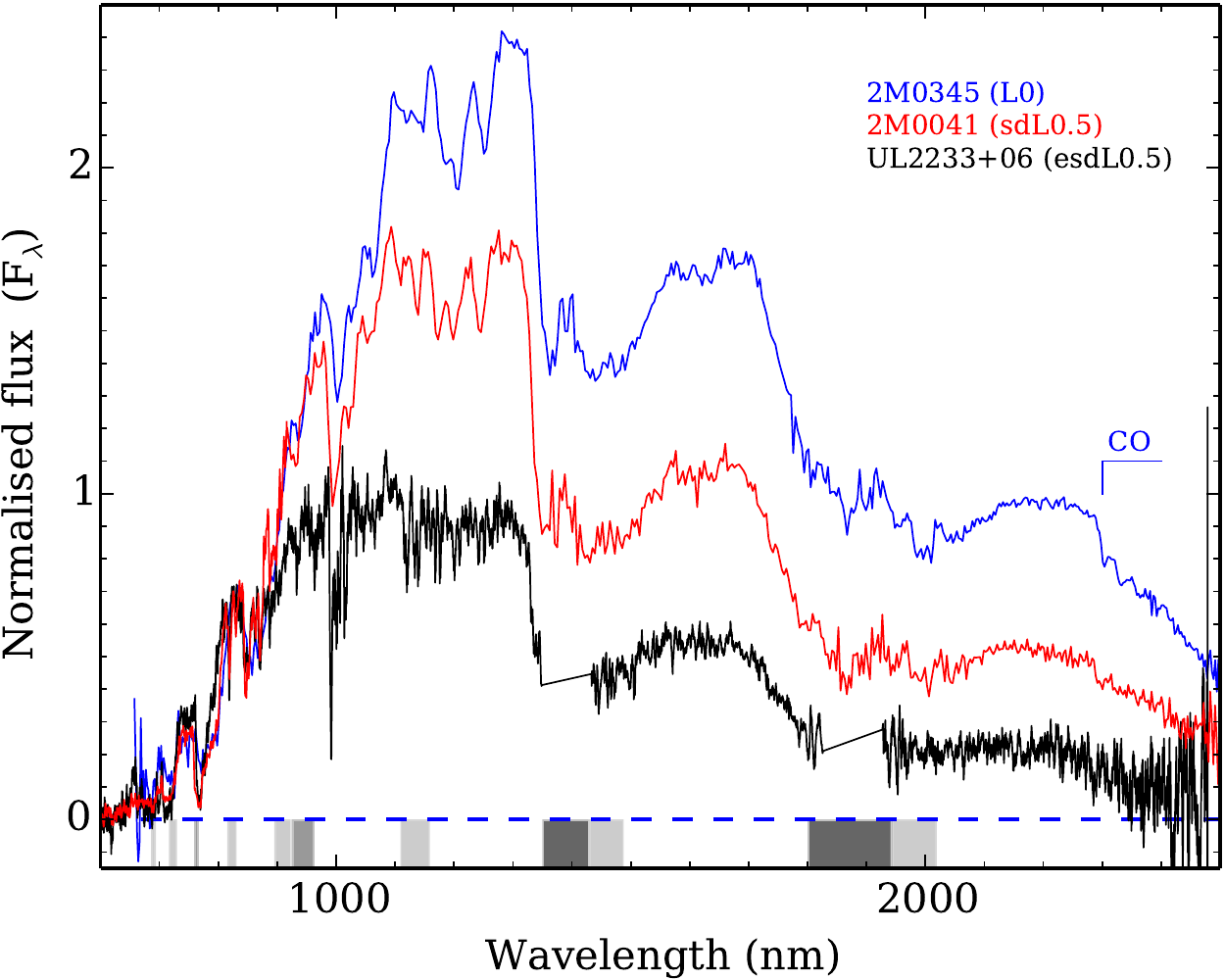}
\caption[]{The X-shooter spectrum of UL2233+06 compared with optical to NIR spectra of 2M0041 (sdL0.5) and SD0345 (L0). NIR and optical spectra of 2M0041 are from \citet{bur04a} and Paper I, respectively. The spectrum of SD0345 is from \citet{bur06a}. Telluric absorption is indicated with grey shading and has been corrected in our VLT spectra. Lighter and thicker shaded bands indicate regions with weaker and stronger telluric effects.}
\label{ful2233}
\end{center}
\end{figure}

\begin{figure}
\begin{center}
  \includegraphics[width=\columnwidth]{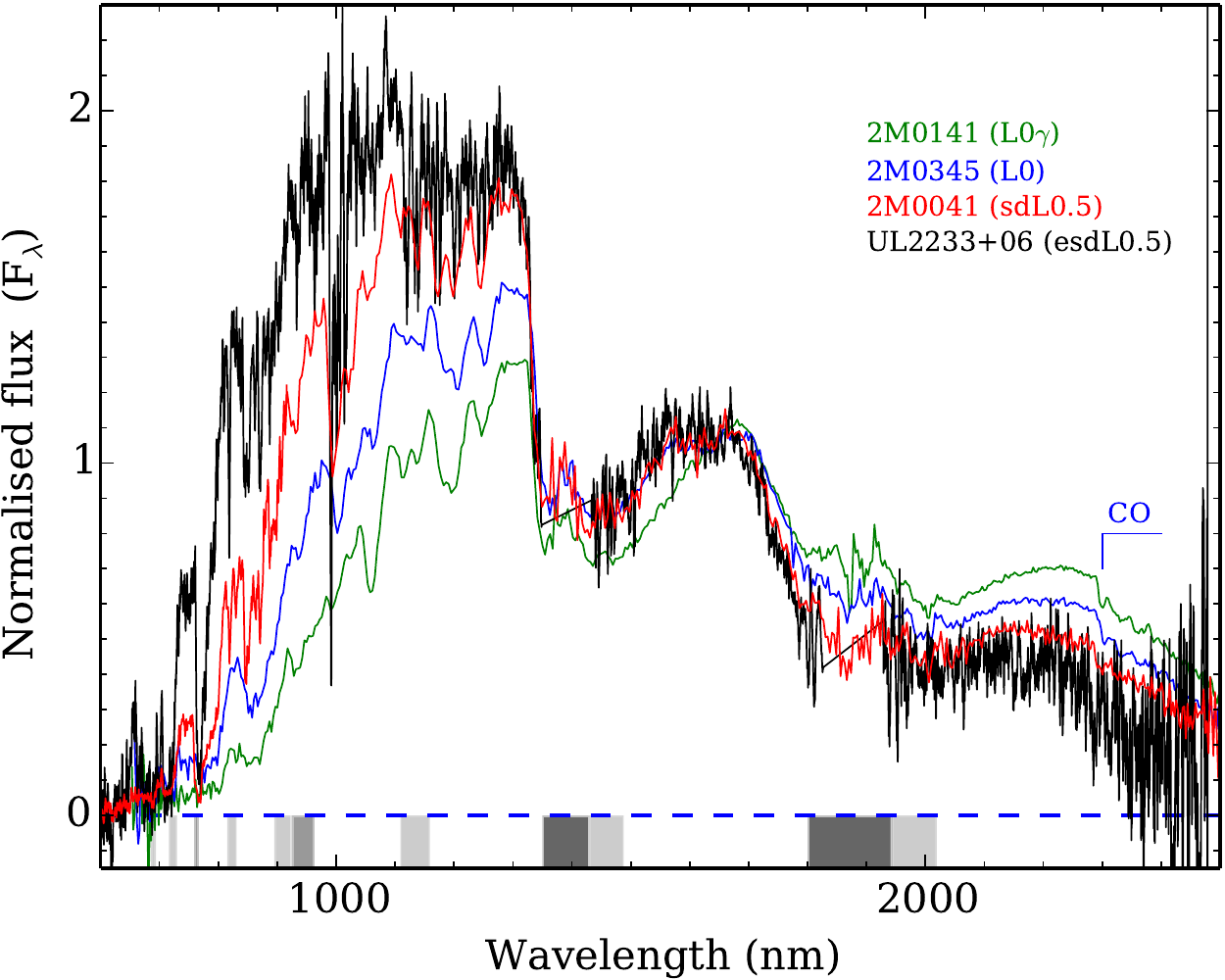}
\caption[]{NIR spectra of UL2233+06, 2M0041 and 2M0345 normalized in the $H$ band. The young L0$\gamma$ dwarf 2M0141 is also plotted for comparison.}
\label{ful2233nh}
\end{center}
\end{figure}

\begin{figure}
\begin{center}
  \includegraphics[width=\columnwidth]{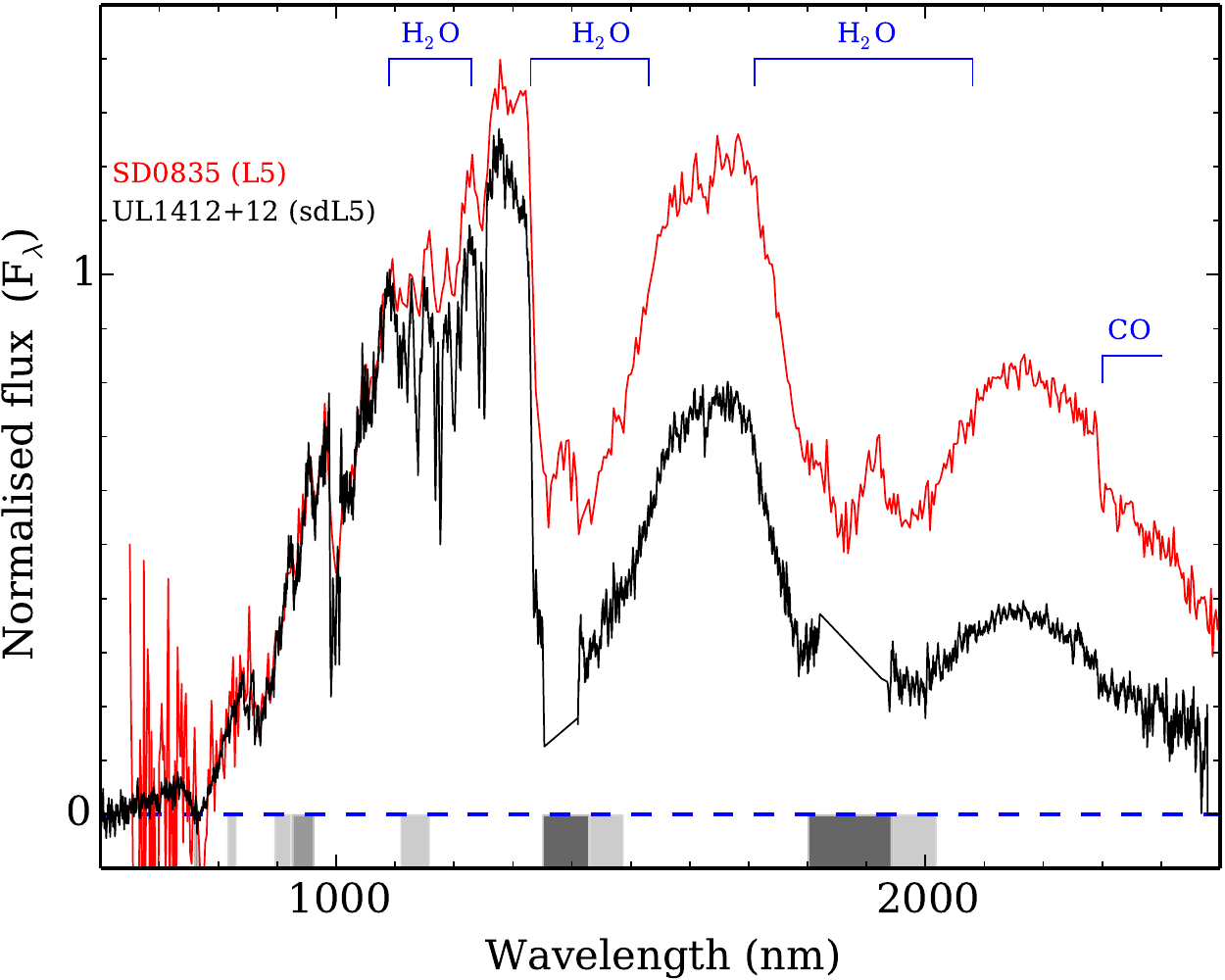}
\caption[]{The optical--NIR spectrum of UL1412+12 compared to that of an L5 standard (SD0835) from \citet{chiu06}.}
\label{ul1412}
\end{center}
\end{figure}

\begin{figure}
\begin{center}
 \includegraphics[width=\columnwidth]{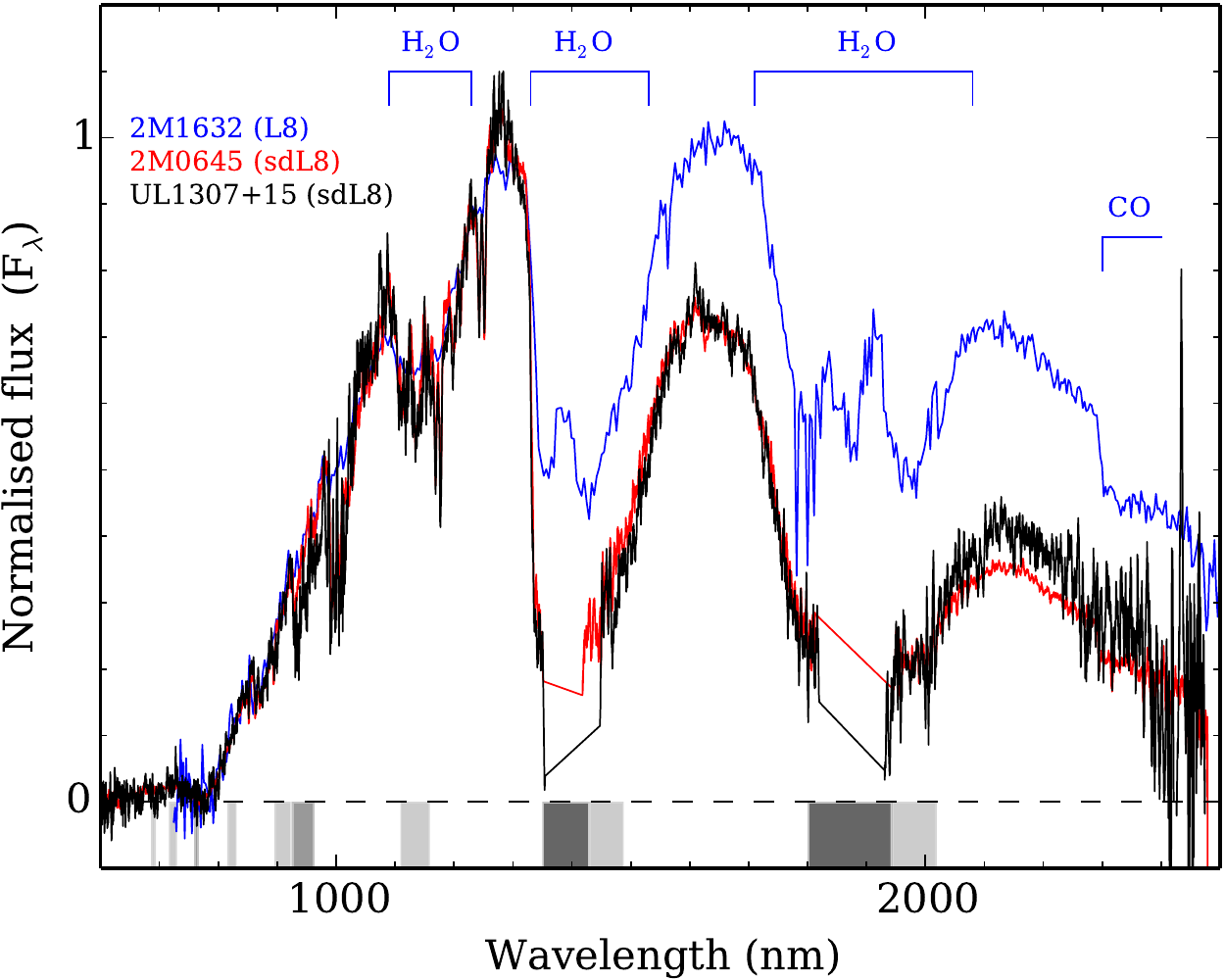}
\caption[]{Optical--NIR spectrum of UL1307+15 compared to the L8 2M1632 and the sdL8  2M0645. The X-shooter spectra of UL1307+15 and 2M0645 were smoothed by 101 (VIS) and 51 (NIR) pixels. The spectrum of  2M1632 is from \citet{bur07b}.}
\label{ul1307fit}
\end{center}
\end{figure}

\section{Observation}
\label{sobs}
\subsection{Candidate selection}
The L subdwarf candidates in our programme were generally selected from the UKIDSS Large Area Survey (LAS) and the SDSS, following the selection criteria/procedure described in Paper I. In addition two extra candidates without SDSS detections were selected from a ULAS proper motion catalogue, covering 1500 deg$^2$ to an approximate 5$\sigma$ depth of $J=19.6$ \citep{smit14}. These two objects (ULAS J130710.22+151103.4 and ULAS J144151.55+043738.5) both have relatively blue $J-K$ colour and proper motion higher than 0.4 arcsec per year. 

Our programmatic candidate sample consisted of 64 objects, of which five had been confirmed as L subdwarfs by other studies. We have spectroscopically followed up the remaining 59 candidates, which has in total resulted in the confirmation/study of 34 L subdwarfs (an overall 61\% confirmation rate). Eight of these were published in Papers I and III, and the remaining 27 are reported here. Table \ref{tsdlm} presents optical and NIR photometry for all known L subdwarfs, which includes the 34 studied by our programme plus 31 from elsewhere in the literature. The 24 programmatic candidates that are not L subdwarfs consist of fourteen late M dwarfs, one L1 dwarf, six late-type sdM subdwarfs, 1 esdM7 subdwarf, a probable L+T unresolved binary (see Section 3.4), and one galaxy.

\subsection{GTC spectroscopy}
Candidates have been followed up with optical spectroscopy using the Optical System for Imaging and low-Resolution Integrated Spectroscopy \citep[OSIRIS;][]{cepa00} instrument on the Gran Telescopio Canarias (GTC) since 2012. Table \ref{tgtc} shows a summary of our GTC spectroscopic observations. The spectra were mostly obtained using the R500R grism, with three making use of the R300R grism. These provide resolving power of around 500 and 300, respectively. The spectra were reduced using standard procedures with {\scriptsize IRAF}\footnote{IRAF is distributed by the National Optical Observatory, which is operated by the Association of Universities for Research in Astronomy, Inc., under contract with the National Science Foundation.}. The HgAr, Ne, and Xe arcs were used on the wavelength calibration of our spectra. Spectral flux calibrations were achieved with standard stars \citep{bohl95,hamu94,oke74,oke90}. Spectral features caused by telluric absorptions in these spectra were not corrected. Five objects were observed twice to gain additional exposure time, and for the purposes of this paper we combined the spectra of these objects from both epochs.

\subsection{VLT spectroscopy}
We have also obtained optical-NIR spectroscopy of five of our candidates, a known sdL8 subdwarf and three radial velocity standard L dwarfs using the X-shooter spectrograph \citep{ver11} on the Very Large Telescope (VLT) since 2014. Table \ref{tvltob} shows a summary of our VLT spectroscopic observations. These X-shooter spectra were observed in an ABBA nodding mode with a 1.2 arcsec slit providing a resolving power of 6700 in the VIS arm and 4000 in the NIR arm. These spectra were reduced with ESO Reflex in a lazy mode \citep{freu13}. The flat field used for the science data was also used for flat-fielding the flux standard star spectrum (by set flat strategy to false). Telluric corrections were achieved  for both VIS and NIR arms using telluric standards that were observed right after or before our targets at a very close airmass. To extract telluric spectra for correction, we fitted and normalized spectra of telluric standards stars to their continuum and removed non-telluric features (e.g. hydrogen absorption lines) with {\scriptsize IRAF SPLOT}. These X-shooter spectra were smoothed by 101 and 51 pixels in the VIS and NIR arms, respectively, for display in figures, which reduced the resolving power to about 600 and increased their signal-to-noise ratio (SNR per pixel) by 10 and 7 times. 

\section{Classification}
\label{scla}
In Paper I, we classified L subdwarfs into sdL, esdL, and usdL subclasses. The subtypes of L subdwarfs are based on the comparison of their optical spectra to those of L dwarfs. The subclasses of L subdwarfs are based on their metallicity sensitive spectral features in the optical and NIR.

\subsection{Optical classification}
To assign spectral type/class to our candidates, we compared their optical spectra to those of established L subdwarf standards and identified the ones that provided the closest match to the overall spectral profile and to known spectral features sensitive to low metallicity. We also compared optical spectra of some candidates to those of L dwarf standards, when we could not find a close match to available L subdwarf standards.

We used TiO and VO absorption bands in the optical region to distinguish L subdwarfs from L dwarfs, taking into account the standard/model spectral comparisons shown in fig. 10 of Paper I. This figure shows that the TiO absorption bands at around 710, 775, and 850 nm are getting stronger from the dL to sdL subclasses (from [Fe/H] = 0.0 to $-$0.5), and then getting gradually weaker from the sdL to usdL subclasses (from [Fe/H] = $-$0.5 to $-$2.0). The strong TiO absorption bands are a signature of early-type sdL, which is not the case for early type L dwarfs. The VO absorption band at 788--810 nm gets weaker from early-type dL to sdL subclasses and disappears in the esdL subclass ([Fe/H] $\la -$1.0). Furthermore, a characteristic difference between early-type sdL and early-type esdL subclasses can be seen in the 770--810 nm range, where early-type esdL spectra follow a sloping straight line but early-type sdL spectra follow a dipping curve due to stronger TiO and weaker VO absorption.

Fig. \ref{f6sdl} shows six of our new L subdwarfs that have stronger metal-poor spectral features with very weak or absent VO absorption bands at around 800 nm. Five of these have been classified as esdL, with one (UL0212+06) classified as sdL (but close to the sdL/esdL border). The VO absorption band strength of these subdwarfs is similar to the synthetic spectral models with [Fe/H] $\leq -1.0$. By comparison, Fig. \ref{f20sdl} shows 20 of our new L subdwarfs whose spectral features are indicative of slightly higher metallicity than those in Fig. \ref{f6sdl}. The spectra of these objects compared well with known sdL subdwarfs or showed typical sdL spectral features (strong TiO absorption bands).

\subsubsection{Six L subdwarfs with strong  metal-poor features}

\textit{ULAS J111429.54+072809.5} (UL1114+07) was classified esdL0, since it compares well with ULAS J033350.84+001406.1 \citep[UL0333;][]{lod12}, an esdL0 classified in Paper I.

\textit{ULAS J135216.31+312327.0} (UL1352+31), \textit{ULAS J145234.65+043738.4} (UL1452+04), and \textit{ULAS J223302.03+062030.8} (UL2233+06) were all classified esdL05 since they compare well with ULAS J124425.90+102441.9 \citep[UL1244;][]{lod12}, an esdL0.5 classified in Paper I. We also note that UL2233+06 compared best to UL1244 at 600--840 nm. And UL1352+31 has slightly more flux than UL1244 at 770--810 nm, which is indicative of a slightly lower metallicity. Since UL1244 has [Fe/H] $\approx -1.5$ (Paper I), UL1352+31 is close to the esdL/usdL boundary.

\textit{ULAS J021258.08+064115.9} (UL0212+06) was classified sdL1 since it compares well to the sdL1 subdwarf SDSS J133348.24+273508.8 (SD1333; Paper I). It may have a flatter plateau over 0.738--0.757 $\mu$m than SD1333, though this could be due to noise or telluric effects. The flux of UL0212+06 beyond 880 nm is not well calibrated due to the lack of second-order correction for the OSIRIS spectrum. Although Ul0212+06 (and SD1333) are classified as sdL they have weaker 800 nm VO absorption than any other sdL dwarfs, and thus lie close to the sdL/esdL border.

\textit{ULAS J231924.35+052524.5} (UL2319+05) has a similar spectral profile to SD1333, but has weaker TiO absorption bands and slightly more flux over 770--810 nm (with a sloping straight line morphology) indicative of a lower metallicity. We therefore classified this object as esdL1.

\subsubsection{Eighteen sdL0--8 subdwarfs}
Thirteen of our objects in Fig. \ref{f20sdl} compare well with the sdL0 subdwarf SDSS J134749.74+333601.7 (SD1347; Paper I) and were thus classified sdL0. These objects are  \textit{ULAS J011840.73+084424.7}, \textit{ULAS J023803.12+054526.1}, \textit{ULAS J075335.23+200622.4}, \textit{ULAS J082206.61+044101.8}, \textit{ULAS J123142.99+015045.4}, \textit{ULAS J124104.75$-$000531.4}, \textit{ULAS J125226.62+092920.1}, \textit{ULAS J134852.93+101611.8},  \textit{ULAS J135359.58+011856.7},  \textit{ULAS J141832.35+025323.0},  \textit{ULAS J225902.14+115602.1},  \textit{ULAS J230256.53+121310.2}, and  \textit{ULAS J230443.30+093423.9}.

\textit{ULAS J134206.86+053724.9} (UL1342+05) was classified sdL0.5, since it compares well with 2MASS J00412179+3547133 \citep[2M0041;][]{bur04a}, an sdL0.5 classified in Paper I.
 
\textit{ULAS J223440.80+001002.6} (UL2234+00) was classifed sdL1, since it compares well to the sdL1 subdwarf 2MASS J17561080+2815238 \citep[2M1756;][]{kir10}, and shows stronger TiO absorption at around 710, 780 and 850 nm than the L1 dwarf standard 2MASS J14392836+1929149 \citep[2M1439;][]{kir99a}.
 
 \textit{ULAS J134423.98+280603.8} (UL1344+28) and \textit{ULAS J144151.55+043738.5} (UL1441+04) compared well to the L4 standard 2MASS J1155009+230706 \citep[2M1155;][]{kir99a}. However, both have somewhat stronger TiO absorption at 850 nm (than this L4 dwarf) that is a signature of the sdL subclass. UL1344+28 and UL1441+04 compared well with the sdL4 subdwarf ULAS J021642.97+004005.6 (UL0216; Paper I) and were thus classified sdL4. We note that UL1344+28 has slightly weaker 850 nm TiO absorption than UL0216, and thus a slightly higher metallicity than UL0216. UL1441+04 was previously reported as a blue L1 dwarf by an independent search for L and T dwarfs \citep{maro15}.

\textit{ULAS J130710.22+151103.4} (UL1307+15) compares well to the L8 dwarf standard 2MASS J16322911+1904407 \citep[2M1632;][]{kir99a} at wavelengths below 860 nm. The metal-poor sensitivity of stronger TiO absorption bands below 860 nm become insignificant in the spectra of late-type sdL subclasses. However, metal hydride (CrH, FeH) absorption for late-type sdL subdwarfs is stronger than that in late-type L dwarfs. UL1307+15 has stronger 870 nm FeH absorption than 2M1632, indicating a lower metallicity. UL1307+15 compares well with the sdL8 subdwarf 2MASS J06453153$-$6646120 \citep[2M0645;][]{kir10}, particularly in the red optical (including the 870 nm FeH band), and it was thus classified sdL8.

\subsubsection{Three sdL subdwarfs with new subtypes}
Three objects did not closely compare with any of the L subdwarf or L dwarf standards, but showed clear evidence of L subdwarf nature.

\textit{ULAS J154638.34$-$011213.0} (UL1546$-$01) compares reasonably to the optical spectral profile of the L3 dwarf standard 2MASS J03554191+2257016 \citep[2M0355;][]{kir99a}, but has stronger TiO absorption bands at around 710  and 850 nm. Therefore, we classified this object as sdL3.

\textit{ULAS J141203.85+121609.9} (UL1412+12) and \textit{ULAS J151649.84+083607.1} (UL1516+08) both compare reasonably to the L5 dwarf standard 2MASS J15074769$-$1627386 \citep[2M1507;][]{rei00}, but have stronger TiO absorption at around 850 nm. We therefore classified these objects as sdL5. UL1516+08 has a slightly weaker 850 nm TiO band compared to UL1412+12, and likely has somewhat higher metallicity. These objects were both previously reported as blue L4 and L5 dwarfs respectively, by \citet{maro15}.

We propose UL1412+12 as an sdL5 standard, since we have obtained good-quality optical and NIR spectra with the GTC and VLT (Section \ref{snir}). Note that a previously classified sdL5 object, 2MASS J06164006$-$6407194 \citep[2M0616;][]{cus09}, was re-classified as esdL6 in Paper I. Also, ULAS J135058.85+081506.8 \citep{lod10} was originally classified as sdL5 but was re-classified as sdL3.5--4 in \citet{lod17} and as usdL3 in Paper I and III. And a suspected sdL5 subdwarf, WISEA J135501.90$-$825838.9 \citep[WI1355;][]{kir10}, seems less secure as a spectroscopic standard, since its $J-H$ and $J-K$ colours are similar to those of L5 dwarfs.

From Table \ref{tsdlm}, we can see that we still missing some spectral subtypes of L subdwarfs. These missing subtypes including sdL2, sdL6, sdL9; esdL2, esdL5, esdL8--9; and usdL2 and usdL5--9. Most of them are late L subtypes as late L subdwarfs are fainter and more difficult to identify with current facilities.

\begin{table*}
 \centering
  \caption[]{Summary of the characteristics of the spectroscopic observations made with the OSIRIS on the GTC. R500R and R300R grisms cover a wavelength range of 480-1020 nm. A 0.8 arcsec slit used for all observations. }
\label{tgtc}
  \begin{tabular}{c c l c c c r l c}
\hline
    ULAS Name  & Proposal ID  & UT date & Seeing(\arcsec)  & Airmass &  Grism  & $T_{\rm int}$ (s)   & SpT  & Std (SpT)$^{Ref.}$   \\
\hline
J111429.54+072809.5  & GTC39-12B & 2013-01-17 & 0.67 & 1.24 & R500R & 900 $\times$ 1 & esdL0 & GD 140 (DA2.2)$^3$\\
J135216.31+312327.0  & GTC46-14A & 2014-03-12 & 0.80 & 1.15 & R500R & 600 $\times$ 1 & esdL0.5  & GD 153 (DA1.2)$^1$ \\
J145234.65+043738.4  & GTC39-12B & 2013-01-20 & 1.00 & 1.14 & R500R & 900 $\times$ 1 & esdL0.5 & G191-B2B (DA.8)$^3$ \\
J223302.03+062030.8  & GTC39-12B & 2012-09-02 & 0.70 & 1.11 & R500R & 1800 $\times$ 1 & esdL0.5 & G191-B2B (DA.8)$^3$\\
J231924.35+052524.5  & GTC39-12B & 2012-09-02 & 0.60 & 1.13 & R500R & 900 $\times$ 1 & esdL1 & G191-B2B (DA.8)$^3$ \\
------------\texttt{"}------------ & GTC46-14A & 2014-07-06 & 0.90 & 1.40 & R500R & 900 $\times$ 1 & ---\texttt{"}--- & GD 248 (DA5)$^4$ \\
\hline
J011840.73+084424.7  & GTC63-13A & 2013-08-01 & 0.80 & 1.09 & R500R & 1200 $\times$ 3 & sdL0 &  Ross 640 (DZA5.5)$^3$ \\
J021258.08+064115.9  & GTC39-12B & 2012-09-01 & 0.70 & 1.08 & R500R & 1800 $\times$ 1 & sdL0.5 & Ross 640 (DZA5.5)$^3$ \\
J023803.12+054526.1  & GTC80-15A & 2015-08-23 & 0.70 & 1.18 & R300R & 900 $\times$ 1 & sdL0 & Ross 640 (DZA5.5)$^3$ \\
J075335.23+200622.4  & GTC39-12B & 2012-12-06 & 0.75 & 1.15 & R500R &  900 $\times$ 1 & sdL0 & G191-B2B (DA.8)$^3$ \\
J082206.61+044101.8  & GTC39-12B & 2012-12-06 & 0.75 & 1.11 & R500R & 900 $\times$ 1 & sdL0.5 & G191-B2B (DA.8)$^3$ \\
J123142.99+015045.4  & GTC80-15A & 2015-03-16 & 0.90 & 1.52 & R300R & 900 $\times$ 1 & sdL0 & Hilt 600 (B1)$^2$ \\
J124104.75$-$000531.4  & GTC63-13A & 2013-05-08 & 0.70 & 1.68 & R500R & 1200 $\times$ 3 & sdL0 & GD 190 (DB2) \\
J125226.62+092920.1  & GTC80-15A & 2015-03-16 & 0.70 & 1.21 & R300R & 900 $\times$ 1 & sdL0.5 & Hilt 600 (B1)$^2$ \\
J130710.22+151103.4  & GTC46-14A & 2014-07-26 & 0.50 & 1.80 & R500R & 1200 $\times$ 3 & sdL8 & GD 153 (DA1.2)$^1$ \\
J134206.86+053724.9  & GTC46-14A & 2014-07-25 & 0.80 & 1.56 & R500R & 1200 $\times$ 2 & sdL1  & GD 153 (DA1.2)$^1$ \\
J134423.98+280603.8  & GTC63-13A & 2013-04-08 & 0.90 & 1.06 & R500R & 1200 $\times$ 3 & sdL4 & GD 190 (DB2)$^3$ \\
J134852.93+101611.8  & GTC63-13A & 2013-03-19 & 0.90 & 1.37 & R500R &  900 $\times$ 1 & sdL0 &  L 1363-3 (DQ6)$^3$\\
J135359.58+011856.7  & GTC46-14A & 2014-07-28 & 0.90 & 1.73 & R500R & 900 $\times$ 3 & sdL0  & GD 153 (DA1.2)$^1$ \\
J141203.85+121609.9  & GTC39-12B & 2013-01-20 & 1.00 & 1.09 & R500R & 1800 $\times$ 1 & sdL5 & G191-B2B (DA.8)$^3$ \\
J141832.35+025323.0  & GTC39-12B & 2013-01-17 & 0.67 & 1.12 & R500R & 900 $\times$ 1 & sdL0 & GD 140 (DA2.2)$^3$ \\
------------\texttt{"}------------  & GTC46-14A & 2014-03-12 & 0.80 & 1.19 & R500R & 600 $\times$ 1 & ---\texttt{"}---  & GD 153 (DA1.2)$^1$\\
J144151.55+043738.5  & GTC46-14A & 2014-07-19 & 0.60 & 1.34 & R500R & 1200 $\times$ 3 & sdL4  & GD 153 (DA1.2)$^1$ \\
J151649.84+083607.1  & GTC46-14A & 2014-07-25 & 0.70 & 1.36 & R500R & 1200 $\times$ 3 & sdL5  & GD 153 (DA1.2)$^1$ \\
J154638.34$-$011213.0  & GTC63-13A & 2013-05-07 & 0.90 & 1.17 & R500R & 1200 $\times$ 3 & sdL3 &  L 1363-3 (DQ6)$^3$ \\
J223440.80+001002.6  & GTC63-13A & 2013-07-17 & 0.80 & 1.15 & R500R & 1200 $\times$ 3 & sdL1 &  L 1363-3 (DQ6)$^3$ \\
J225902.14+115602.1  & GTC63-13A & 2013-07-18 & 0.95 & 1.18 & R500R & 900 $\times$ 2 & sdL0  & GD 140 (DA2.2)$^3$ \\
------------\texttt{"}------------  & GTC46-14A & 2014-07-20 & 0.60 & 1.06 & R500R & 1200 $\times$ 2 & ---\texttt{"}---  & GD 153 (DA1.2)$^1$ \\
J230256.53+121310.2  & GTC63-13A & 2013-07-18 & 0.80 & 1.08 & R500R & 1350 $\times$ 2 & sdL0 & GD 190 (DB2)$^3$ \\
J230443.30+093423.9  & GTC63-13A & 2013-07-13 & 0.70 & 1.08 & R500R & 1350 $\times$ 2 & sdL0 & LDS 749B (DB4)$^4$ \\
\hline
J233227.03+123452.0  & GTC46-14A & 2014-08-24 & 0.80 & 1.07 & R500R & 1200 $\times$ 3 & L6p+T4p  & G158-100 (DC)$^4$ \\
\hline
J135122.15+141914.9  & GTC63-13A & 2013-05-08 & 0.70 & 1.53 & R500R &  900 $\times$ 1 & esdM7 & GD 190 (DB2)$^3$ \\
J002009.35+160451.2  & GTC63-13A & 2013-07-31 & 0.80 & 1.69 & R500R & 900 $\times$ 1 & sdM9 & GD 190 (DB2)$^3$\\
J010756.85+100811.3  & GTC63-13A & 2013-08-09 & 0.90 & 1.54 & R500R &  900 $\times$ 2 & sdM7 & GD 140 (DA2.2)$^3$ \\
J020628.22+020255.6  & GTC39-12B & 2012-08-28 & 0.60 & 1.12 & R500R & 900 $\times$ 1 &  sdM7 & Ross 640 (DZA5.5)$^3$ \\
J024035.36+060629.3  & GTC39-12B & 2012-09-01 & 0.70 & 1.09 & R500R & 1800 $\times$ 1 & sdM7 & Ross 640 (DZA5.5)$^3$ \\
J143154.18$-$004114.3  & GTC46-14A & 2014-07-13 & 0.65 & 1.23 & R500R & 900 $\times$ 1 & sdM9 & Ross 640 (DZA5.5)$^3$ \\
J143517.18$-$014713.1  & GTC63-13A & 2013-05-08 & 0.70 & 1.67 & R500R & 900 $\times$ 1 & sdM9 & GD 190 (DB2)$^3$\\
\hline
J001747.60+130757.1  & GTC63-13A & 2013-07-30 & 0.70 & 1.15 & R500R & 1500 $\times$ 2 & galaxy & --- \\
J001837.37+020015.7  & GTC63-13A & 2013-07-30 & 0.90 & 1.13 & R500R & 1200 $\times$ 3 & M9 &  L 1363-3 (DQ6)$^3$ \\
J001931.33+063111.0  & GTC39-12B & 2012-08-26 & 0.65 & 1.23 & R500R & 900 $\times$ 3 & M9 & G191-B2B (DA.8)$^3$ \\
J004602.85+091131.2  & GTC63-13A & 2013-08-01 & 0.80 & 1.48 & R500R & 1500 $\times$ 2 & M9 & Ross 640 (DZA5.5)$^3$ \\
J004716.65+161242.4  & GTC63-13A & 2013-07-31 & 0.70 & 1.60 & R500R & 1350 $\times$ 2 & M9 & GD 190 (DB2)$^3$ \\
J011711.98$-$005213.4  & GTC63-13A & 2013-07-29 & 0.90 & 1.18 & R500R &  900 $\times$ 1 & M7 & GD 190 (DB2)$^3$ \\
J125938.50+301500.2  & GTC63-13A & 2013-05-08 & 0.70 & 1.48 & R500R & 900 $\times$ 1 & M9 & GD 190 (DB2)$^3$\\
J134505.85+342441.8  & GTC63-13A & 2013-03-17 & 0.80 & 1.16 & R500R & 900 $\times$ 1 & M9 &  L 1363-3 (DQ6)$^3$\\
------------\texttt{"}------------ & GTC46-14A & 2014-07-13 & 0.60 & 1.20 & R500R & 900 $\times$ 1 & --\texttt{"}-- & Ross 640 (DZA5.5)$^3$ \\
J205721.89+005628.7  & GTC63-13A & 2013-05-07 & 0.70 & 1.33 & R500R & 1350 $\times$ 2 & M7 & GD 140 (DA2.2)$^3$ \\
J214816.13+012225.1  & GTC63-13A & 2013-07-16 & 0.80 & 1.13 & R500R & 1500 $\times$ 2 & M7 & GD 140 (DA2.2)$^3$ \\
J223123.44+010025.1  & GTC63-13A & 2013-07-13 & 0.60 & 1.15 & R500R & 900 $\times$ 1 & M6 & LDS 749B (DB4)$^4$ \\
J223623.17+034344.5  & GTC63-13A & 2013-07-05 & 0.80 & 1.39 & R500R & 1200 $\times$ 3 & M9 & GD 190 (DB2)$^3$ \\
J224054.61+030902.0  & GTC63-13A & 2013-07-05 & 0.60 & 1.11 & R500R & 1200 $\times$ 1 & M7 & GD 140 (DA2.2)$^3$ \\
J224749.77+053207.9  & GTC46-14A & 2014-07-19 & 0.70 & 1.12 & R500R & 1200 $\times$ 3 & L1  & GD 153 (DA1.2)$^1$ \\
J231949.36+044559.5  & GTC63-13A & 2013-07-16 & 0.80 & 1.10 & R500R & 1200 $\times$ 1 & M7 & GD 140 (DA2.2)$^3$\\
------------\texttt{"}------------  & GTC46-14A & 2014-07-20 & 0.60 & 1.21 & R500R & 1200 $\times$ 2 & --\texttt{"}--  & GD 153 (DA1.2)$^1$ \\
J233211.22+045554.2  & GTC63-13A & 2013-08-08 & 0.80 & 1.18 & R500R & 900 $\times$ 2 & M6  & Ross 640 (DZA5.5)$^3$ \\
\hline
\end{tabular}
\begin{list}{}{}
\item[Notes. References for flux calibration standard stars are 1. \citet{bohl95}; 2. \citet{hamu94}; 3. \citet{oke74}; 4. \citet{oke90}.] 
\end{list}
\end{table*}

\begin{table*}
 \centering
  \caption[]{Summary of the characteristics of the spectroscopic observations made with X-shooter.  Wavelength ranges for the VIS and NIR arms are 530--1020 and 990--2480 nm. 1.2 arcsec slits are used for both VIS and NIR arms. DENIS-P J025503.3$-$470049, 2MASS J08354256$-$0819237 and DENIS-P J144137.3$-$4094559 were observed as RV standards.}
\label{tvltob}
  \begin{tabular}{l l c c c r r c c}
\hline
    Name  & SpT &  UT date & Seeing & Airm & $T_{\rm int}$(VIS) &    $T_{\rm int}$(NIR) & Telluric (SpT) & Airm \\ 
\hline
ULAS  J024035.36+060629.3  & sdM7 &  2015-01-17 & 0.97\arcsec & 1.42 &  12$\times$237 s   &  12$\times$250 s & HD 22686 (A0 V) & 1.27 \\
ULAS J141203.85+121609.9  & sdL5 &  2015-02-25  & 0.96\arcsec & 1.26 &   12$\times$287 s  &  12$\times$300 s & HIP 76069 (B9 V) & 1.33 \\
ULAS J233227.03+123452.0  & L6p+T4p &  2015-09-10 & 1.03\arcsec & 1.28 &   12$\times$285 s  & 12$\times$300 s & HIP 117927 (B9V) & 1.20  \\
ULAS J223302.03+062030.8  & esdL0.5 &  2015-09-11 & 1.20\arcsec & 1.24 &   12$\times$285 s   &  12$\times$300 s & HIP 105164 (B7 III) & 1.09  \\
ULAS J130710.22+151103.4  & sdL8 &  2018-04-23 & 1.03\arcsec & 1.34 &   12$\times$285 s  & 12$\times$300 s & HIP 61257 (B9.5 V) & 1.15  \\
2MASS J06453153$-$6646120  & sdL8 &  2016-02-19 & 1.09\arcsec & 1.35 &    8$\times$290 s &  8$\times$300 s  & HD 77281 (A3 IV) & 1.26 \\
DENIS-P J025503.3$-$470049 & L9 & 2015-08-12  & 1.57\arcsec & 1.44 &   8$\times$140 s  & 8$\times$150 s & HIP 105164 (B7 III) & 1.36\\
2MASS J08354256$-$0819237 & L5 &  2015-10-31 & 0.62\arcsec & 1.31 &   8$\times$140 s & 8$\times$150 s & HIP 38734 (B9 V) & 1.33 \\
DENIS-P J144137.3$-$094559 & L0.5 &  2016-03-19 & 0.66\arcsec & 1.11 &   8$\times$140 s  & 8$\times$150 s & HIP 76836 (B9.5 V) & 1.04  \\
\hline
\end{tabular}
\end{table*}

\subsection{NIR constraints}
\label{snir}
L dwarfs/subdwarfs have rather diverse NIR spectral features because they are distributed into a wide range of metallicity and gravity. Objects that are more metal-poor have lower opacity, thus have more flux at shorter wavelength, and tend to be more compact (i.e. higher gravity). NIR spectral fluxes are suppressed gradually as metallicity decreases due to enhanced CIA H$_2$ (see fig. 9 in Paper I). The optical-NIR morphology of L subdwarfs is thus very characteristic. And within the NIR itself, it is relatively easier to distinguish sdL from dL subclass using NIR spectral features (which are more significant than optical features such as TiO absorptions for sdL subclasses). NIR spectra are also very useful for distinguishing usdL from esdL subclasses because the spectral variations caused by the metallicity differences are larger in the NIR than in the optical.

UL2233+06 was observed under our X-shooter follow-up programme. The spectrum of this object has SNR of $\sim$8 at 820 nm and $\sim$6 at 1300 nm, and was smoothed by 101 (VIS) and 51 (NIR) pixels for display. Fig. \ref{ful2233} shows the optical to NIR spectrum of UL2233+06 compared (via optical nomalization) to those of the sdL0.5 subdwarf 2MASS J00412179+3547133 (2M0041; \citealt{bur04a}; Paper I) and the L0 dwarf 2MASP J0345432+254023 \citep[SD0345;][]{kir99a}. UL2233+06 has much stronger NIR flux suppression than the sdL0.5 subdwarf, and its CO absorption is absent. This is entirely consistent with the esdL0.5 classification of UL2233+06, and the dL--sdL--esdL sequence clearly shows the gradually increasing level of NIR suppression.

Note that Fig. \ref{ful2233} does not show the true relative flux of these subclasses. Fig. 16 in Paper I shows that esdL and dL subclasses with close spectral type have similar $H$ band absolute magnitudes. Thus relative flux is best shown by normalising spectra in the $H$ band, as in Fig. \ref{ful2233nh}. This figure shows the same three objects from Fig. \ref{ful2233} but with the additional L0$\gamma$ dwarf 2MASS J01415823$-$4633574 \citep[2M0141;][]{kir06}. Although these objects all have spectral subtype L0--L0.5 (classified in optical) they have very different physical parameters. UL2233+06 is older with lower metallicity, but relatively higher mass and $T_{\rm eff}$. While at the other extreme the L0$\gamma$ dwarf is young with higher metallicity, cooler $T_{\rm eff}$, and much lower mass and gravity \citep{fili15,fahe16}.

UL1412+12 was also observed by our X-shooter follow-up programme. The original spectrum has SNR of $\sim$6 at 910 nm and $\sim$30 at 1300 nm, and was smoothed by 101 (VIS) and 51 (NIR) pixels for display. Fig. \ref{ul1412} shows that UL1412+12 has largely suppressed NIR flux and a weaker CO absorption band compared to the L5 dwarf standard SDSS J083506.16+195304.4 \citep[SD0835;][]{chiu06,kir10}. This is as expected for the sdL5 classification (from the optical).

We obtained the optical to NIR spectra of UL1307+15 and 2M0645 under our X-shooter follow-up programme. The original spectrum of UL1307+15 has SNR of $\sim$ 4 at 900 nm and $\sim$ 7 at 1300 nm. The original spectrum of 2M0645 has SNR of $\sim$20 at 900 nm and $\sim$45 at 1300 nm. Fig. \ref{ul1307fit} shows that UL1307+15 has a similar spectral profile to 2M0645 from optical to $H$ band. We note that UL1307+15 has more $K$ band flux and slightly narrower $H$ band spectral profile than 2M0645. UL1307+15 is worth of further investigation (e.g. on binarity).
 
\begin{figure*}
\begin{center}
   \includegraphics[width=\textwidth]{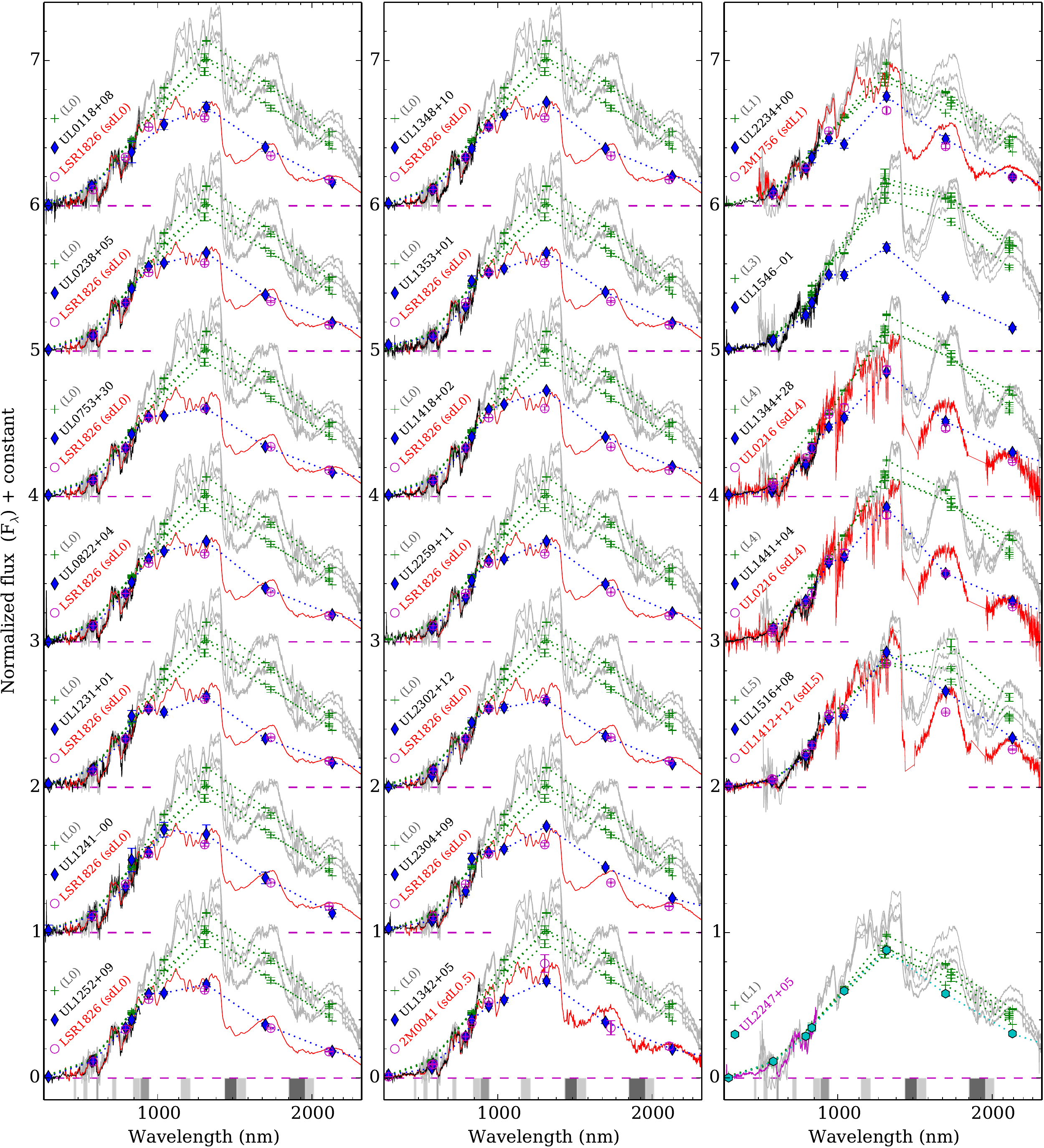}
\caption[]{Optical--NIR photometric spectral energy distributions of 19 new L subdwarfs (black lines and blue diamonds) and an L1 dwarf (UL2247+05; magenta line and cyan hexagons), compared to those of L dwarfs (grey lines and green crosses) and known L subdwarfs (red lines and magenta open circles). The spectrum of LSR1826 is from \citet{bur04a}. The spectrum of 2M1756 is from \citet{kir10}. The spectrum of UL0216 is from Paper I. Spex spectra of L0--L5 dwarfs used for comparison are of 2MASS J13313310+3407583 (L0), 2MASS J01340281+0508125 \citep[L0;][]{kir10}; 2MASSI J2107316$-$030733 (L0), SDSS J202820.32+005226.5 (L3), 2MASSI J1104012+195921 \citep[L4;][]{bur04a}; 2MASSW J0228110+253738 (L0), 2MASS J12212770+0257198 (L0), 2MASSW J0208183+254253 (L1), SDSS J104842.84+011158.5 \citep[L1;][]{bur08a}; 2MASS J20343769+0827009 (L1), 2MASS J08234818+2428577 (L3), 2MASSW J1146345+223053 (L3), 2MASS J22425317+2542573 (L3), 2MASS J13571237+1428398 (L4), 2MASS J13571237+1428398 (L4), 2MASS J17461199+5034036 \citep[L5;][]{burg10}; SDSSp J053951.99$-$005902.0 [L5; M. Cushing (unpublished); \citealt{fan00}]; SD0835 \citep[L5;][]{chiu06}.}
\label{fsdlsed}
\end{center}
\end{figure*}

To confirm the subdwarf status of the other 19 L subdwarfs (Fig. \ref{f20sdl}) without NIR spectra, we instead compared their optical to NIR photometric spectral energy distributions (SEDs) to those of known L dwarfs and subdwarfs with the same optical spectral subtypes (Fig. \ref{fsdlsed}). These 19 L subdwarfs all show significantly suppressed NIR SEDs compared to those of the L dwarfs. 

The 13 new L0 subdwarfs have similar SEDs to LSR J182611.3+301419.1 \citep[LSR1826;][]{lep02}, which was classified as sdL0 in Paper I. UL1342+05 shows a similar SED to the sdL0.5 2M0041. Although there is not sdL3 subdwarf to compare with UL1546$-$01, it can be seen that the UL1546$-$01 SED is more suppressed relative to the L3 dwarfs, and by a level similar to the suppression amongst the sdL0s (when compared to the L0 dwarfs). UL1344+28 and UL1441+04 have similar SEDs to the sdL4 Ul0216. UL1412+12 has a slightly stronger NIR suppression than UL1516+08 indicating a slightly lower metallicity. And this is also supported by its slightly stronger 850 nm TiO absorption band when compared to UL1516+08. 

\subsection{Astrometry and radial velocity}
\label{sam}
One of our new L subdwarfs (UL0753+20) was observed by the {\sl Gaia} \citep{gaia18} astrometric survey. Eleven of our new L subdwarfs (including UL0753+20) were observed by the UKIDSS second epoch survey, and are therefore included in the UKIDSS proper motion catalogue of \citet{smit14}. Proper motions of the rest of our objects were measured from SDSS, UKIDSS and PS1 epochs following the procedure described in \citet{zha09}. We estimated spectroscopic distances for these 27 new L subdwarfs based on the relationship between spectral type and $H$ band absolute magnitude, which is less sensitive to metallicity/subclass for L subdwarfs.  The relationship was derived from 20 L subdwarfs observed by the {\sl Gaia} mission \citep[][see Section \ref{sgaia}]{gaia16}. 
Table \ref{tam} shows the proper motions, distances and tangential velocities of these 27 new L subdwarfs. They are distributed at distances from $\sim$ 50 to 250 pc. Seventeen of them have tangential velocities higher than 100 km s$^{-1}$. Seven of them have tangential velocities above 200 km s$^{-1}$, which includes these six L subdwarfs with the strongest metal-poor features shown in Fig. \ref{f6sdl}.

We note that UL0212 (the sdL in Figure \ref{f6sdl} that is close to the sdL/esdL border) has tangential velocity of 329$^{+71}_{-59}$ km s$^{-1}$, which is similar to the space motions of the five esdL subdwarfs in Fig. \ref{f6sdl}. Thus the most metal-poor objects of the sdL subclass (UL0212 and SD1333) both have typical halo kinematics.

We measured the radial velocities (RV) of UL2233+06, UL1412+12, UL1307+15, and 2M0645 based on their X-shooter spectra. To measure the RVs of L subdwarfs in our X-shooter follow-up programme, we also observed three RV standard L dwarfs [DENIS-P J144137.3$-$094559 \citep{mart99}, an L0.5 dwarf with an RV of $-27.9\pm1.2$ km s$^{-1}$ \citep{bail04}; 2MASS J08354256$-$0819237 \citep{cruz03}, an L5 dwarf with an RV of $29.89\pm0.06$ km s$^{-1}$ \citep{blak10}; and DENIS-P J025503.3$-$470049 \citep{mart99}, an L9 dwarf with an RV of $17.5\pm2.8$ km s$^{-1}$ \citep{zapa07}]. All X-shooter spectra were smoothed by 21 and 11 pixels in the VIS and NIR respectively, to increase their SNR. We first measured the RV differences between our objects and the RV standards with the closest spectral subtype using cross-correlation on their strong absorption lines (Na I and Cs I in the optical and K I in the NIR). Measured RV differences were then corrected for barycentric effects, and converted into final RV using the known RVs of the standards. The final RVs of UL2233+06, UL1412+12, UL1307+15, and 2M0645 are $-164\pm15$, $-57\pm12$, $-60\pm$18, and $-33\pm10$ km s$^{-1}$, respectively.

The esdL+usdL and the sdL subclasses are approximately kinematically associated with the halo population and thick disc populations, respectively. The ratio between the number of esdL+usdL and sdL subdwarfs in our UKIDSS-SDSS sample is much higher than that between halo and thick disc stars measured by \citet{redd06}. This is likely due to our selection bias, as esdL or usdL subdwarfs can more easily be picked out by their extreme colours. There are likely a lot more sdL subdwarfs observed by existing photometric surveys, but not yet confirmed by spectroscopy. This arises because sdL subdwarf identification is more problematic than esdL and usdL, due to contamination from scattered M and L dwarfs in colour--colour diagrams. There are 66 L subdwarfs known to-date, including 7 usdL, 20 esdL, and 39 sdL subdwarfs. Meanwhile, there are a few thousand photometric candidates \citep[e.g.][]{skrz16} and spectroscopically confirmed L dwarfs to-date \citep[e.g.][]{best18}. The ratio between known esdL+usdL subdwarfs and L dwarfs is comparable to the fraction of halo population in the solar neighbourhood \citep[e.g. 0.6\%;][]{redd06}.

\begin{figure}
\begin{center}
   \includegraphics[width=\columnwidth]{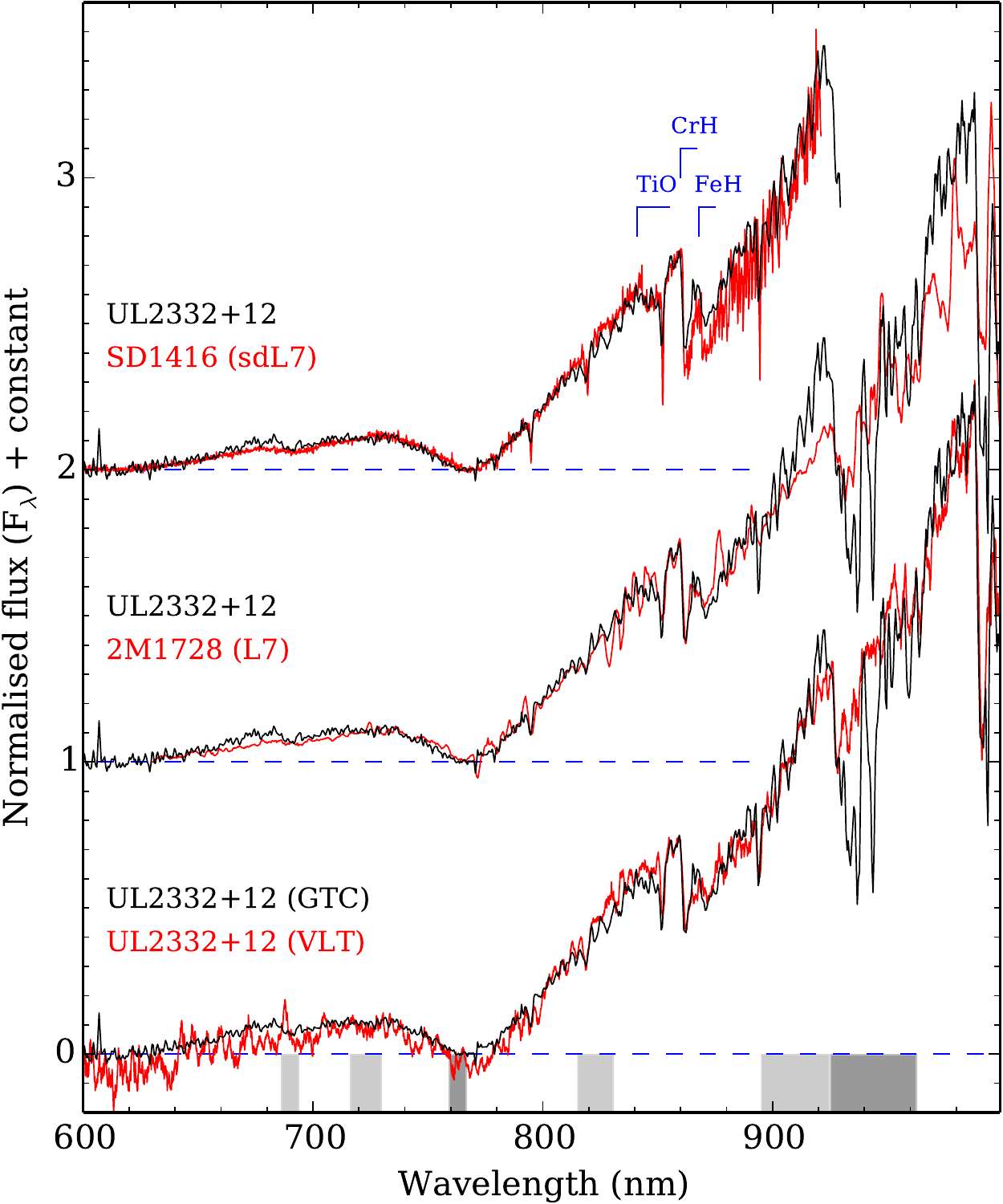}
\caption[]{Optical spectra of UL2332+12 observed with GTC and VLT compared to those of SD1416 (from SDSS) and 2M0850 \citep[from][]{kir99a}. Spectra are normalized at 910 nm.}
\label{ul2332op}
\end{center}
\end{figure}

\begin{figure}
\begin{center}
  \includegraphics[width=\columnwidth]{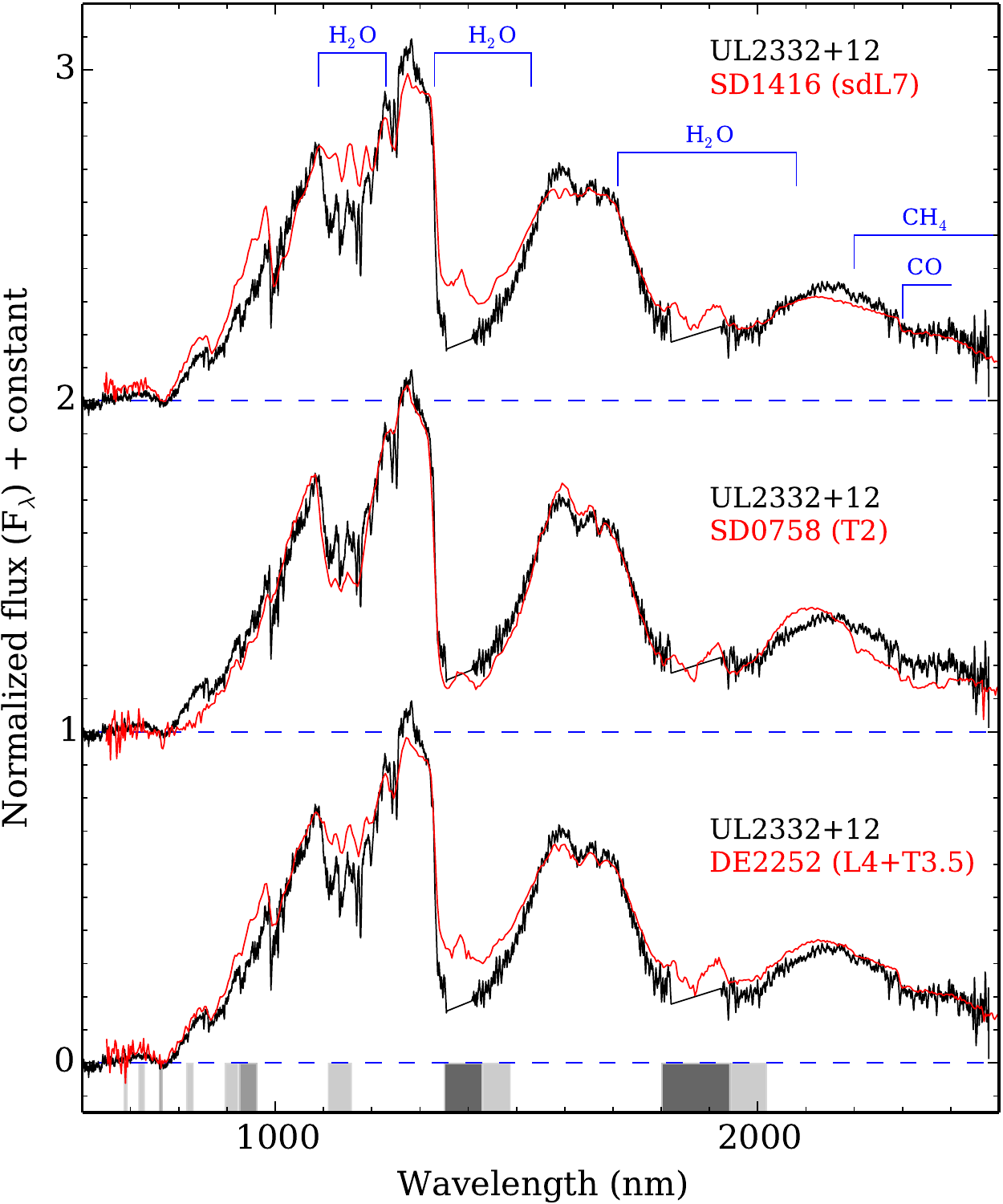}
\caption[]{Optical--NIR spectrum of UL2332+12 compared to those of SD1416 \citep[from][]{sch10}, SD0758 \citep[from][]{bur08c} and DE2252 \citep[from][]{reid06}.
 Spectra are normalized at 1300 nm. }
\label{ul2332xs}
\end{center}
\end{figure}

\begin{figure}
\begin{center}
  \includegraphics[width=\columnwidth]{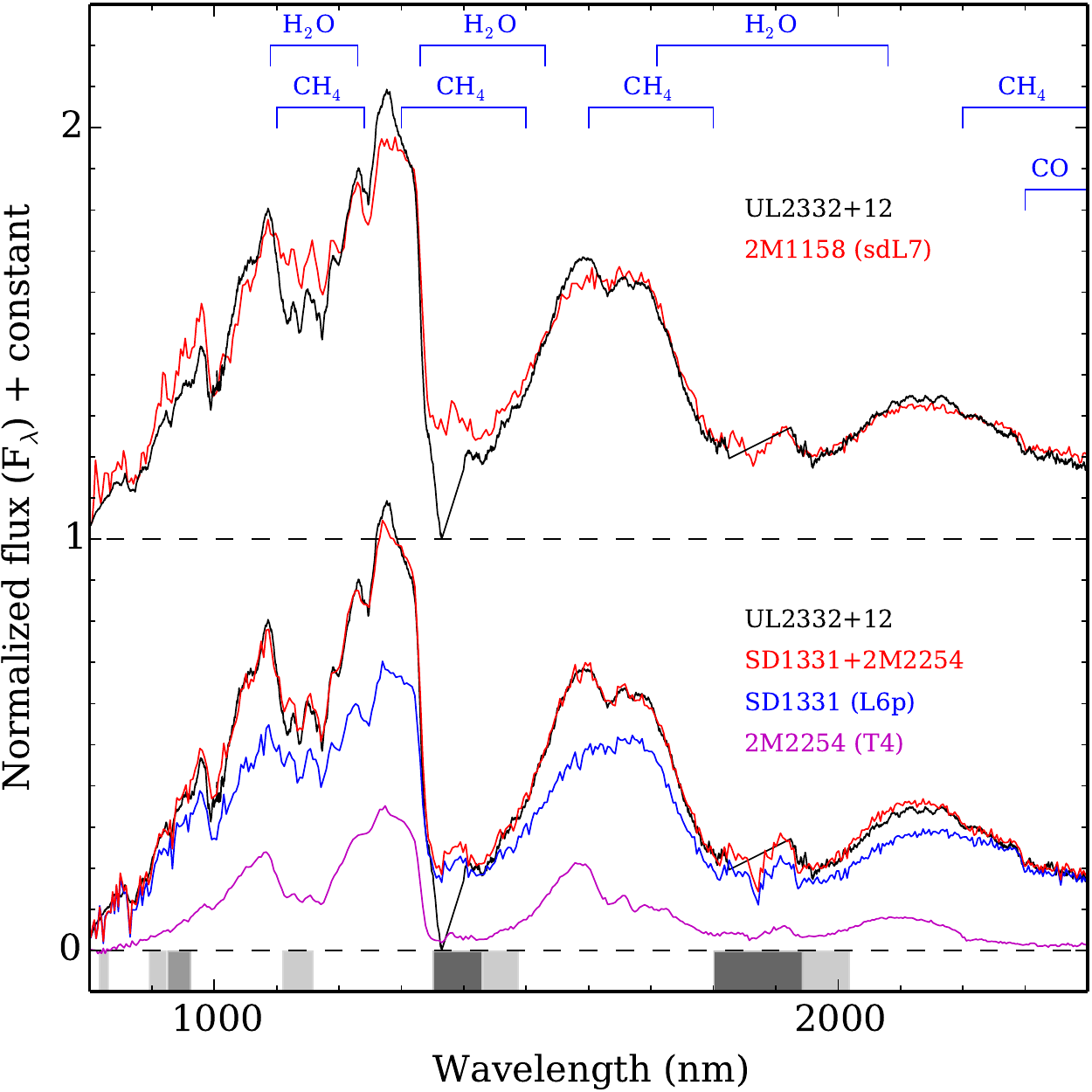}
\caption[]{The closest matching single object to UL2332+12 (top panel), and the best-fitting synthesized spectroscopic binary combination (bottom panel). The X-shooter spectrum of UL2332+12 (black) were smoothed by 401 (VIS) and 201 (NIR) pixels. The red spectrum in the bottom panel is synthesized from the L6 (blue) and T4 (magenta) dwarfs. Telluric absorption regions are excluded in our fits (indicated with grey bands). The spectrum of 2MASS J11582077+0435014 is from \citet{kir10}. The spectrum of SDSS J133148.92$-$011651.4 is from \citet{burg10}. The spectrum of 2MASSI J2254188+312349 (2M2254) is from \citet{bur04a}.}
\label{ul2332fit}
\end{center}
\end{figure}

\begin{figure}
\begin{center}
  \includegraphics[width=\columnwidth]{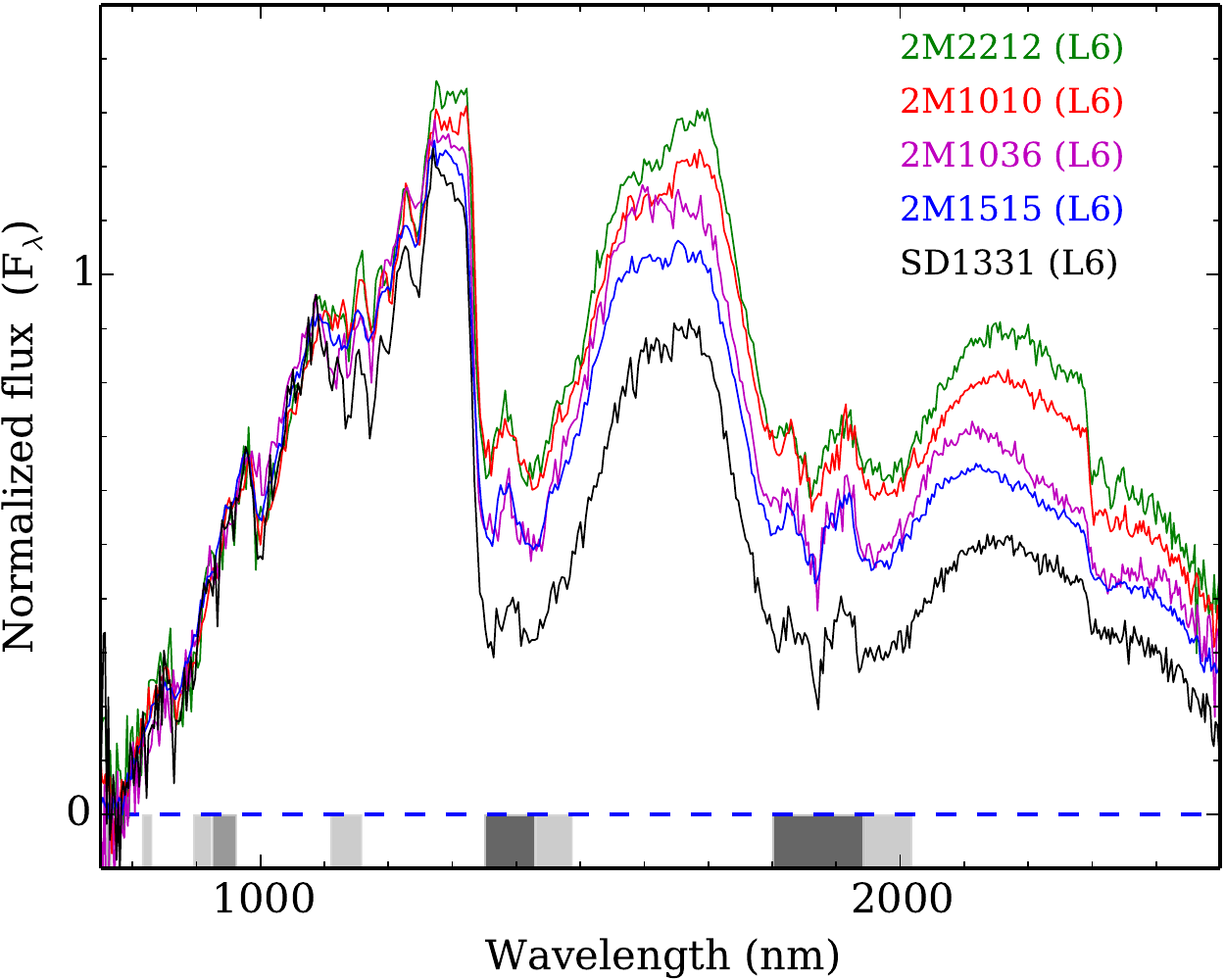}
\caption[]{NIR spectrum of SD1331 compared to those of the L6 dwarfs from \citet[][2MASS J22120703+3430351; 2MASS J10365305$-$3441380; 2MASS J15150083+4847416]{burg10} and \citet[][2MASS J10101480$-$0406499]{reid06}.}
\label{sd1331}
\end{center}
\end{figure}

\begin{table}
 \centering
  \caption[]{Proper motions, distances and tangential velocities of our 27 new L subdwarfs.   }
 \label{tam}
  \begin{tabular}{l r r r r}
\hline
    Name  & $\mu_{\rm RA}$~ & $\mu_{\rm Dec}$~  & Distance & $V_{\rm tan}$~~ \\
      & (mas/yr)  & (mas/yr)  & (pc)~~ & (km s$^{-1}$) \\
\hline
UL0118+03 & 14$\pm$12 & $-$80$\pm$12 & 209$^{+20}_{-18}$ & 80$^{+18}_{-18}$ \\
UL0212+06 & 2$\pm$4 & $-$429$\pm$5 & 139$^{+13}_{-12}$ & 282$^{+27}_{-25}$ \\
UL0238+05 & 186$\pm$18 & $-$145$\pm$15 & 95$^{+9}_{-8}$ & 106$^{+15}_{-14}$ \\
UL0753+20 & $-$36$\pm$1 & $-$191$\pm$1 & 74$^{+7}_{-6}$ & 68$^{+7}_{-6}$ \\
UL0822+04 & 35$\pm$7 & $-$154$\pm$5 & 92$^{+9}_{-8}$ & 69$^{+8}_{-7}$ \\
UL1114+07 & $-$17$\pm$9 & $-$306$\pm$6 & 211$^{+20}_{-18}$ & 306$^{+31}_{-29}$ \\
UL1231+01 & $-$225$\pm$6 & 22$\pm$4 & 164$^{+16}_{-14}$ & 176$^{+18}_{-16}$ \\
UL1241$-$00 & $-$61$\pm$9 & $-$45$\pm$5 & 245$^{+23}_{-21}$ & 89$^{+15}_{-14}$ \\
UL1252+09 & $-$299$\pm$7 & 6$\pm$6 & 117$^{+11}_{-10}$ & 166$^{+17}_{-15}$ \\
UL1307+15 & $-$391$\pm$14 & $-$124$\pm$13 & 86$^{+8}_{-8}$ &168$^{+18}_{-17}$ \\
UL1342+05 & $-$106$\pm$8 & $-$212$\pm$5 & 142$^{+14}_{-12}$ & 160$^{+17}_{-15}$ \\
UL1344+28 & $-$345$\pm$11 & 120$\pm$10 & 88$^{+8}_{-8}$ & 153$^{+16}_{-15}$ \\
UL1348+10 & $-$283$\pm$11 & $-$179$\pm$9 & 109$^{+10}_{-10}$ & 174$^{+18}_{-17}$ \\
UL1352+31 & 542$\pm$8 & $-$111$\pm$29 & 148$^{+14}_{-13}$ & 388$^{+42}_{-40}$ \\
UL1353+01 & $-$93$\pm$12 & $-$37$\pm$10 & 142$^{+14}_{-12}$ & 67$^{+12}_{-12}$ \\
UL1412+12 & $-$227$\pm$10 & $-$3$\pm$10 & 54$^{+5}_{-5}$ & 58$^{+7}_{-6}$ \\
UL1418+02 & $-$3$\pm$7 & $-$277$\pm$6 & 79$^{+8}_{-7}$ & 103$^{+10}_{-10}$ \\
UL1441+04 & $-$264$\pm$8 & $-$508$\pm$8 & 98$^{+9}_{-9}$ & 266$^{+26}_{-24}$ \\
UL1452+04 & $-$46$\pm$9 & $-$291$\pm$11 & 157$^{+15}_{-14}$ & 220$^{+23}_{-22}$ \\
UL1516+08 & $-$171$\pm$8 & 44$\pm$5 & 79$^{+8}_{-7}$ & 66$^{+7}_{-7}$ \\
UL1546$-$01 & $-$45$\pm$7 & $-$115$\pm$6 & 122$^{+12}_{-11}$ & 71$^{+9}_{-8}$ \\
UL2233+06 & 320$\pm$6 & $-$80$\pm$5 & 236$^{+22}_{-20}$ & 370$^{+36}_{-33}$ \\
UL2234+00 & $-$34$\pm$9 & $-$96$\pm$7 & 145$^{+14}_{-13}$ & 70$^{+11}_{-10}$ \\
UL2259+11 & 34$\pm$19 & $-$115$\pm$13 & 126$^{+12}_{-11}$ & 72$^{+15}_{-15}$ \\
UL2302+12 & $-$123$\pm$7 & $-$191$\pm$6 & 153$^{+15}_{-13}$ & 165$^{+17}_{-16}$ \\
UL2304+09 & $-$21$\pm$13 & $-$46$\pm$4 & 133$^{+13}_{-12}$ & 32$^{+9}_{-9}$ \\
UL2319+05 & 509$\pm$11 & $-$401$\pm$11 & 167$^{+16}_{-14}$ & 513$^{+50}_{-46}$ \\
\hline
\end{tabular}
\end{table}

\begin{table}
 \centering
  \caption[]{Properties of a spectral binary UL2332+12. Note the spectral type is from synthesized fit.}
\label{prop}
  \begin{tabular}{l c  }
\hline
Parameter & UL2332+12   \\	
\hline 
 $\alpha$ (J2000) & $23^h32^m27\fs03$  \\
 $\delta$ (J2000) &  $+12\degr34\arcmin52\farcs0$  \\
Epoch & 2009-09-04   \\
SDSS $i$  & 22.64$\pm$0.27  \\
SDSS $z$ &  19.91$\pm$0.11 \\
PS1 $i$ & 22.00$\pm$0.17  \\
PS1 $z$ & 20.23$\pm$0.04  \\
PS1 $y$ & 19.13$\pm$0.03   \\
UKIDSS $Y$  & 18.10$\pm$0.04  \\
UKIDSS $J$  &   16.90$\pm$0.02   \\
UKIDSS $H$ & 16.39$\pm$0.03  \\
UKIDSS $K$ & 15.88$\pm$0.03  \\
$WISE ~ W1$  & 15.18$\pm$0.04  \\
$WISE ~ W2$ &  14.77$\pm$0.07   \\
Spectral type & L6p+T4p  \\
Distance (pc) &  58$^{+12}_{-10}$  \\ 
$\mu_{\rm RA}$(mas/yr) &  420$\pm$10   \\
$\mu_{\rm Dec}$(mas/yr) &  $-$107$\pm$13   \\
$V_{tan}$ (km s$^{-1}$) &  118$^{+25}_{-21}$  \\
RV (km s$^{-1}$) &  $-35\pm8$   \\
$U$ (km s$^{-1}$) & $-85\pm18$  \\ 
$V$ (km s$^{-1}$) & $-86\pm14$   \\ 
$W$ (km s$^{-1}$) & $-28\pm13$   \\ 
\hline
\end{tabular}
\end{table}

\begin{figure}
\begin{center}
  \includegraphics[width=\columnwidth]{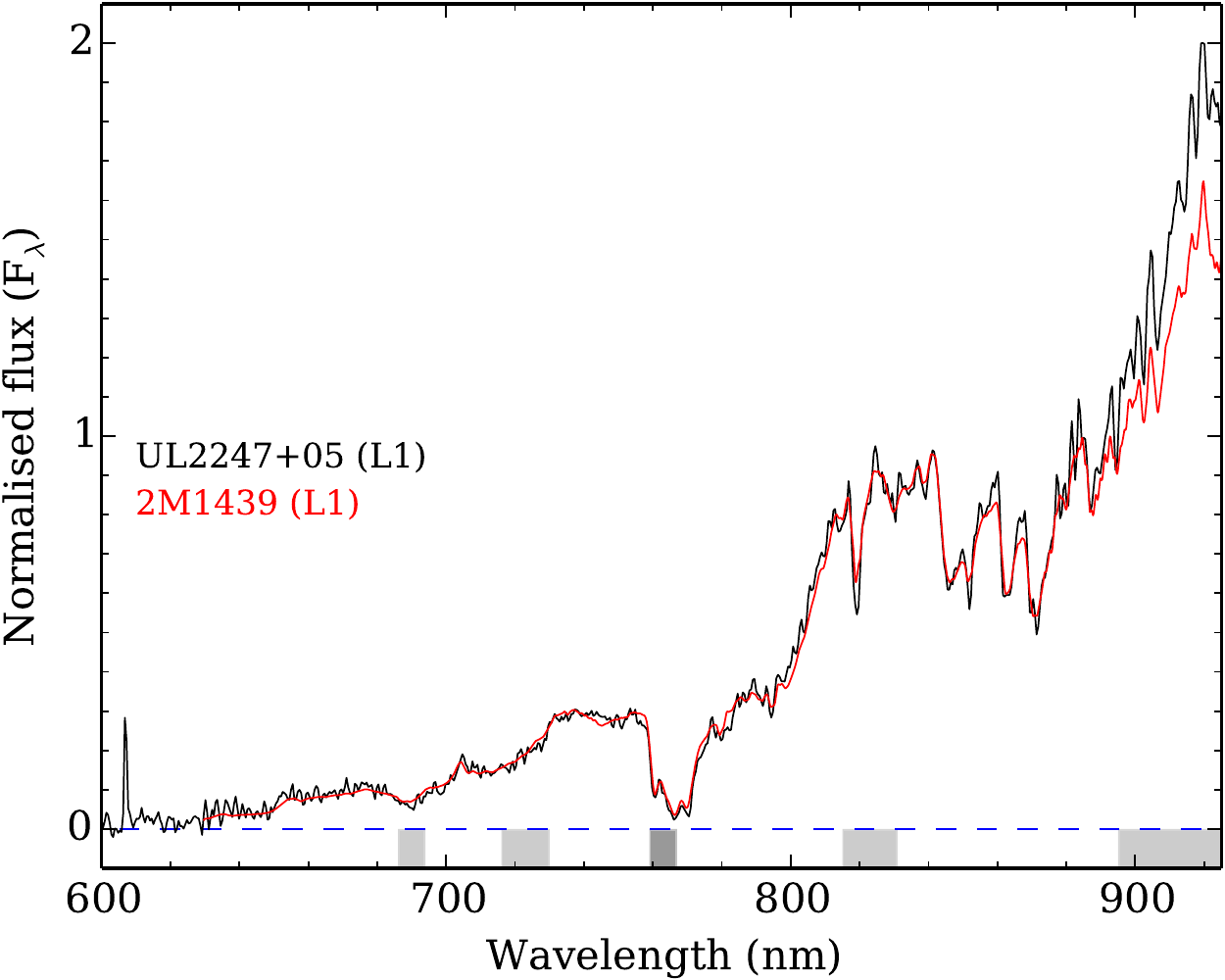}
\caption[]{The optical spectrum of UL2247+05 compared to the L1 2M1439 \citep{kir99a}.  }
\label{ul2247op}
\end{center}
\end{figure}

\begin{figure*}
\begin{center}
   \includegraphics[width=\textwidth]{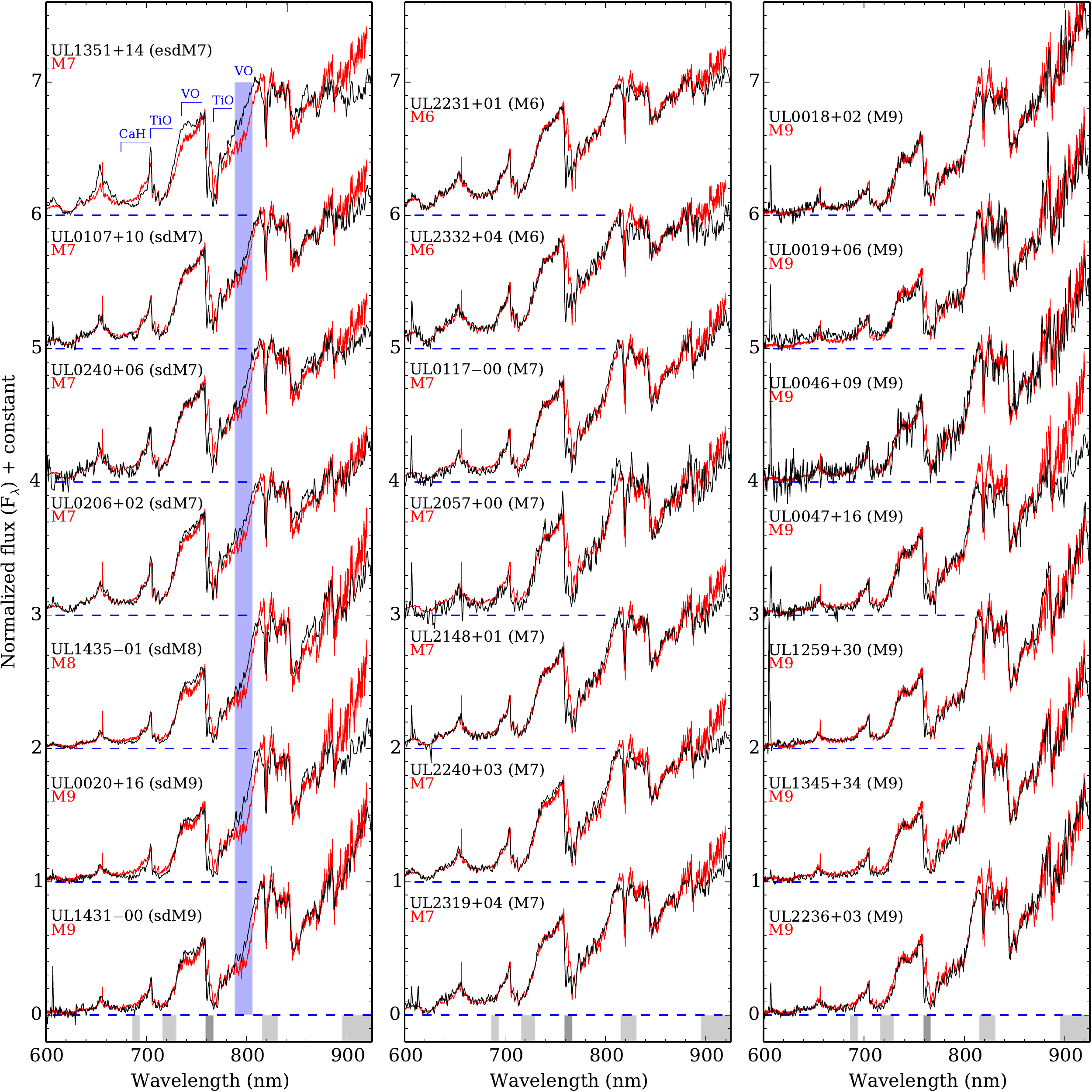}
\caption[]{Optical spectra of newly discovered M subdwarfs (left panel) and dwarfs (middle and right panels) compared to SDSS M dwarf template spectra \citep{boch07}. The VO absorption band at around 800 nm is highlighted with a blue band in the left-hand panel. Spectra are normalized at 825 nm.}
\label{f20dmp}
\end{center}
\end{figure*}

\begin{figure}
\begin{center}
  \includegraphics[width=\columnwidth]{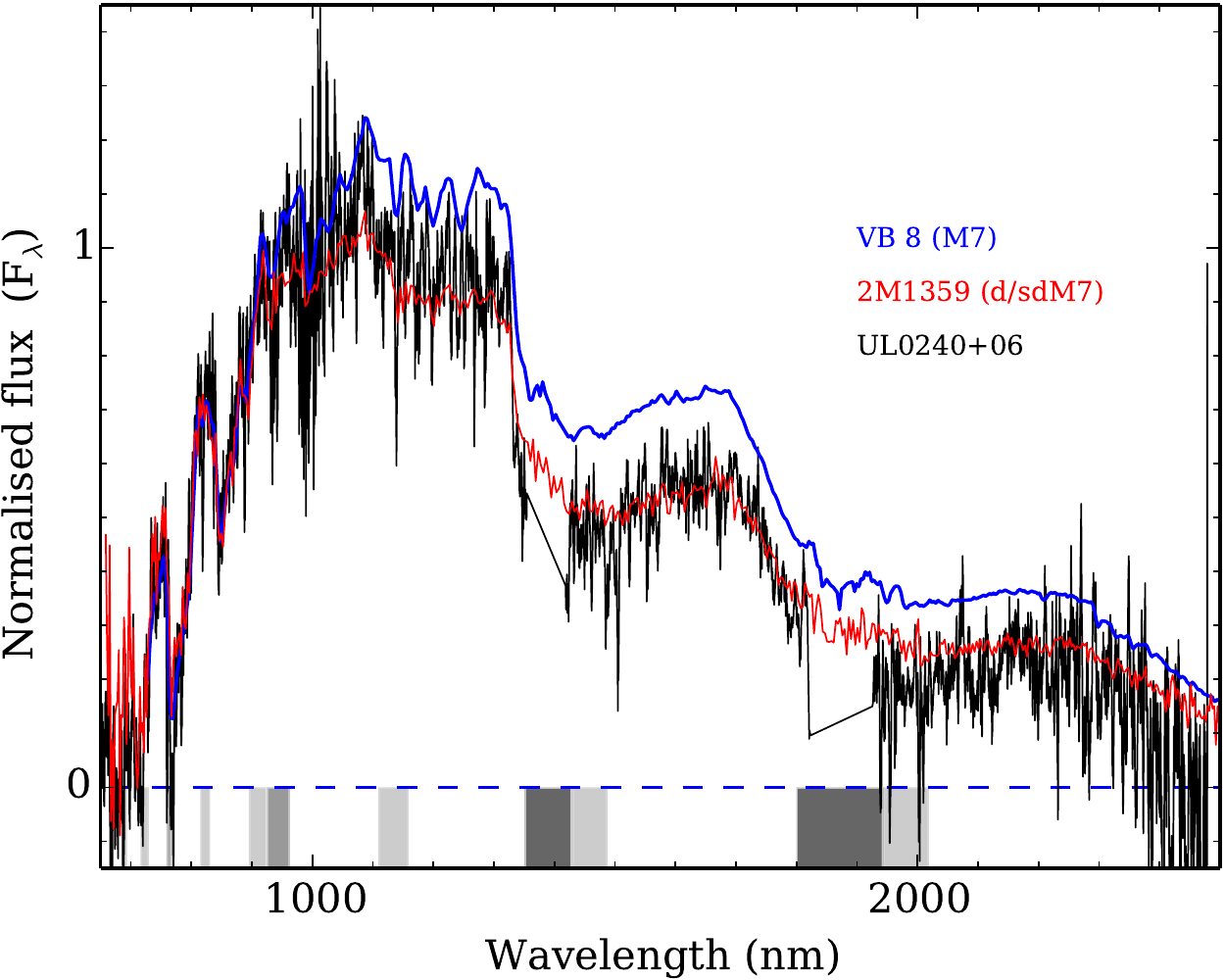}
\caption[]{The X-shooter spectrum of UL0240+06 compared to those of 2M1359 \citep[d/sdM7;][]{bur04a} and VB 8 \citep[M7;][]{kir91,bur08a}.}
\label{ul0240}
\end{center}
\end{figure}

\subsection{A blue BD binary}
\label{sbbdb}
\textit{ULAS J233227.03+123452.0} (UL2332+12) was selected as one of our ultra-cool subdwarf candidates. This object has previously been identified by \citet{skrz16}, who flagged it as a candidate binary from SED fitting. Their subsequent spectroscopic assessment classified it as a T1.5p, and following a more detailed NIR spectroscopic analysis they concluded it was more likely to be a metal-poor T0 dwarf. We re-assess the nature of this object based on our full optical-NIR spectroscopy.

Fig. \ref{ul2332op} shows that the optical spectrum of UL2332+12 compares well with the L7 dwarf standard 2MASS J08503593+1057156 \citep[2M0850;][]{kir99a} below 860 nm, but has relatively more flux at longer wavelength. In the red optical UL2332+12 compares better to the sdL7 subdwarf SD1416. However, UL2332+12 and SD1416 are quite different in the NIR.

UL2332+12 was observed by our X-shooter follow-up programme, and Fig. \ref{ul2332xs} shows the full optical--NIR spectrum. The original spectrum of UL2332+12 has SNR of $\sim$20 at 900 nm and $\sim$45 at 1300 nm, and was smoothed by 101 (VIS) and 51 (NIR) pixels for display. UL2332+12 has a similar NIR spectral profile to the sdL7 subdwarf SD1416, but has deeper H$_2$O absorption bands (a feature characteristic of later T dwarfs). UL2332+12 has a similar NIR spectrum to the T2 standard SDSS J075840.33+324723.4 \citep[SD0758;][]{knap04}; however, it has more flux at around 850 nm and no pronounced CH$_4$ absorption. Indeed, the full optical--NIR spectrum of UL2332+12 does not compare very well to any single L dwarf/subdwarf or T dwarf, but does have much more in common with the spectrum of DENIS-P J225210.73$-$173013.4 \citep[DE2252;][]{kend04,reid06}, an L4+T3.5 spectroscopic binary. L dwarfs with unresolved T dwarf companions can have bluer NIR colours \citep[e.g. fig. 14 of][]{zha10} and peculiar spectral features compared to normal L dwarfs \citep{burg10,bard14,maro15,manj16}.

We compared the NIR spectrum of UL2332+12 to those of synthesized spectral binaries constructed using L and T spectra from the Spex library, as described in \citet{burg10}. Fig. \ref{ul2332fit} shows best-fitting synthesized spectrum, which combines the L dwarf SDSS J133148.92$-$011651.4 \citep[SD1331;][]{hawl02} with the T4 dwarf \citep[2MASSI J2254188+312349;][]{bur02}. SD1331 is optically classified L6 \citep{hawl02} but also has a NIR classification of L8p \citep{knap04}, a peculiar blue L dwarf. This synthesized combination is a much better match than the best-fitting single object \citep[2MASS J11582077+0435014(sdL7);][]{kir10}. We assessed our 78 best-fitting synthesized spectra to constrain spectral type estimates for primary (L6.1$\pm$0.5) and the secondary (T4.0$\pm$1.7) components. And we note that these best-fitting combinations all include SD1331 (or other blue L dwarfs) as the primary. 

Fig. \ref{sd1331} compares the optical--NIR spectrum of SD1331 to that of other optically classified L6 dwarfs, demonstrating the `blue'  peculiarity of this object \citep[see also][]{maro15}. By direct comparison UL2332+12 has bluer $J-K$ and redder $i-J$ colours than SD1331, consistent with an additional unresolved mid-T companion. In colour-space SD1331 lies $\sim$mid-way between the dL and sdL sequences (e.g., it is indicated by a magenta open pentagon in Fig. \ref{ijk}, a figure that will be fully discussed in Section \ref{sccp}), and is on the BD side of the stellar-substellar boundary. Objects like SD1331 are thus likely to be mildly metal-poor BDs, and our best-fitting explanation for UL2332+12 is that it is a mildly metal-poor L6+T4 blue BD binary. Note that there is not mildly metal-poor mid-T dwarf in the Spex library available for our synthesized spectroscopic fitting. However, this does not affect our conclusion, as the affect of slightly lower metallicity on the shape and flux ratio of $J$ and $H$ band spectra of T dwarfs is negligible \citep[e.g.][]{pin12}. UL2332+12 has slightly less $K$ band flux than the synthesized spectrum. This indicates that the T type companion of UL2332+12 is slightly more metal-poor than the T4 dwarf used in the synthesized spectrum.

The $H$ band absolute magnitudes (MKO) of L6 and T4 dwarfs are around 12.90 and 14.10 mag according to \citet{dup12}. The combined $H$ band absolute magnitude of an L6+T4 binary is thus around 12.59 mag, suggesting a distance of 58$^{+12}_{-10}$ pc for UL2332+12 as a binary. A mildly metal-poor L6 dwarf could have a slightly brighter ($\sim$ 0.2 mag) $H$ band absolute magnitude compared to a normal L6 dwarf, which would lead to a slightly increased distance. However, such an increase ($\sim$5 pc) is much less than our estimated distance uncertainty and is thus not significant. 

We measured the proper motion and RV of UL2332+12 following the procedure described in Section \ref{sam}. UL2332+12 has proper motion $\mu_{\rm RA} = 420\pm10$ mas/yr and $\mu_{\rm Dec} = -107\pm13$ mas/yr, which is measured from its UKIDSS and PS1 epochs with a baseline of 2.74 yr. The tangential velocity of UL2332+12 is thus 118$^{+25}_{-21}$ km s$^{-1}$. The RV of UL2332+12 was measured from its X-shooter spectrum and an L9 type RV standard (DENIS-P J025503.3$-$470049), yielding an RV of $-35\pm8$ km s$^{-1}$. The space motion of UL2332+12 is thus $U = -85\pm18$, $V = -86\pm14$ and $W = -28\pm13$ km s$^{-1}$. We calculated the halo membership probabilities of UL2332+12 based on kinematics and stellar population fractions of the thin disc (0.93), thick disc (0.07) and halo (0.006) in the solar neighbourhood \citep{redd06}. The thin disc and thick disc membership probabilities of UL2332+12 are 22 and 77 per cent, respectively.

A summary of the measured and best-fitting properties of UL2332+12 are presented in Table \ref{prop}.

\begin{table*}
 \centering
  \caption[]{Photometric properties of new M--L dwarfs/subdwarfs identified in this work. } 
\label{tdmp}
  \begin{tabular}{c r c c c c c c}
\hline
    Name  & SpT & SDSS \emph{i} & SDSS \emph{z} & UKIDSS \emph{Y} & UKIDSS \emph{J} & UKIDSS \emph{H} & UKIDSS \emph{K}    \\
\hline
ULAS  J135122.15+141914.9 &  esdM7 & 19.24$\pm$0.03 & 18.02$\pm$0.03 & 17.01$\pm$0.01 & 16.31$\pm$0.01 & 15.92$\pm$0.01 & 15.63$\pm$0.02 \\
ULAS  J002009.35+160451.2 &  sdM9 & 20.81$\pm$0.06 & 19.21$\pm$0.05 & 18.07$\pm$0.03 & 17.32$\pm$0.02 & 16.90$\pm$0.04 & 16.65$\pm$0.04 \\
ULAS  J010756.85+100811.3 &  sdM7 & 21.29$\pm$0.09 & 20.00$\pm$0.11 & 18.80$\pm$0.06 & 18.08$\pm$0.07 & 17.76$\pm$0.08 & 17.50$\pm$0.10 \\
ULAS  J020628.22+020255.6 &  sdM7 & 20.79$\pm$0.04 & 19.48$\pm$0.05 & 18.45$\pm$0.04 & 17.65$\pm$0.03 & 17.28$\pm$0.07 & 17.10$\pm$0.10 \\
ULAS  J024035.36+060629.3 &  sdM7 & 21.07$\pm$0.15 & 19.82$\pm$0.08 & 18.66$\pm$0.05 & 17.99$\pm$0.04 & 17.65$\pm$0.06 & 17.48$\pm$0.12 \\
ULAS  J143154.18$-$004114.3 &  sdM9 & 20.86$\pm$0.06 & 19.25$\pm$0.05 & 17.92$\pm$0.03 & 17.19$\pm$0.03 & 16.84$\pm$0.04 & 16.48$\pm$0.05 \\
ULAS  J143517.18$-$014713.1 &  sdM8 & 19.79$\pm$0.04 & 18.31$\pm$0.03 & 17.20$\pm$0.02 & 16.39$\pm$0.01 & 15.97$\pm$0.02 & 15.62$\pm$0.02 \\
\hline
ULAS  J001837.37+020015.7 &  M9 & 21.98$\pm$0.16 & 20.19$\pm$0.11 & 19.10$\pm$0.08 & 17.99$\pm$0.06 & 17.58$\pm$0.07 & 17.15$\pm$0.08 \\
ULAS  J001931.33+063111.0 &  M9 & 21.79$\pm$0.13 & 19.76$\pm$0.08 & 18.47$\pm$0.05 & 17.53$\pm$0.04 & 17.07$\pm$0.04 & 16.71$\pm$0.05 \\
ULAS  J004602.85+091131.2 &  M9 & 22.32$\pm$0.26 & 20.28$\pm$0.16 & 19.07$\pm$0.07 & 18.14$\pm$0.05 & 17.66$\pm$0.08 & 17.29$\pm$0.10 \\
ULAS  J004716.65+161242.4 &  M9 & 21.86$\pm$0.16 & 20.76$\pm$0.28 & 19.22$\pm$0.08 & 18.44$\pm$0.10 & 17.99$\pm$0.12 & 18.22$\pm$0.26 \\
ULAS  J011711.98$-$005213.4 & M7 & 20.67$\pm$0.04 & 19.45$\pm$0.06 & 18.21$\pm$0.03 & 17.53$\pm$0.03 & 17.13$\pm$0.05 & 16.92$\pm$0.07 \\
ULAS  J125938.50+301500.2 &  M9 & 19.62$\pm$0.03 & 17.87$\pm$0.02 & 16.41$\pm$0.01 & 15.57$\pm$0.01 & 15.18$\pm$0.01 & 14.74$\pm$0.01 \\
ULAS  J134505.85+342441.8 &  M9 & 20.64$\pm$0.04 & 18.91$\pm$0.04 & 17.61$\pm$0.03 & 16.77$\pm$0.02 & 16.45$\pm$0.02 & 16.04$\pm$0.03 \\
ULAS  J205721.89+005628.7 &  M7 & 21.41$\pm$0.09 & 19.95$\pm$0.08 & 18.90$\pm$0.08 & 17.98$\pm$0.06 & 17.65$\pm$0.08 & 17.26$\pm$0.08 \\
ULAS  J214816.13+012225.1 &  M7 & 21.74$\pm$0.13 & 20.30$\pm$0.13 & 19.34$\pm$0.10 & 18.39$\pm$0.08 & 17.97$\pm$0.11 & 18.07$\pm$0.20 \\
ULAS  J223123.44+010025.1 &  M6 & 20.84$\pm$0.05 & 19.73$\pm$0.07 & 18.47$\pm$0.05 & 17.53$\pm$0.04 & 17.46$\pm$0.06 & 17.29$\pm$0.10 \\
ULAS  J223623.17+034344.5 &  M9 & 22.22$\pm$0.16 & 20.25$\pm$0.11 & 19.03$\pm$0.06 & 18.17$\pm$0.05 & 17.91$\pm$0.06 & 17.51$\pm$0.10 \\
ULAS  J224054.61+030902.0 &  M7 & 21.04$\pm$0.05 & 19.87$\pm$0.08 & 18.85$\pm$0.07 & 18.02$\pm$0.06 & 17.59$\pm$0.07 & 17.46$\pm$0.10 \\
ULAS J224749.77+053207.9 & L1 & 22.44$\pm$0.23 & 20.12$\pm$0.12 & 18.60$\pm$0.04 & 17.46$\pm$0.03 & 16.90$\pm$0.04 & 16.43$\pm$0.04  \\
ULAS  J231949.36+044559.5 &  M7 & 21.34$\pm$0.13 & 19.95$\pm$0.13 & 18.94$\pm$0.06 & 18.16$\pm$0.06 & 17.70$\pm$0.10 & 17.81$\pm$0.18 \\
ULAS  J233211.22+045554.2 &  M6 & 21.28$\pm$0.17 & 20.04$\pm$0.14 & 18.95$\pm$0.08 & 18.16$\pm$0.07 & 17.98$\pm$0.08 & 17.66$\pm$0.10 \\
\hline
\end{tabular}
\end{table*}

\subsection{M subdwarfs and M--L dwarfs amongst our candidates}
About one-third (22) of our candidate L subdwarfs were identified as M6--L1 dwarfs or M7--M9 subdwarfs through spectroscopic follow-up. Table \ref{tdmp} shows spectral types and photometry for these objects.

ULAS J224749.77+053207.9 (UL2247+05) has $i-J$ and $J-K$ colours similar to those of mid-type sdL. However, its optical spectrum compares well with the L1 dwarf 2M1349 (see Fig. \ref{ul2247op}). Its $i_{\rm P1}-J$ and $J-K$ colours are consistent with early-type L dwarfs, and we note that its SDSS $i$ band magnitude has a much larger uncertainty than the PS1 $i_{\rm P1}$ band magnitude. Fig. \ref{fsdlsed} shows that UL2247+05 has a slightly bluer SEDs than other L1 dwarfs. UL2247+05 has a proper motion of $\mu_{\rm RA} = -56\pm14$ mas/yr and $\mu_{\rm Dec} = -56\pm15$ mas/yr (measured using the method described in Section \ref{sam}). Its distance is 146$^{+30}_{-25}$ pc based on spectral type and $H$ band magnitude, leading to a tangential velocity of 54$^{+18}_{-17}$ km s$^{-1}$.

When considering the M subdwarfs, there are several classification schemes for the late-types. In the most recent sdM scheme \citep{lep07}, the metallicity consistency of the latest sdM subclasses has not been fully tested across all subtypes, due to the lack of wide binaries containing both early- and late-type M subdwarfs. We have also noticed that the strength of the metallicity features are usually under estimated for late-type M subdwarfs compared to early-type M subdwarfs (section 4.1 of Paper I). As a result, late-type M stars with the same metallicity as early-type esdM subdwarfs can be classified as sdM, and some late-type M stars with the same metallicity as early-type sdM subdwarfs can be classified as M dwarfs. \citet{zha13} noticed that $\sim$ 18 per cent of SDSS M dwarfs classified on the scheme of \citet{lep07} have thick disc or halo like kinematics ($V < -100$ km s$^{-1}$). Furthermore, different subclass names are used for objects with similar metallicity in different classification schemes. The late-type d/sdM subclass of \citet{bur07} would be expected to have a similar metallicity range as the early-type sdM subclass of \citet{lep07}.

We classified our M subdwarfs using their optical metal-poor spectral features, such as weak VO absorption near 800 nm. This represents an extension of our L subdwarf scheme (Paper I) into the late-type sdMs, and allows us to consistently carry out sdM/sdL classification of our programmatic sample. Fig. \ref{f20dmp} shows the spectra of these 21 late-type M stars compared to stacked SDSS M6--M9 spectral templates \citep{boch07}. Seven of these objects show weaker VO absorption bands than the M dwarf templates. ULAS J135122.15+141914.9 has the strongest metal-poor features with the absence of 800 nm VO absorption, and was classified as esdM7. ULAS  J002009.35+160451.2, ULAS  J020628.22+020255.6, and ULAS  J024035.36+060629.3 (UL0240+06) have a weaker 800 nm VO absorption bands than the M dwarf tamplates, and were classified as sdM9, sdM7, and sdM7, respectively. ULAS  J010756.85+100811.3, ULAS  J143154.18$-$004114.3 and ULAS  J143517.18$-$014713.1 have weaker 800 nm VO absorption bands than the M dwarf templates, but slightly stronger than UL0240+06. We classified these objects as sdM subdwarfs, although they could be classified as dM or d/sdM using existing classification schemes. The other 14 objects were classified as late-type M dwarfs since they compare well with M dwarf templates.

Fig. \ref{ul0240} shows the X-shooter spectrum of UL0240+06. The original spectrum of UL0240+06 has SNR of $\sim$2 at 910 nm and $\sim$4 at 1300 nm, and was smoothed by 101 (VIS) and 51 (NIR) pixels for display. UL0240+06 shows suppressed NIR flux compared to the M7 dwarf VB 8 \citep{kir91}. UL0240+06 has a similar spectral profile to the d/sdM7 object 2MASS J13593574+3031039 \citep[2M1359;][]{bur04a}.

\begin{figure}
\begin{center}
  \includegraphics[width=\columnwidth]{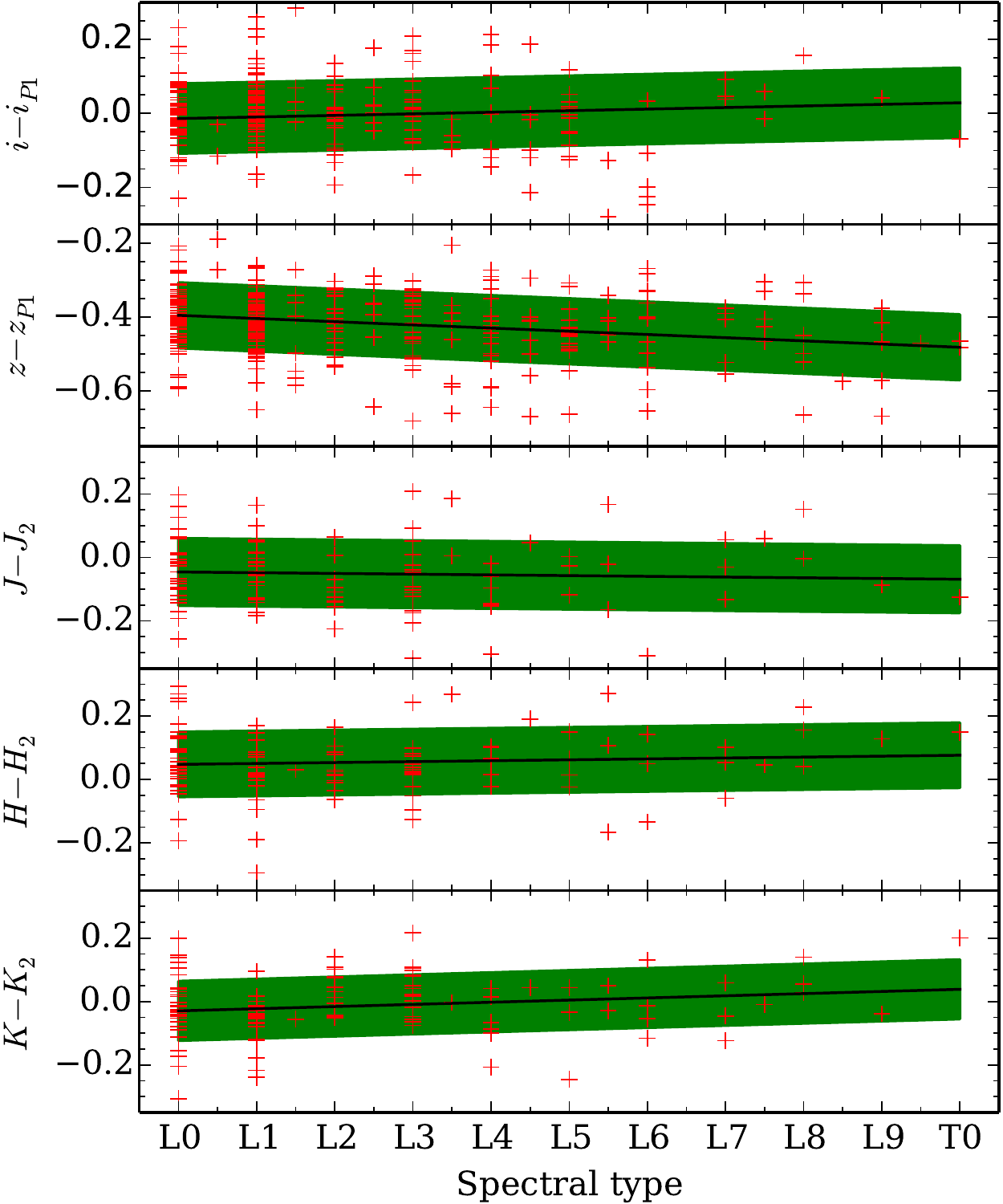}
\caption[]{Polynomial fits to SDSS--PS1 and MKO--2MASS magnitude differences as a function of spectral type, for L0--T0 dwarfs (red crosses). The green shaded area indicates the rms of these fits. L and T dwarfs observed by SDSS and PS1 have $i$ or $z$ band photometric precision better than 0.12 mag. Those observed by 2MASS have $J$-, $H$- and $K$-band photometric precision better than 0.16 mag.}
\label{sptph}
\end{center}
\end{figure}

\begin{table*}
 \centering
  \caption[]{PS1 and {\sl WISE} photometry of L subdwarfs.} 
\label{tsdps}
  \begin{tabular}{l c c cc c c   }
\hline
    Name   & PS1 name &   $i_{\rm P1}$ &  $z_{\rm P1}$  &  $y_{\rm P1}$ & $W1$ & $W2$ \\
\hline
SSSPM J10130734$-$1356204 & J101307.404$-$135632.796 & 17.23$\pm$0.01 & 16.25$\pm$0.01 & 15.92$\pm$0.01 & 13.80$\pm$0.03 & 13.60$\pm$0.03 \\
SDSSS J010448.47+153501.9 & J010448.655+153459.325 & 20.52$\pm$0.02 & 19.49$\pm$0.02 & 19.09$\pm$0.03 & 16.61$\pm$0.08 & 16.36$\pm$0.25 \\
SDSS J125637.13$-$022452.4 & J125636.715$-$022456.107 & 19.45$\pm$0.02 & 18.10$\pm$0.02 & 17.51$\pm$0.01 & 15.21$\pm$0.04 & 15.01$\pm$0.08 \\
ULAS J135058.85+081506.8 & J135058.782+081505.607 & 21.26$\pm$0.17 & 19.76$\pm$0.02 & 19.32$\pm$0.04 & 17.45$\pm$0.20 & $\geq$16.45 \\
2MASS J16262034+3925190 & J162618.820+392522.418 & 17.92$\pm$0.01 & 16.34$\pm$0.01 & 15.79$\pm$0.01 & 13.48$\pm$0.02 & 13.14$\pm$0.03 \\
WISEA J213409.15+713236.1 & J213409.144+713236.852 & 19.09$\pm$0.01 & 17.99$\pm$0.02 & 17.58$\pm$0.02 & 15.23$\pm$0.03 & 14.95$\pm$0.04 \\
ULAS J230711.01+014447.1 & J230711.008+014446.289 & 21.69$\pm$0.16 & 20.16$\pm$0.04 & 19.57$\pm$0.08 & 17.47$\pm$0.18 & $\geq$16.79 \\
WISEA J001450.17$-$083823.4 & J001450.175$-$083823.247 & 17.40$\pm$0.06 & 16.26$\pm$0.01 & 15.77$\pm$0.01 & 13.43$\pm$0.03 & 13.20$\pm$0.03 \\
WISEA J020201.25$-$313645.2 & J020201.198$-$313649.894 & 18.48$\pm$0.01 & --- & --- & 14.31$\pm$0.03 & 13.96$\pm$0.04 \\
ULAS J020858.62+020657.0 & J020858.628+020656.920 & 21.39$\pm$0.03 & 19.99$\pm$0.03 & 19.50$\pm$0.06 & 17.17$\pm$0.13 & 16.63$\pm$0.29 \\
ULAS J033351.10+001405.8 & J033351.342+001405.465 & 19.22$\pm$0.01 & 18.05$\pm$0.01 & 17.54$\pm$0.01 & 15.08$\pm$0.04 & 14.77$\pm$0.07 \\
WISEA J030601.66$-$033059.0 & J030601.672$-$033059.676 & 18.02$\pm$0.02 & 16.57$\pm$0.01 & 15.99$\pm$0.01 & 13.43$\pm$0.03 & 13.18$\pm$0.03 \\
WISEA J043535.82+211508.9 & J043535.926+211507.487 & 18.60$\pm$0.01 & 17.15$\pm$0.01 & 16.56$\pm$0.01 & 14.00$\pm$0.03 & 13.67$\pm$0.04 \\
2MASS J05325346+8246465 & J053309.378+824622.210 & 20.31$\pm$0.06 & 18.09$\pm$0.01 & 16.95$\pm$0.01 & 13.82$\pm$0.03 & 13.26$\pm$0.03 \\
2MASS J06164006$-$6407194 & --- & --- & --- & --- & 15.65$\pm$0.03 & 15.18$\pm$0.04 \\
ULAS J111429.54+072809.5 & J111429.525+072809.033 & 20.69$\pm$0.02 & 19.55$\pm$0.02 & 19.06$\pm$0.03 & 16.84$\pm$0.12 & 16.85$\pm$0.39 \\
SDSS J124410.11+273625.8 & J124410.063+273624.064 & 20.40$\pm$0.02 & 19.32$\pm$0.01 & 18.94$\pm$0.03 & 16.68$\pm$0.09 & 16.94$\pm$0.40 \\
ULAS J124425.75+102439.3 & J124425.617+102436.751 & 19.47$\pm$0.03 & 18.21$\pm$0.01 & 17.71$\pm$0.01 & 15.45$\pm$0.04 & 15.14$\pm$0.09 \\
ULAS J135216.31+312327.0 & J135216.314+312326.963 & 19.97$\pm$0.03 & 18.81$\pm$0.01 & 18.34$\pm$0.02 & 16.14$\pm$0.06 & 15.89$\pm$0.13 \\
SDSS J141405.74$-$014202.7 & J141405.632$-$014204.911 & 19.88$\pm$0.01 & 18.73$\pm$0.01 & 18.20$\pm$0.02 & --- & --- \\
SSSPM J144420.67$-$201922.2 & J144418.155$-$201946.066 & 16.08$\pm$0.01 & 14.69$\pm$0.01 & 14.08$\pm$0.02 & 11.47$\pm$0.02 & 11.21$\pm$0.02 \\
ULAS J145234.65+043738.4 & J145234.651+043737.276 & 20.76$\pm$0.04 & 19.45$\pm$0.01 & 18.95$\pm$0.04 & 16.15$\pm$0.05 & 16.23$\pm$0.18 \\
ULAS J151913.03$-$000030.0 & J151913.024$-$000031.791 & 21.36$\pm$0.07 & 19.63$\pm$0.02 & 18.86$\pm$0.02 & 16.25$\pm$0.06 & 15.72$\pm$0.14 \\
2MASS J16403197+1231068 & J164031.759+123104.908 & 19.15$\pm$0.01 & 17.96$\pm$0.02 & 17.39$\pm$0.01 & 15.03$\pm$0.04 & 14.85$\pm$0.07 \\
WISEA J204027.30+695924.1 & J204027.336+695924.279 & 16.89$\pm$0.01 & 15.65$\pm$0.01 & 15.10$\pm$0.01 & 12.68$\pm$0.02 & 12.46$\pm$0.02 \\
ULAS J223302.03+062030.8 & J223302.105+062030.621 & 21.15$\pm$0.04 & 19.89$\pm$0.03 & 19.38$\pm$0.02 & 16.92$\pm$0.12 & $\geq$16.45 \\
ULAS J231924.35+052524.5 & J231924.396+052524.161 & 20.78$\pm$0.06 & 19.51$\pm$0.03 & 18.89$\pm$0.02 & 16.44$\pm$0.08 & 16.12$\pm$0.19 \\
2MASS J00412179+3547133 & J004121.680+354712.791 & 20.21$\pm$0.04 & 18.70$\pm$0.02 & 17.82$\pm$0.01 & 14.74$\pm$0.03 & 14.45$\pm$0.05 \\
WISEA J005757.65+201304.0 & --- & --- & --- & --- & 14.32$\pm$0.03 & 13.87$\pm$0.04 \\
WISEA J011639.05$-$165420.5 & --- & --- & --- & --- & 14.42$\pm$0.03 & 13.99$\pm$0.04 \\
ULAS J011824.89+034130.4 & J011824.899+034130.329 & 21.20$\pm$0.05 & 20.19$\pm$0.04 & 19.55$\pm$0.10 & --- & --- \\
WISEA J013012.66$-$104732.4 & --- & --- & --- & --- & 14.42$\pm$0.03 & 14.00$\pm$0.04 \\
ULAS J021258.08+064115.9 & J021258.090+064113.734 & 21.00$\pm$0.07 & 19.60$\pm$0.01 & 19.03$\pm$0.02 & 16.31$\pm$0.06 & 15.98$\pm$0.18 \\
ULAS J021642.96+004005.7 & J021642.932+004005.180 & 21.77$\pm$0.12 & 20.30$\pm$0.02 & 19.25$\pm$0.02 & 15.96$\pm$0.05 & 15.75$\pm$0.13 \\
SDSS J023803.12+054526.1 & J023803.161+054525.696 & 20.46$\pm$0.02 & 18.91$\pm$0.01 & 18.08$\pm$0.01 & 15.30$\pm$0.04 & 15.13$\pm$0.09 \\
2MASS J06453153$-$6646120 & --- & --- & --- & --- & 13.76$\pm$0.02 & 13.31$\pm$0.02 \\
ULAS J075335.23+200622.4 & J075335.224+200621.615 & 19.71$\pm$0.01 & 18.23$\pm$0.01 & 17.46$\pm$0.01 & 14.60$\pm$0.03 & 14.27$\pm$0.05 \\
ULAS J082206.61+044101.8 & J082206.621+044100.926 & 20.33$\pm$0.03 & 18.80$\pm$0.03 & 17.98$\pm$0.02 & 15.10$\pm$0.03 & 14.93$\pm$0.10 \\
WISEA J101329.72$-$724619.2 & --- & --- & --- & --- & 14.59$\pm$0.03 & 14.17$\pm$0.04 \\
2MASS J11582077+0435014  & J115821.180+043451.812 & 20.73$\pm$0.03 & 18.65$\pm$0.01 & 17.57$\pm$0.01 & 13.70$\pm$0.03 & 13.36$\pm$0.03 \\
ULAS J123142.99+015045.4 & J123142.963+015045.403 & 21.32$\pm$0.04 & 19.93$\pm$0.02 & 19.19$\pm$0.02 & 16.15$\pm$0.06 & 15.81$\pm$0.18 \\
ULAS J124104.75$-$000531.4 & J124104.737$-$000531.675  & --- & 21.00$\pm$0.06 & 20.16$\pm$0.06 & --- & --- \\
ULAS J124947.04+095019.8 & J124946.950+095018.252 & 20.41$\pm$0.01 & 19.01$\pm$0.01 & 18.36$\pm$0.02 & 15.96$\pm$0.05 & 15.71$\pm$0.13 \\
ULAS J130710.22+151103.4 & J130710.063+151102.511 & --- & 21.27$\pm$0.15 & --- & 16.25$\pm$0.07 & 15.99$\pm$0.18 \\
ULAS J133348.27+273505.5 & J133348.294+273503.885 & 20.45$\pm$0.02 & 18.94$\pm$0.01 & 18.27$\pm$0.02 & --- & --- \\
ULAS J133836.97$-$022910.7 & J133836.963$-$022911.152 & --- & 20.55$\pm$0.05 & 19.54$\pm$0.05 & 15.78$\pm$0.05 & 15.62$\pm$0.12 \\
ULAS J134206.86+053724.9 & J134206.849+053724.420 & 21.55$\pm$0.04 & 20.05$\pm$0.04 & 19.23$\pm$0.04 & 16.26$\pm$0.06 & 16.27$\pm$0.21 \\
ULAS J134423.98+280603.8 & J134423.954+280603.887 & 22.08$\pm$0.09 & 20.36$\pm$0.04 & 19.30$\pm$0.05 & 15.61$\pm$0.04 & 15.44$\pm$0.10 \\
ULAS J134749.79+333601.7 & J134749.789+333601.640 & 19.85$\pm$0.01 & 18.35$\pm$0.01 & 17.53$\pm$0.01 & 14.69$\pm$0.03 & 14.33$\pm$0.05 \\
ULAS J134852.93+101611.8 & J134852.814+101610.787 & 20.69$\pm$0.02 & 19.22$\pm$0.02 & 18.46$\pm$0.03 & 15.47$\pm$0.04 & 15.14$\pm$0.08 \\
ULAS J125226.62+092920.1 & J125226.519+092920.150 & 20.74$\pm$0.03 & 19.25$\pm$0.02 & 18.47$\pm$0.01 & 15.80$\pm$0.05 & 15.52$\pm$0.14 \\
ULAS J135359.58+011856.7 & J135359.573+011856.654 & 21.49$\pm$0.04 & 19.94$\pm$0.03 & 19.08$\pm$0.03 & 15.87$\pm$0.05 & 15.77$\pm$0.13 \\
WISEA J135501.90$-$825838.9 & --- & --- & --- & --- & 14.12$\pm$0.03 & 13.55$\pm$0.03 \\
ULAS J141203.85+121609.9 & J141203.779+121609.818 & 21.35$\pm$0.03 & 19.44$\pm$0.02 & 18.39$\pm$0.03 & 14.77$\pm$0.03 & 14.49$\pm$0.05 \\
SDSS J141624.12+134827.4 & J141624.149+134827.768 & 18.35$\pm$0.01 & 16.30$\pm$0.01 & 15.21$\pm$0.01 & 11.36$\pm$0.02 & 11.03$\pm$0.02 \\
ULAS J141832.35+025323.0 & J141832.360+025322.309 & 20.10$\pm$0.01 & 18.56$\pm$0.01 & 17.69$\pm$0.01 & 14.86$\pm$0.03 & 14.43$\pm$0.05 \\
ULAS J144151.55+043738.5 & J144151.466+043735.872 & 21.71$\pm$0.14 & 20.21$\pm$0.03 & 19.29$\pm$0.02 & 15.92$\pm$0.05 & 15.54$\pm$0.10 \\
ULAS J151649.84+083607.1 & J151649.778+083607.405 & --- & 20.65$\pm$0.05 & 19.57$\pm$0.05 & 15.67$\pm$0.05 & 15.43$\pm$0.11 \\
ULAS J154638.34$-$011213.0 & J154638.320$-$011213.746 & 21.93$\pm$0.19 & 20.37$\pm$0.07 & 19.32$\pm$0.05 & --- & --- \\
2MASS J17561080+2815238 & J175610.126+281517.780 & 19.03$\pm$0.01 & 17.38$\pm$0.01 & 16.41$\pm$0.01 & 13.38$\pm$0.02 & 13.07$\pm$0.03 \\
LSR J182611.3+301419.1 & J182608.778+301410.509 & 15.48$\pm$0.01  & 13.97$\pm$0.01	 & 13.22$\pm$0.01  & 10.40$\pm$0.02 & 10.04$\pm$0.02 \\
ULAS J223440.80+001002.6 & J223440.785+001002.087 & 21.83$\pm$0.06 & 20.51$\pm$0.03 & 19.64$\pm$0.08 & 16.47$\pm$0.08 & 16.59$\pm$0.33 \\
ULAS J225902.14+115602.1 & J225902.146+115601.795 & 21.16$\pm$0.02 & 19.65$\pm$0.03 & 18.76$\pm$0.01 & 15.96$\pm$0.06 & 15.81$\pm$0.15 \\
ULAS J230256.53+121310.2 & J230256.527+121309.986 & 21.26$\pm$0.04 & 19.84$\pm$0.02 & 19.08$\pm$0.04 & --- & --- \\
ULAS J230443.30+093423.9 & J230443.305+093423.871 & 21.52$\pm$0.07 & 19.96$\pm$0.10 & 19.02$\pm$0.02 & 15.82$\pm$0.05 & 15.39$\pm$0.10 \\
\hline
\end{tabular}
\end{table*}

\begin{table*}
 \centering
  \caption[]{Coefficients of third-order polynomial fits of colours as a function of spectral types ($SpT$) and second-order fits of absolute magnitudes for L0--L9 dwarfs in Figs \ref{fsptc} and \ref{fsptmag}. The fits are defined as  $Colour / M = \sum_{i=0}^3 c_i \times (SpT)^{i}$,  
  $SpT$ = 10 for L0, and $SpT$ = 19 for L9.  The rms of polynomial fits are listed in the last column. The $i$ and $z$ photometry is assumed to be on the SDSS (AB) system, and the $J$, $H$, and $K$ photometry is assumed to be on the MKO (Vega) system.}
 \label{tcolour}
  \begin{tabular}{l r r r c c}
\hline
    $Colour / M$  & $c_0~~$ & $c_1~~$ & $c_2~~~~~~~~$ & $c_3$ & rms  \\
\hline
$i-z$ & 9.049 & $-$1.544 & $1.025\times10^{-1}$ & $-1.979\times10^{-3}$ & 0.174 \\
$i-J$ & 9.434 & $-$1.227 & $9.063\times10^{-2}$ & $-1.888\times10^{-3}$ & 0.203 \\
$z-J$ & 3.583 & $-$0.408 & $4.134\times10^{-2}$ & $-1.175\times10^{-3}$ & 0.157 \\
$Y-J$ & $-$2.951 & 0.757 & $-4.367\times10^{-2}$ & $~~8.076\times10^{-4}$ & 0.106 \\
$J-H$ & 2.564 & $-$0.528 & $4.487\times10^{-2}$ & $-1.134\times10^{-3}$ & 0.093 \\
$J-K$ & 4.205 & $-$0.852 & $7.452\times10^{-2}$ & $-1.940\times10^{-3}$ & 0.146 \\
$Y-W1$ & 4.176 &  $-$0.820 &  $9.090\times10^{-2}$ & $-2.592\times10^{-3}$ & 0.285 \\
$J-W1$ & 5.271 & $-$1.169  &  $9.641\times10^{-2}$ & $-2.388\times10^{-3}$ & 0.145 \\
$H-W1$ & 7.892 & $-$1.722  &  $1.437\times10^{-1}$ & $-3.589\times10^{-3}$ & 0.213 \\
$K-W1$ & 3.985 & $-$0.915 &  $7.167\times10^{-2}$ & $-1.703\times10^{-3}$ & 0.098 \\
$J-W2$ & 11.065 & $-$2.324 & $1.837\times10^{-1}$ & $-4.418\times10^{-3}$ & 0.258 \\
$W1-W2$ & 2.045 & $-$0.371 & $2.416\times10^{-2}$ & $-4.647\times10^{-4}$ & 0.092 \\
$H-W2$ & 8.883 & $-$1.879 & $1.447\times10^{-1}$ & $-3.419\times10^{-3}$ & 0.192 \\
$W2-W3$ & 12.055 & $-$2.088 & $1.314\times10^{-1}$ & $-2.658\times10^{-3}$ & 0.903 \\
$i_{\rm P1}-z_{\rm P1}$ & 5.857 & $-$0.836 & $4.562\times10^{-2}$ & $-5.525\times10^{-4}$ & 0.202 \\
$i_{\rm P1}-y_{\rm P1}$ & 5.333 & $-$0.560 & $2.874\times10^{-2}$ & $-2.007\times10^{-4}$ & 0.169 \\
$z_{\rm P1}-y_{\rm P1}$ & 1.654 & $-$0.192 & $1.600\times10^{-2}$ & $-3.962\times10^{-4}$ & 0.070 \\
$i_{\rm P1}-J$ & 10.030 & $-$1.310 & $9.259\times10^{-2}$ & $-1.875\times10^{-3}$ & 0.392 \\
$z_{\rm P1}-J$ & 1.456 & 0.065 & $1.232\times10^{-2}$ & $-5.777\times10^{-4}$ & 0.177 \\
$y_{\rm P1}-J$ & 0.481 & 0.116 & $5.786\times10^{-3}$ & $-3.881\times10^{-4}$ & 0.127 \\
$i_{\rm P1}-H$ & 9.785 & $-$1.220 & $9.273\times10^{-2}$ & $-1.951\times10^{-3}$ & 0.412 \\
$y_{\rm P1}-H$ & 1.724 & $-$0.138 & $3.165\times10^{-2}$ & $-1.087\times10^{-3}$ & 0.189 \\
$z_{\rm P1}-H$ & 2.847 & $-$0.221 & $4.053\times10^{-2}$ & $-1.332\times10^{-3}$ & 0.228 \\
$y_{\rm P1}-K$ & 2.444 & $-$0.283 & $4.933\times10^{-2}$ & $-1.623\times10^{-3}$ & 0.259 \\
$M_G$ ({\sl Gaia})  & 11.264 & 0.546 & $-3.690\times10^{-3}$ &--- & 0.428 \\
$M_{BP}$ ({\sl Gaia}) & 12.819 & 0.587 & $ -6.725\times10^{-3}$ &--- & 0.944 \\
$M_{BP}$ ({\sl Gaia}) & 9.218 & 0.622 & $ -7.554\times10^{-3}$ &--- & 0.421 \\
$M_i$ (PS1) & 13.490 & 0.106 & $ 1.459\times10^{-2}$ &--- & 0.400 \\
$M_z$ (PS1) & 9.250 & 0.602 & $ -7.531\times10^{-3}$ &--- & 0.369 \\
$M_y$ (PS1) & 8.581 & 0.566 & $ -6.708\times10^{-3}$ &--- & 0.361 \\
\hline
\end{tabular}
\end{table*}

\begin{figure*}
\begin{center}
  \includegraphics[angle=0,width=\textwidth]{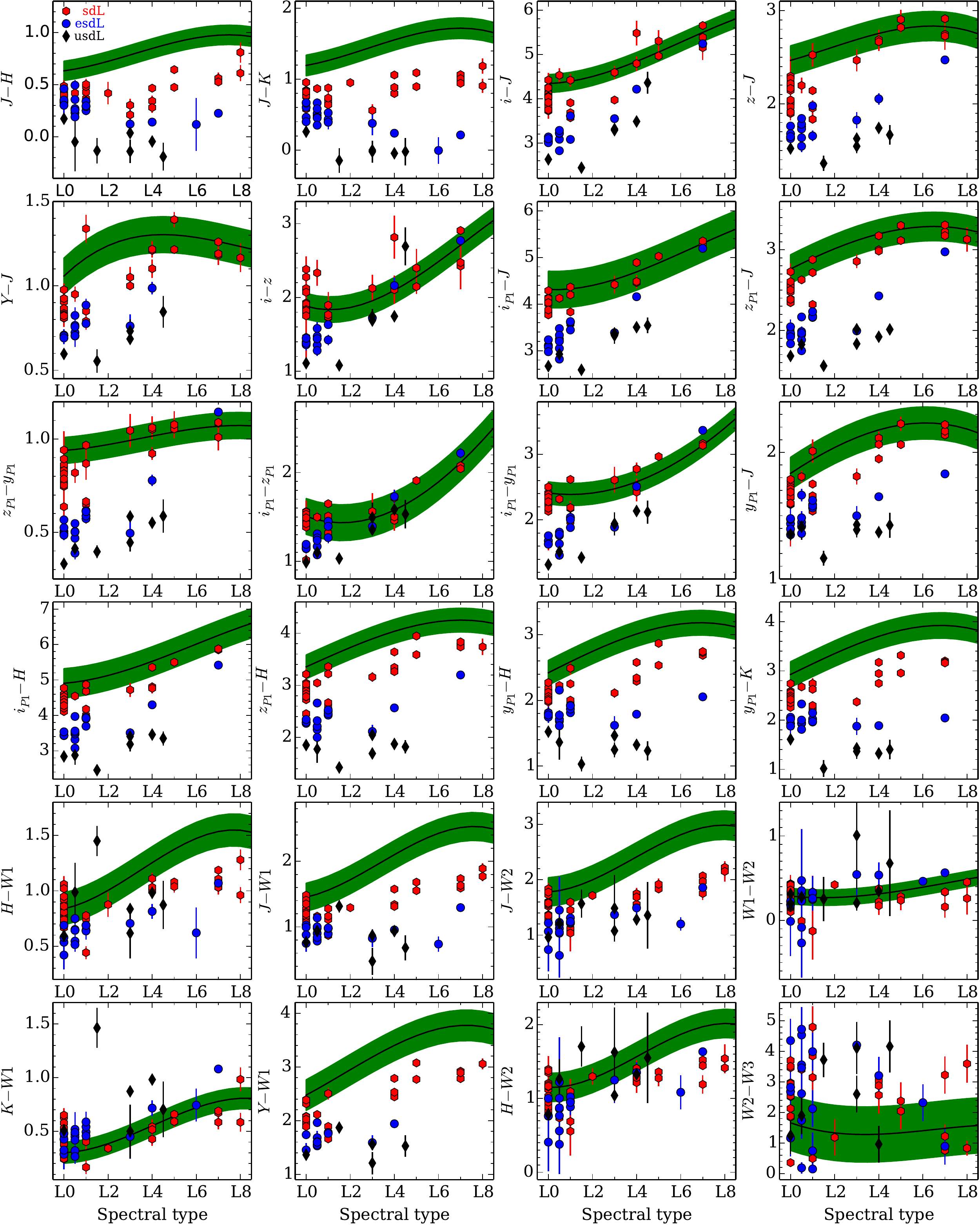}
\caption[]{Optical to infrared colours of L subdwarfs. Red hexagons, blue circles, and black diamonds represent sdL, esdL, and usdL subclasses. Third-order polynomial fits (black lines) and their rms (green shaded areas) for colour--spectral type relations of L dwarfs are also plotted for comparison.}
\label{fsptc}
\end{center}
\end{figure*}

\begin{figure*}
\begin{center}
   \includegraphics[angle=0,width=\textwidth]{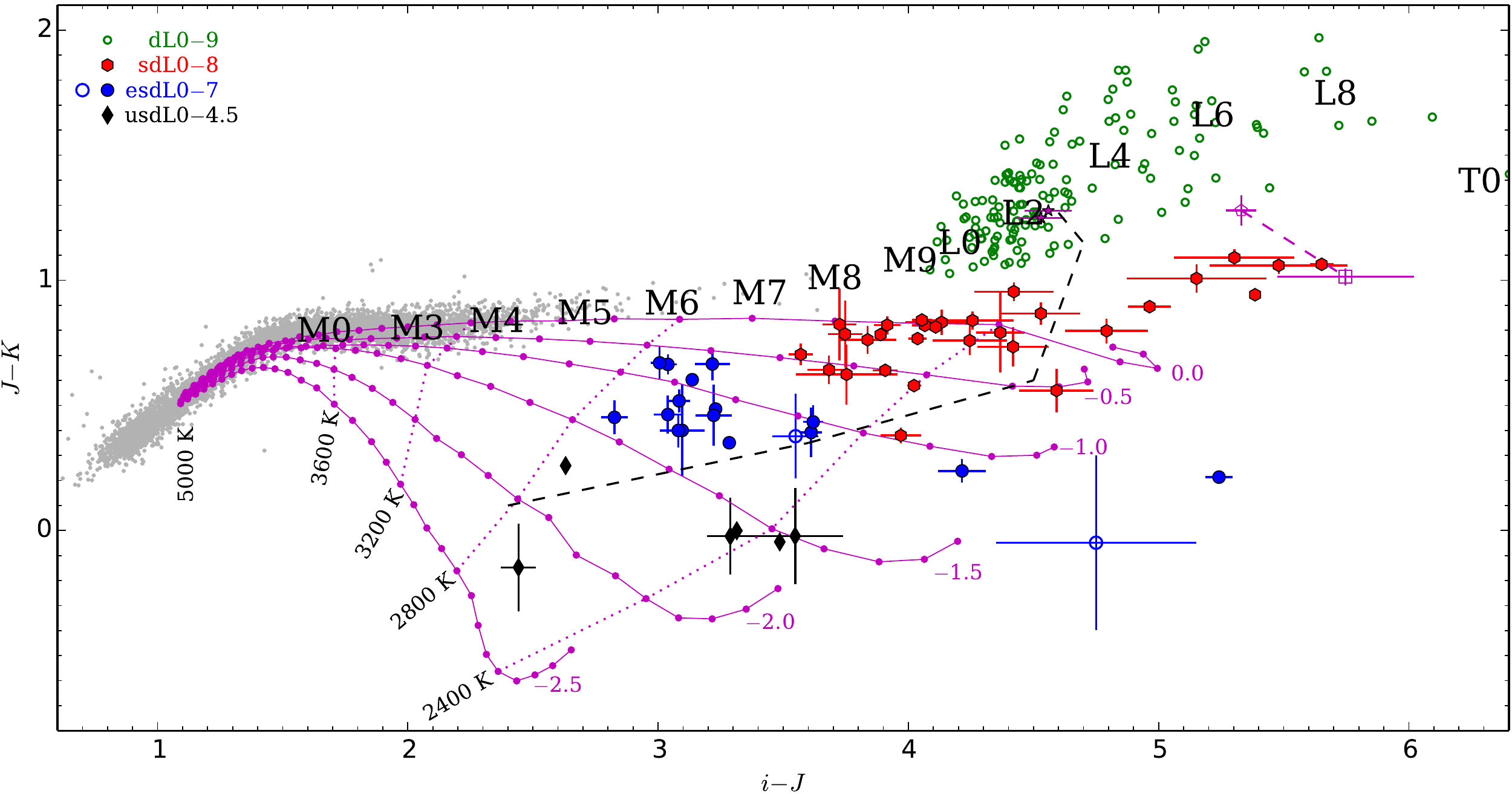}
\caption[]{The $i-J$ versus $J-K$ colours of L subdwarfs. Green open circles, red hexagons, blue circles, and black diamonds represent dL, sdL, esdL and usdL populations, respectively. Error bars are smaller than the symbol size for some objects. Grey dots are 5000 point sources selected from a 10 deg$^2$ area of ULAS-SDSS-PS1-{\sl WISE} sky with $14<J<16$, which represents main-sequence stars (mostly with spectral types of FGK and M0--M4). The broken dashed line indicates an empirical stellar--substellar boundary (Paper II). The $i$-band magnitudes of UL2307 and VLMS (magenta filled five-pointed stars) in the local field \citep{diet14} have been converted from $i_{\rm P1}$ with equation (\ref{ei}). We converted 2MASS magnitudes to MKO magnitudes for L subdwarfs not observed in UKIDSS. The blue open circle on the right indicates 2M0616 with an estimated $i-J$ colour. The blue open circle in the middle indicates UL0208 with uncertain $K$ band photometry. The BT-Dusty model grid with log $g$ = 5.5 is plotted for comparison (with $T_{\rm eff}$ and [Fe/H] indicated). M and L subtypes are marked at corresponding locations by their average colours. UL2332+12 (magenta open square) and SD1331 (magenta open pentagon) are joined by a magenta dashed line (see Section \ref{sbbdb}).}
\label{ijk}
\end{center}
\end{figure*}

\begin{figure*}
\begin{center}
   \includegraphics[angle=0,width=\textwidth]{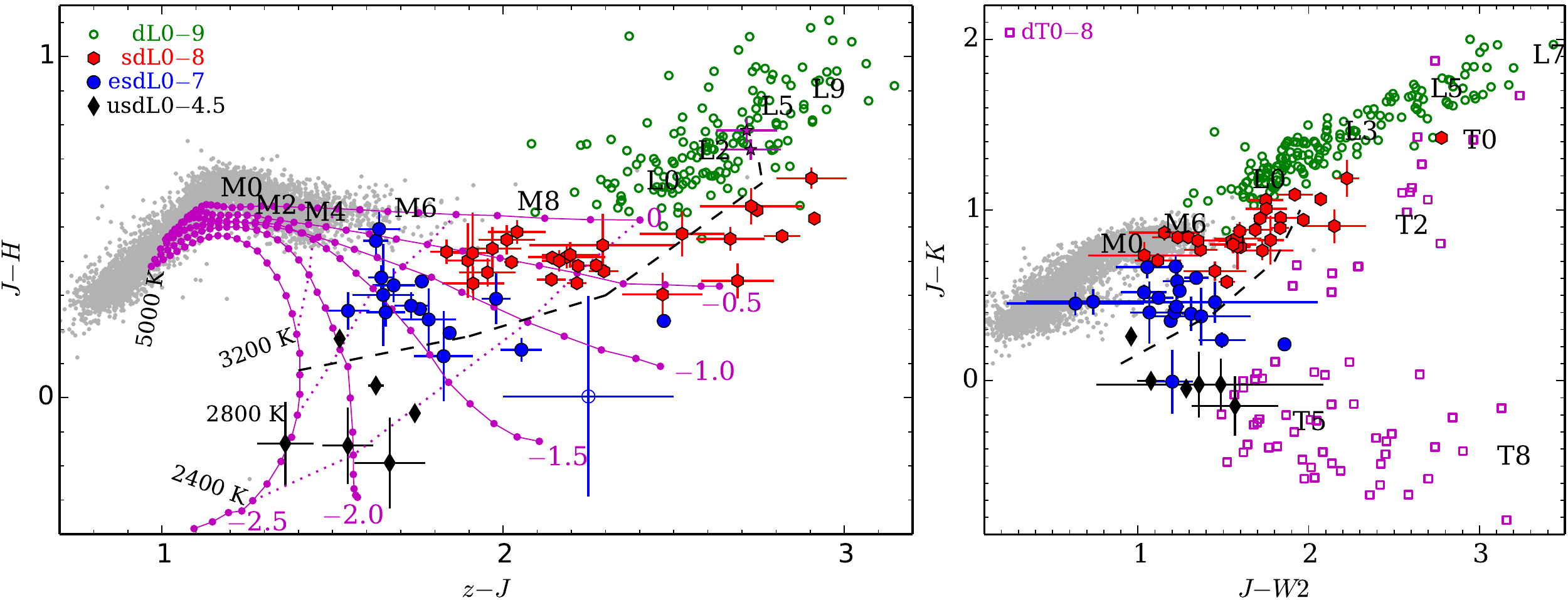}
\caption[]{The $z-J$ versus $J-H$ (left) and $J-W2$ versus $J-K$ (right) colour--colour plots for the L subdwarfs. Symbols are as described in Fig. \ref{ijk}, with additional magenta open squares to represent T dwarfs.}
\label{zjh}
\end{center}
\end{figure*}

\begin{figure*}
\begin{center}
  \includegraphics[angle=0,width=\textwidth]{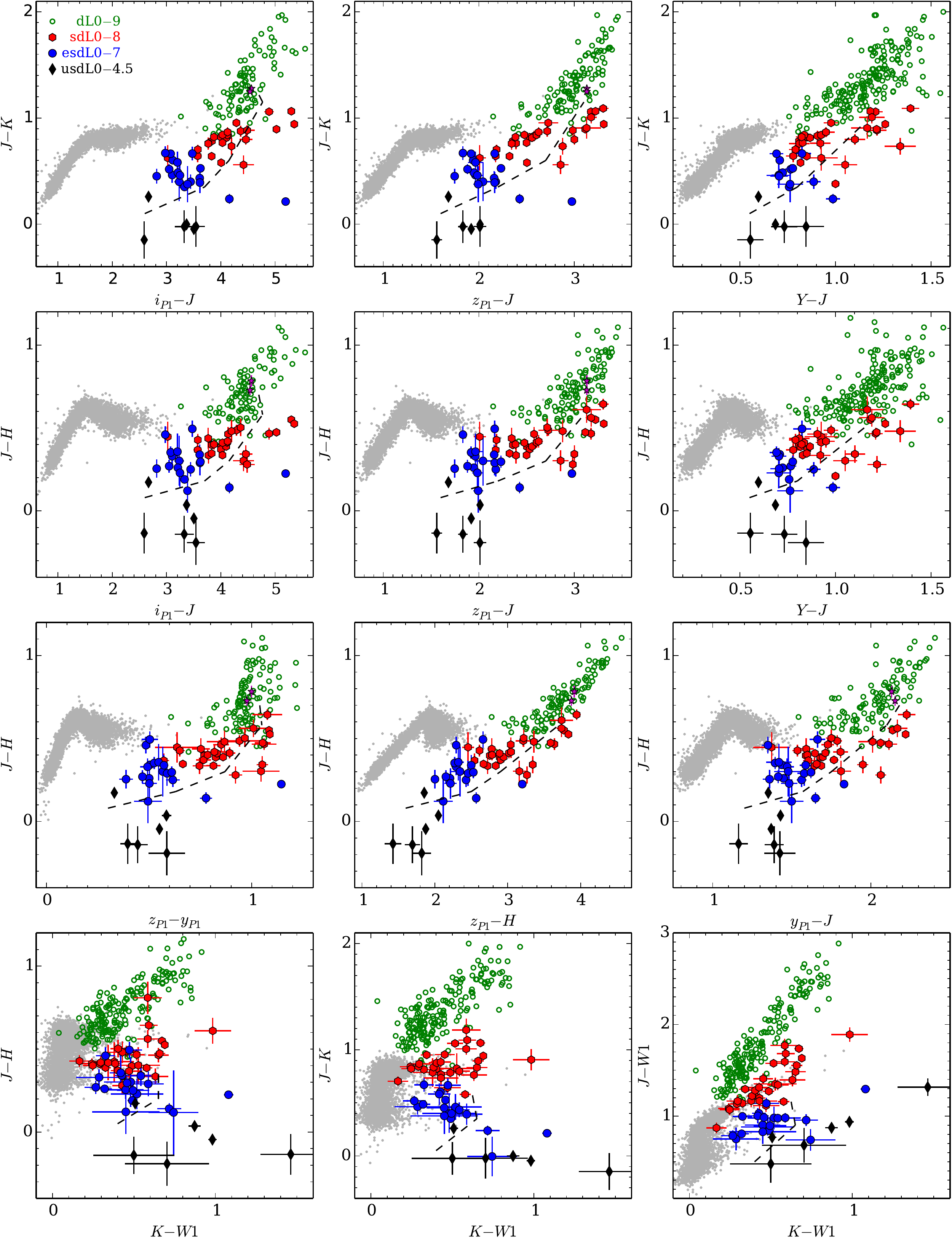}
\caption[]{Optical to infrared colour--colour plots for L subdwarfs. Symbols are as described in Fig. \ref{ijk}.}
\label{fallc}
\end{center}
\end{figure*}

\section{L subdwarf photometry and stellar/substellar mix}
\label{spop}
The label `subdwarf' originally acknowledged the location of these objects below the stellar main-sequence in the Hertzsprung-Russell diagram (HRD), due to their lower opacity and bluer colour resulting from subsolar metallicity. However, the three metallicity subclasses of L subdwarfs exhibit a range of different characteristics, sometimes showing quite substantial differences in spectral energy distribution from their L dwarf counterparts (see figs 14 and 15 of Paper I). In this section we explore the optical to infrared colours of L subdwarfs across the different subtypes and subclasses, compare them with L dwarf and main-sequence star populations, and assess the stellar/substellar mix amongst L subdwarfs.

\subsection{A photometric sample of L subdwarfs}
To study the photometric diversity of L subdwarfs, we joined our UKIDSS-SDSS sample with L subdwarfs collected from the literature to form a larger sample of 66 L subdwarfs. We gathered their photometry: SDSS $i$, $z$, UKIDSS (MKO) $Y, J, H, K$. We also used $i$ and $z$ photometry from the VLT Survey Telescope's (VST) ATLAS survey \citep{shan15} for a few objects that are not covered by SDSS. We used NIR photometry from the Visible and Infrared Survey Telescope for Astronomy (VISTA) surveys for a few objects that were not observed by UKIDSS. We also gathered $W1, W2$, and $W3$ photometry from the all-sky {\sl WISE} data release, and  $i_{\rm P1}, z_{\rm P1}$, and $y_{\rm P1}$ photometry from Pan-STARRS \citep[PS1;][]{cham16}. This compilation of optical to infrared photometry of 66 L subdwarfs is presented in Tables \ref{tsdlm} and \ref{tsdps}.

Some known L subdwarfs that were discovered by 2MASS and {\sl WISE} were not observed by UKIDSS or VISTA's Hemisphere Survey (VHS) and Kilo-Degree Infrared Galaxy Survey (VIKING). To convert 2MASS photometry ($J_2$, $H_2$, and $K_2$) into the MKO system, we determined polynomial relationships between MKO and 2MASS photometric differences and spectral type (SpT) for known L0--T0 dwarfs. Ideally, these relations would be based on L subdwarf measurements, but data availability requires us to take the next best approach, and use L dwarfs:
\begin{eqnarray}
\label{ej}
J - J_2 = -0.0230 - 0.0023 \times SpT; ~~~ (0.1062), \\ 
\label{eh}
H - H_2  = 0.0188 + 0.0029 \times SpT; ~~~ (0.1027), \\ 
\label{ek}
K - K_2  = -0.0977 + 0.0068 \times SpT; ~~~ (0.0938). 
\end{eqnarray}
where SpT is 10 for L0, 15 for L5, and 20 for T0. The root mean square (rms) of these fits are given after each equation.
 
Some L subdwarfs were not in the SDSS footprint but were observed by PS1. We thus studied the correlation between SDSS and PS1 photometry. We found that SDSS and PS1 photometry are similar in the $i$-band ($<$ 0.04 for main-sequence stars), but correlate with spectral type in the $z$-band. This is presumably because the SDSS and PS1 $i$-band filters have similar wavelength coverage and transmission profiles, though their $z$-band filters are different. These differences between SDSS and PS1 $z$-band magnitudes ($z - z_{\rm P1}$) are around $-$0.05 for FGK dwarfs, $-$0.11 for M0--M3 dwarfs, and $-$0.43 for L dwarfs. We determined polynomial relationships between SDSS--PS1 photometric differences and spectral type for known L0--T0 dwarfs:
\begin{eqnarray}
\label{ei}
i - i_{\rm P1} = -0.0565 + 0.0043 \times SpT; ~~~ (0.0944), \\
\label{ez}
z - z_{\rm P1}  = -0.3092 - 0.0086 \times SpT; ~~~ (0.0886). 
\end{eqnarray}
Fig. \ref{sptph} shows our relationships between SDSS--PS1 and MKO--2MASS photometric differences and spectral types for L0--T0 dwarfs.

Four of 66 known L subdwarfs (2M0041, WISEA J005757.65+201304.0, UL1241$-$00, and WI1355) are excluded from our sample due to problematic photometry or ambiguous spectral type. We thus have 62 L subdwarfs in our photometric sample.

\subsection{L subdwarf colour--spectral type relations}
To identify the best colours for distinguishing and characterizing L subdwarfs with different spectral types and subclasses, we assessed various optical to near- and mid-infrared colours. Fig. \ref{fsptc} shows optical to infrared colours of sdL, esdL and usdL subdwarfs compared to third-order polynomial fits of spectral type -- colour relations for L dwarfs. The coefficients of these polynomial fits and rms are presented in Table \ref{tcolour}. L dwarfs used for these polynomial fits are collected from DwarfArchives.org, and cross-matched within the SDSS, UKIDSS, PS1, and {\sl WISE} data bases. Note that a few objects near the {\sl WISE} detection limit have photometric uncertainties of 0.6 mag.

The $J-H$, $J-K$, $J-W2$, $y_{\rm P1} - H$, $z_{\rm P1} - H$, $y_{\rm P1}-K$, $Y-W1$, and $J-W1$ are the best metallicity indicators, as they can be used to separate sdL, esdL, and usdL subclasses from L dwarfs. The $i-J$, $z-J$, $i_{\rm P1} - y_{\rm P1}$, $z_{\rm P1} - y_{\rm P1}$, $i_{\rm P1} - J$, $y_{\rm P1} - J$, and $i_{\rm P1} - H$ can be used to separate esdL and usdL from L dwarfs, but cannot distinguish sdL from L dwarfs well. 

The esdL subclass has bluer $H-W1$ and $H-W2$ colours compared to L dwarfs, but the usdL subclass tends to have relatively redder $H-W1$ and $H-W2$ colours. The most metal-poor usdL subdwarf, SD0104+15 (Paper II), shows a redder $H-W2$ and much redder $H-W1$ compared to L dwarfs. The esdL and usdL subclasses are redder in $K-W1$ than the L dwarfs. Note that SD0104+15 is much redder in $K-W1$ than the L dwarfs.

The $i-z$, $i_{\rm P1} - z_{\rm P1}$, $H-W2$, $W1-W2$, and $W2-W3$ colours are not good metallicity indicators. The spectral subtypes of L subdwarfs are based on red optical spectra ($i$ and $z$ bands) by comparison to L dwarfs (Paper I). Therefore, L subdwarfs have similar $i-z$ colours to L dwarfs with the same subtype. However, early-type esdL and usdL subdwarfs have a different optical spectral profile to early-type L dwarfs, and have bluer $i-z$ and $i_{\rm P1} - z_{\rm P1}$ colours compared to L dwarfs. L subdwarfs also appear to be slightly redder in $W2-W3$ than the L dwarfs.

\begin{figure}
\begin{center}
  \includegraphics[angle=0,width=\columnwidth]{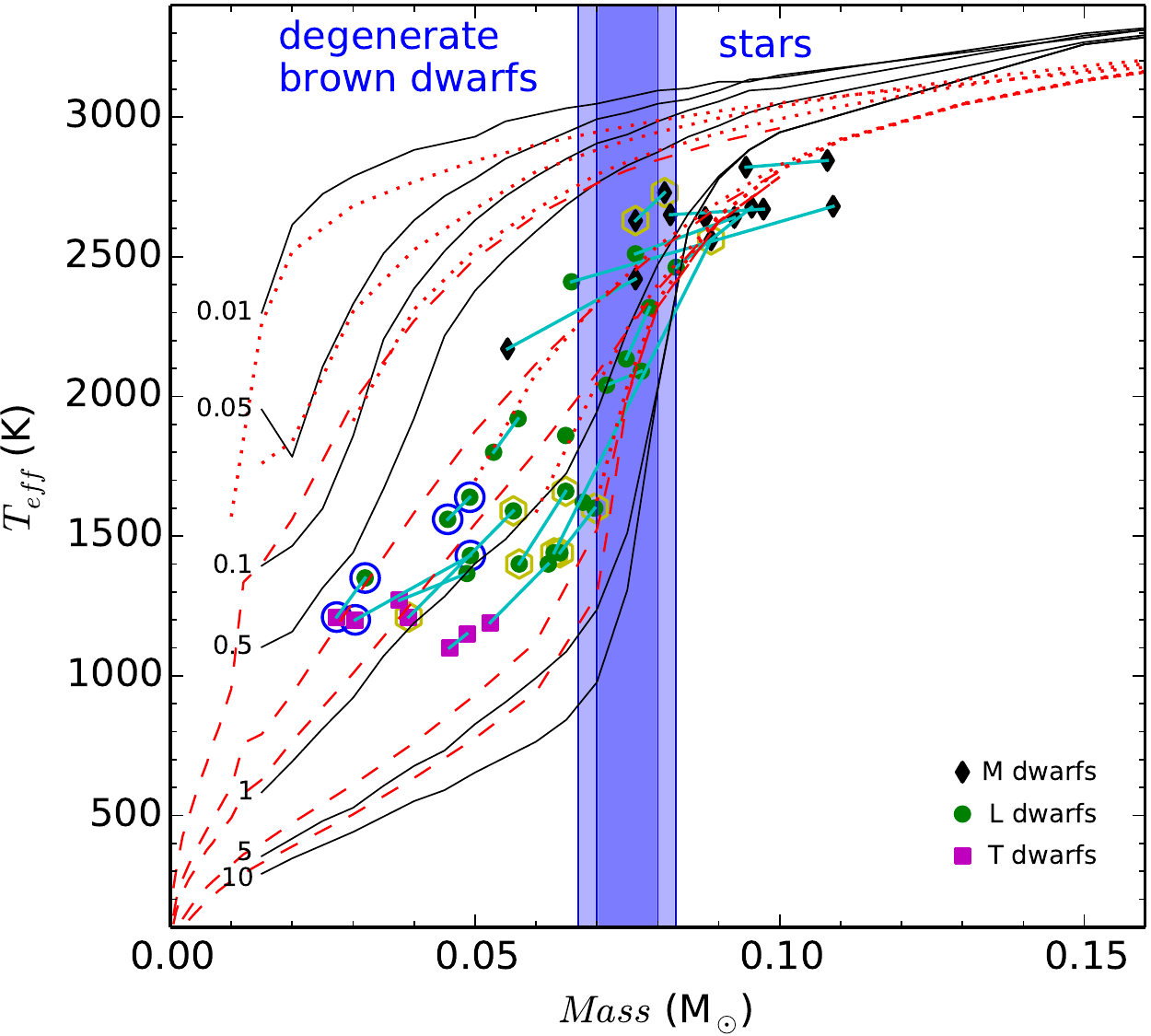}
\caption[]{Solar metallicity isochrones for VLMS and BDs at 0.01, 0.05, 0.1, 0.5, 1, 5, and 10 Gyr. The black, red dashed and dotted lines are isochrones from \citet{burr97} and \citet{bara03,bara15}, respectively. A blue shaded band indicates the BD transition-zone at solar metallicity, which has a small shift between different model predictions. M (black diamonds), L (green circles), and T (magenta squares) dwarfs in binary systems (joined with cyan lines) with dynamical mass measurements \citep{dupu17,lazo18} are also over plotted. Blue open circles and yellow open hexagons indicate objects with and without lithium detection in their spectra, respectively. }
\label{fisoc}
\end{center}
\end{figure}

\begin{figure*}
\begin{center}
  \includegraphics[angle=0,width=\textwidth]{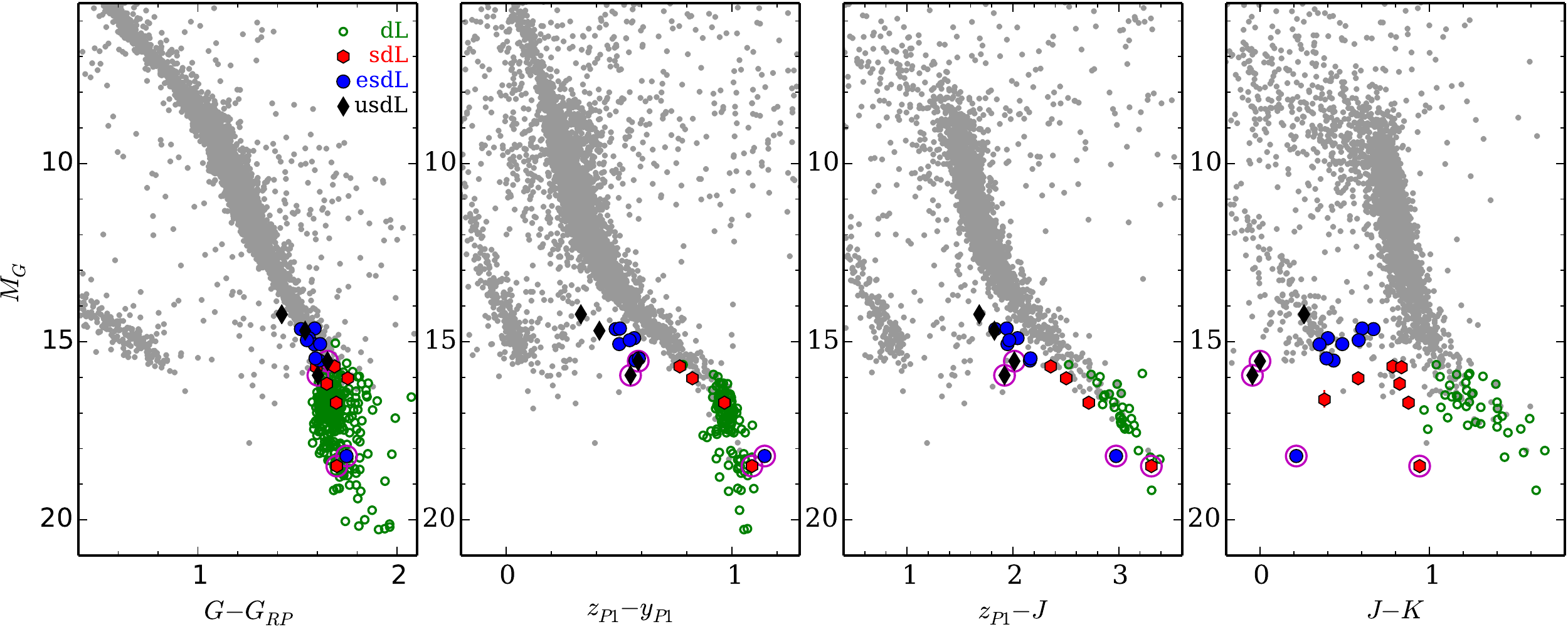}
\caption[]{The Hertzsprung-Russell diagram for L subdwarfs in comparison to field objects. Symbols are as described in Fig. \ref{ijk}. Transitional BDs are indicated with magenta circles. Grey dots are objects selected from {\sl Gaia} DR2, PS1, and LAS with distance $<$ 100 pc, $180\degr < RA < 220\degr$ and $0\degr < Dec. < 20\degr$. The two grey sequences are white dwarfs (left) and main-sequence stars (right). Some field stars are scattered mostly because they are too bright in the PS1 and UKIDSS fields. }
\label{fhrd}
\end{center}
\end{figure*}

\begin{figure*}
\begin{center}
  \includegraphics[angle=0,width=\textwidth]{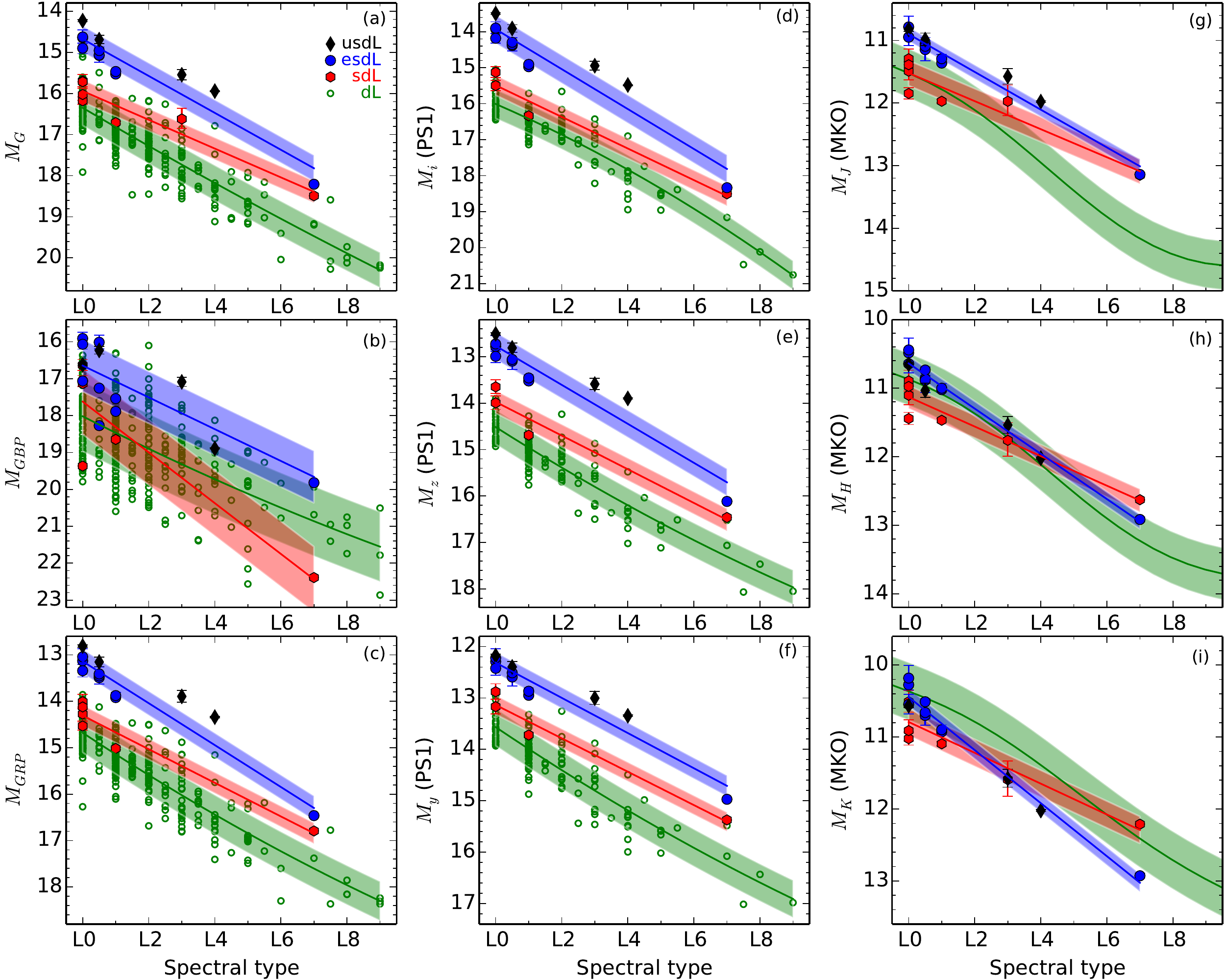}
\caption[]{Correlations between spectral types and absolute magnitudes. The blue, red, and green lines indicate our polynomial fits for esdL/usdL, sdL, and dL subclasses. Their rms are indicated by shaded areas. The polynomial fits for field dwarfs in $J, H$ and $K$ bands are from \citet{dup12}. Other symbols are as described in Fig. \ref{ijk}.}
\label{fsptmag}
\end{center}
\end{figure*}

\begin{figure}
\begin{center}
  \includegraphics[angle=0,width=\columnwidth]{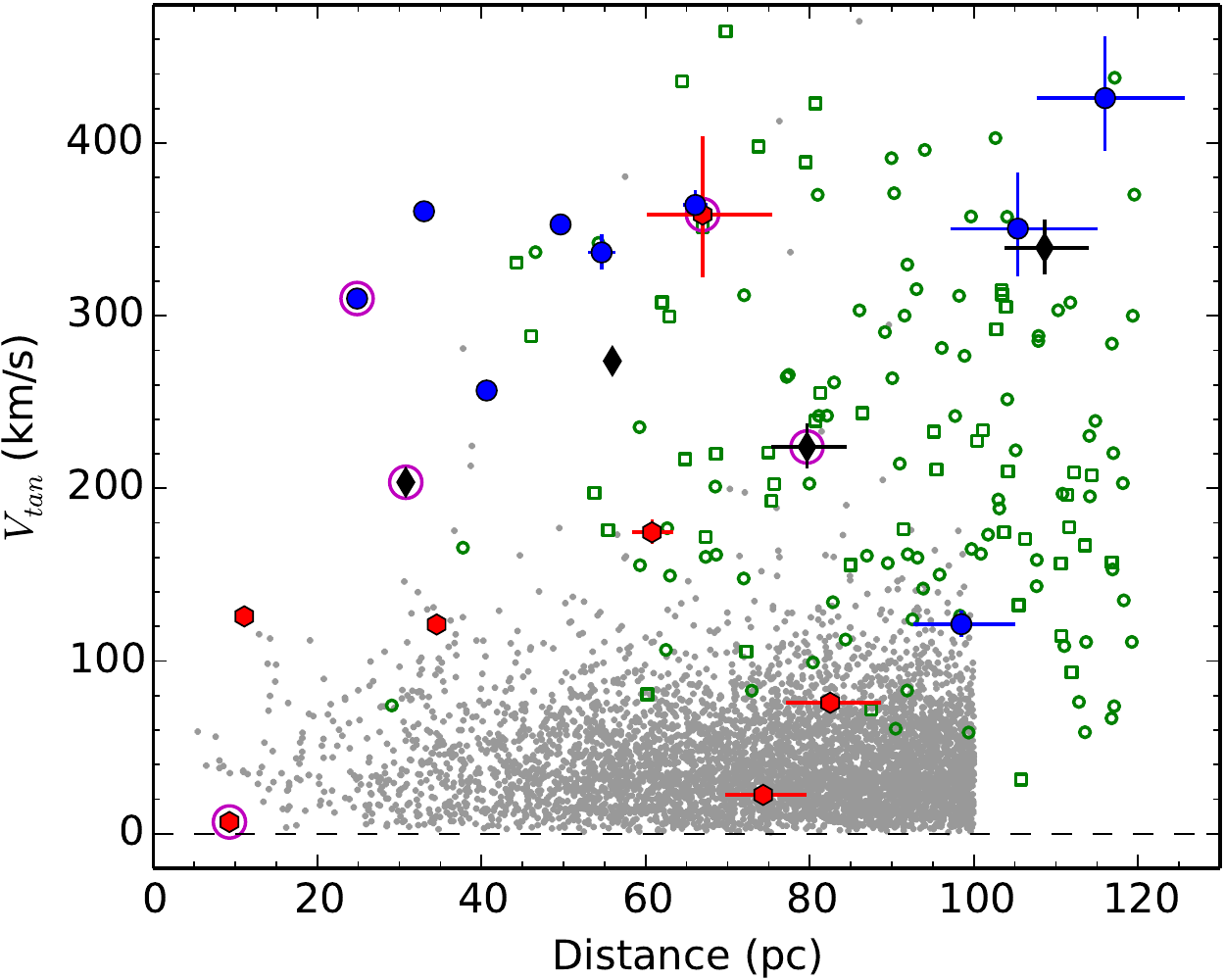}
\caption[]{Tangential velocities of L subdwarfs (symbols are as described in Fig. \ref{ijk}) compared to field stars (grey dots as in Fig. \ref{fhrd}). Note, error bars are smaller than the symbol size for some L subdwarfs. Transitional BDs are indicated with magenta circles. SDSS spectroscopic confirmed esdM and usdM subdwarfs in {\sl Gaia} DR2 are indicated as green open circles and squares, respectively. These esdM and usdM have distances of $<$ 200 pc and were shorten by 0.6 times for better comparison in the plot. }
\label{fvtan}
\end{center}
\end{figure}

\subsection{L subdwarfs in colour--colour plots}
\label{sccp}
Colour--colour plots are often used to distinguish celestial objects of different types in large-scale photometric surveys. Particular objects occupy a variety of colour spaces, with varying level of contamination from other populations \citep[e.g. fig. 2 of][]{zha13}.

Fig. \ref{ijk} shows the $i-J$ versus $J-K$ colour--colour plot for sdL, esdL, and usdL subdwarfs compared to L dwarfs and main-sequence stars. The figure also shows the BT-Dusty model predicted colours of L subdwarfs \citep{alla14}. The different metallicity ranges of these four subclasses are well represented by their broad-band optical-NIR colours, and the dL, sdL, esdL, and usdL sequences are quite distinct from late M through the L types. The $i-J$ and $J-K$ colours are very useful for selecting L subdwarf candidates. The $i-J$ colour is a good indicator of $T_{\rm eff}$ and can be used to separate L subdwarfs from main-sequence stars, while the $J-K$ colour is a good indicator of metallicity and can be used to separate L subdwarfs from L dwarfs.

Fig. \ref{zjh} (left) shows the $z-J$ versus $J-H$ colour--colour plot for L subdwarfs and dwarfs. The main-sequence population and BT-Dusty model grid with log $g$ = 5.5 are plotted for comparison. The dL, sdL, esdL, and usdL sequences are revealed in the $z-J$ versus $J-H$ plane, although not as well as in Fig. \ref{ijk}. 
Fig. \ref{zjh} (right) shows the $J-W2$ versus $J-K$ colour-colour plot showing dL, dT, sdL, esdL and usdL populations. The L subdwarf populations are generally separated from each other by their $J-W2$ and $J-K$ colours, although they are very close to the main-sequence. Late-type L subdwarfs begin to merge with the T dwarfs on the right side of the diagram.

To further explore the potential of future L subdwarf searches and characterization, Fig. \ref{fallc} compares our L subdwarf photometric sample to L dwarfs and main-sequence stars in twelve optical/NIR, NIR, and NIR/mid-IR colour-colour diagrams (constructed using the PS1, MKO, and {\sl WISE} photometric systems). $J-K$ or $J-H$ colour are used for 1 of these plots, and are sensitive to metallicity. As for Figs \ref{ijk} and \ref{zjh}, subdwarfs of different subclasses in these eleven plots are generally well separated. 
The longer base-line optical/NIR colours (e.g., $i_{\rm P1}-J$ and $z_{\rm P1}-J$) are relatively better at separating objects with different spectral type. The last three colour--colour plots in Fig. \ref{fallc} use the $K-W1$ colour and are very good at isolating the most metal-poor L subdwarfs (usdLs), which have red $K-W1$ colour (also see Fig. \ref{fsptc}). 

\subsection{The BD transition-zone}
\label{sbdtz} 
The luminosity (or $T_{\rm eff}$) is a decreasing function of metallicity above the hydrogen-burning minimum mass (HBMM), but an increasing function of metallicity below it. A transition-zone between VLMS and degenerate BDs is revealed as a waterfall-like feature on the 10 Gyr mass--$T_{\rm eff}$ isochrones \citep[fig. 5 of][]{burr01}. We used the intersection points between the 10 Gyr mass--$T_{\rm eff}$ isochrones with different metallicities to define the steady HBMM and identified the corresponding $T_{\rm eff}$ of steady HBMM at 10 Gyr as a function of metallicity. We also found that the transition-zone occupied a very large area in the $T_{\rm eff}$ versus [Fe/H] parameter space (fig. 9 of Paper II).  

Fig. \ref{fisoc} shows solar metallicity isochrones of VLMS and substellar objects from 0.01 to 10 Gyr. We also over plotted VLMS and BD binaries with dynamical mass measurements in/near the BD transition-zone. VLMS on the right-hand side of the transition-zone have steady hydrogen fusion and could maintain their luminosity/$T_{\rm eff}$ throughout their very long lifetimes. Meanwhile, degenerate BDs on the left-hand side of the transition-zone have no hydrogen fusion to provide energy, thus cool and dim over time. In other words, the luminosity/$T_{\rm eff}$ distinction between degenerate BD and VLMS populations is extended over time and the transition-zone is the stretched area between them (also see fig. 5 of Paper III).

Transitional BDs in the transition-zone are fully convective and have sporadical unsteady hydrogen fusion in their cores which partially contributes to their luminosity (Paper II \& III). As the initial thermal energy of a transitional BD is slowly dissipating over time, the unsteady hydrogen fusion slowly becomes the dominate energy source to maintain its luminosity. Since the core temperature of a transitional BD declines slowly over time. The efficiency of the fusion also declines very slowly over time. The fusion is very sensitive to the mass of transitional BDs. Therefore, field transitional BDs at a certain age could span in a wide $T_{\rm eff}$ range within a narrow mass range. Note such a $T_{\rm eff}$ range is even wider at older age or lower metallicity. The long-lasting unsteady hydrogen fusion in the cores of transitional BDs slowed down their cooling speed and also made them different from both VLMS and degenerate BDs. 

Most of field degenerate BDs would have evolved into the mid- to late-type T dwarf domain by their ages of $\sim$0.5--5 Gyr. The rest of them with relatively younger age or higher mass would be crossing the L dwarf domain. This is the main reason for that L dwarfs have a much lower number density than T dwarfs in the solar neighbourhood \citep{bur07b,kir12}. Field L dwarfs are composed of VLMS, transitional BDs, and degenerate BDs which are difficult to distinguish by observation without knowing their mass or age. Field transitional BDs are mostly L dwarfs, but could have spectral types of late-type M and early-type T, depending on their mass and age. The BD transition-zone, which has not drawn much attention in the past, further blurred the observational stellar/substellar boundary in addition to the mass/age degeneracy. This is why we have not reached an agreement on the observational stellar/substellar boundary among field population. 

\subsection{The observational stellar/substellar boundary}
\label{sssb} 
Transitional BDs in the field are difficult to identify by observation because of the mass/age degeneracy and relatively small luminosity/temperature distinction \citep[e.g. GD 165B;][]{kir99b}. However, transitional BDs in the halo are distributed in a very wide range of luminosity/temperature after over $\sim$10 Gyr of cooling to reduce their initial thermal energy. Nine known halo transitional BDs identified by their atmospheric parameters are summarised in Paper III. They are located in a very broad halo BD transition-zone in the $T_{\rm eff}$ versus [Fe/H] parameter space. 

To identify an empirical observational boundary between metal-poor stars and transitional BDs, we compared known metal-poor transitional BDs to least-massive stars in the halo, thick disc (in fig. 9 of Paper II), and the solar neighbourhood \citep{diet14} by their relative locations on colour--colour plots presented in Section \ref{sccp}. Fig. \ref{ijk} shows that metal-poor transitional BDs can be well separated from stars by their $i-J$ and $J-K$ colours that are sensitive to subtype (i.e. $T_{\rm eff}$) and subclass (i.e. [Fe/H]), respectively. Therefore we draw an empirical stellar--substellar boundary in Fig. \ref{ijk} that is shown as a black dashed broken line. 

The stellar--substellar boundary was drawn where we can separate known transitional BDs from least-massive stars as much as possible by their $i-J$ and $J-K$ colours. Then we lead this boundary to the right side of the two least-massive field stars suggested by \citet{diet14}. Eight out of these nine known halo transitional BDs in Paper III lie below/redward of this stellar--substellar boundary. However, we note one apparent inconsistency, since UL0208 appears just above the boundary (blue open circle in the middle of Fig. \ref{ijk}). We believe this is likely due to the uncertain $K$ band photometry of this object (giving a larger $J-K$ error), and further note that this object lies below the boundary in Fig. \ref{zjh} as a result of its NIR $J-H$ colour. It can also be seen that there are a significant number of esdL stars clustered just above/blueward of the boundary, and considerably fewer esdL subdwarfs below/redward of it. This dearth of objects in the colour space on the substellar side of the boundary represents a substellar subdwarf gap spanning mid L to early T types. This region (the transition-zone) covers a wide $T_{\rm eff}$ range but a narrow mass range, which results in relatively fewer objects compared to halo VLMS.

We also draw stellar/substellar boundaries on other colour--colour plots composed of $T_{\rm eff}$ and [Fe/H] sensitive colours (Figs \ref{zjh} and \ref{fallc}). The corresponding kink points on these boundaries in different plots have the same value for each colour. So that the boundaries in different colour-colour plots are consistent. Metal-poor transitional BDs are well separated from stars across the sdL, esdL and usdL populations in Fig. \ref{zjh} and these first nine plots of Fig. \ref{fallc}. 
The last three plots in Fig. \ref{fallc} use the $K-W1$ colour could only separate transitional BDs of esdL/usdL subclasses and invalid for sdL subclass.

Using the stellar/substellar boundaries from these plots in Figs \ref{ijk}--\ref{fallc}, we estimate that about 22 of these 62 L subdwarfs in our photometric sample are likely metal-poor transitional BDs based on their spectral subtype/subclass and multiband colours. These substellar subdwarfs generally have spectral types of L3 or later except the usdL1.5 SD0104+15 which has extremely low metallicity corresponding to a higher HBMM. 
 
Note that sdL3/esdL3/usdL1.5 are likely the earliest spectral types for metal-poor transitional BDs which are in the substellar subdwarf gap between VLMS and degenerate BDs. Halo degenerate BDs would have temperature below $\sim$1000 K and have cooled into the T- and Y-type region (Paper II; \citealt{burr01}). Field degenerate BDs could have much earlier spectral types depending on their ages. Degenerate BDs do not have unsteady hydrogen fusion, but the most massive ones above $\sim$0.055 M$_{\sun}$ (Fig. \ref{fisoc}) would be able to fuse lithium. Therefore, ultra-cool dwarfs with lithium absorption lines in their spectra would fall in the degenerate BD domain \citep[e.g. Teide 1;][]{rebo96}. However, lithium absorption lines are not expected in the spectrum of halo degenerate BDs because of their low temperature (section 3.5 of Paper III). 

The least-massive field stars above the HBMM have spectral types of around L2.5  \citep[2MASSI J0523382$-$140302;][]{diet14} and L3 \citep[2MASSI J1017075+130839B;][]{dupu17}. The latest spectral types of stars in our L subdwarf sample are sdL1/esdL1/usdL0. Note we do not have sdL2/esdL2/usdL1 subdwarfs in our sample; thus we do not know if they could be stars, as they would be very close to the stellar boundary if they exist. The subtypes of L subdwarfs are assigned by comparing their optical spectra with those of L dwarf standards. L subdwarfs are hotter, more luminous, and more massive than L dwarfs with the same spectral type. Meanwhile, the HBMM is higher at lower metallicity (Paper II and III). As a consequence of these two facts, the latest spectral types of stars is slightly shifting to earlier subtypes across the dL, sdL, esdL, and usdL subclasses with decreasing metallicity. 

\begin{table*}
 \centering
  \caption[]{{\sl Gaia} DR2 astrometry and photometry of 20 L subdwarfs.} 
\label{tgaia}
  \begin{tabular}{c c  r r r  c c c }
\hline
Name  &  {\sl Gaia} DR2 ID & $\pi$ (mas) &   $\mu_{\rm RA}$(mas/yr) & $\mu_{\rm Dec}$(mas/yr) &  $G$ &  $G_{\rm BP}$ & $G_{\rm RP}$  \\ 
\hline
WISEA J001450.17$-$083823.4 &  2429285424478485504 & 20.14$\pm$0.28 & 1477.22$\pm$0.55 & $-$257.96$\pm$0.28 & 18.13 & 20.54 & 16.61 \\ 
WISEA J011639.05$-$165420.5 &  2358397882610264960 & 16.45$\pm$0.68 & 602.21$\pm$1.34 & 67.57$\pm$1.16 & 20.10 & 21.05 & 18.45 \\ 
WISEA J020201.25$-$313645.2 &  5019582949474822400 & 15.14$\pm$0.34 & $-$155.09$\pm$0.54 & $-$1152.44$\pm$0.36 & 19.18 & 21.36 & 17.59 \\ 
WISEA J030601.66$-$033059.0 &  5183457632811832960 & 24.61$\pm$0.30 & 331.48$\pm$0.51 & $-$1290.79$\pm$0.52 & 18.57 & 20.59 & 16.96 \\ 
ULAS J033351.10+001405.8 &  3264554308968041728 & 8.62$\pm$0.67 & 772.09$\pm$0.93 & $-$65.70$\pm$1.03 & 19.95 & 21.23 & 18.36 \\ 
WISEA J043535.82+211508.9 &  144711230753602048 & 18.30$\pm$0.55 & 871.71$\pm$0.95 & $-$964.06$\pm$0.61 & 19.16 & 21.57 & 17.57 \\ 
2MASS J05325346+8246465 &  558122277038055808 & 40.24$\pm$0.64 & 2038.34$\pm$1.09 & $-$1663.73$\pm$1.23 & 20.19 & 21.80 & 18.44 \\ 
ULAS J075335.23+200622.4 &  673344017522527488 & 12.12$\pm$0.84 & $-$36.37$\pm$1.45 & $-$190.54$\pm$0.97 & 20.28 & 21.22 & 18.59 \\ 
SSSPM J10130734$-$1356204 &  3753081236588280192 & 17.87$\pm$0.22 & 68.07$\pm$0.34 & $-$1029.79$\pm$0.30 & 17.97 & 20.37 & 16.55 \\ 
ULAS J124425.75+102439.3 &  3927262643141395584 & 9.49$\pm$0.80 & $-$441.87$\pm$1.79 & $-$545.11$\pm$0.78 & 20.18 & 21.13 & 18.57 \\ 
SDSS J125637.13$-$022452.4 &  3685444645661181696 & 12.55$\pm$0.72 & $-$511.25$\pm$1.33 & $-$300.64$\pm$0.83 & 20.05 & 21.59 & 18.40 \\ 
VVV J12564163$-$6202039 &  5863122429178232704 & 14.94$\pm$1.68 & $-$1129.81$\pm$3.25 & 21.29$\pm$2.56 & 20.75 & --- & --- \\ 
ULAS J134749.79+333601.7 &  1458522725665649536 & 13.45$\pm$0.90 & 62.43$\pm$1.11 & $-$13.61$\pm$1.14 & 20.38 & 21.42 & 18.63 \\ 
SDSS J141624.12+134827.4 &  1227133699053734528 & 107.56$\pm$0.30 & 85.69$\pm$0.69 & 129.07$\pm$0.47 & 18.33 & 22.23 & 16.63 \\ 
2MASS J16262034+3925190 &  1332410734823626240 & 32.49$\pm$0.23 & $-$1374.66$\pm$0.35 & 237.36$\pm$0.44 & 18.38 & 21.33 & 16.78 \\ 
2MASS J16403197+1231068 &  4460894909281150208 & 10.15$\pm$0.64 & $-$222.25$\pm$1.14 & $-$134.90$\pm$0.91 & 19.87 & 21.04 & 18.31 \\ 
2MASS J17561080+2815238 &  4584405146372926720 & 28.94$\pm$0.37 & $-$613.74$\pm$0.57 & $-$413.40$\pm$0.67 & 19.40 & 21.34 & 17.71 \\ 
LSR J182611.3+301419.1 &  4588438567346043776 & 90.00$\pm$0.11 & $-$2290.54$\pm$0.17 & $-$683.13$\pm$0.18 & 15.95 & 19.60 & 14.35 \\ 
WISEA J204027.30+695924.1 &  2271357312343219456 & 30.30$\pm$0.13 & 1558.38$\pm$0.27 & 1697.36$\pm$0.25 & 17.55 & 20.87 & 16.00 \\ 
WISEA J213409.15+713236.1 &  2272533033868183168 & 9.21$\pm$0.43 & 466.05$\pm$0.68 & 466.05$\pm$0.68 & 19.87 & 21.41 & 18.33 \\ 
\hline
\end{tabular}
\end{table*}

\begin{table}
 \centering
  \caption[]{Coefficients of first-order polynomial fits of  absolute magnitude as a function of spectral types ($SpT$) for L0--L7 subdwarfs in Fig. \ref{fsptmag}. The fits are defined as  $M = c_0 + c_1 \times SpT$,  $SpT$ = 10 for L0 and $SpT$ = 17 for L7. }
 \label{tmabs}
  \begin{tabular}{l r r r r}
\hline
    $M_{abs}$  & $c_0~$ & $c_1~~$  & rms & Subclass \\
\hline
$G$ ({\sl Gaia}) & 12.441 & 0.3501 & 0.267 & sdL \\
$G$ ({\sl Gaia}) & 10.191 & 0.4488 & 0.316 & esdL/usdL \\
$G_{BP}$ ({\sl Gaia}) & 10.745 & 0.6878 & 0.884 & sdL \\
$G_{BP}$ ({\sl Gaia}) & 12.339 & 0.4307 & 0.703 & esdL/usdL \\
$G_{RP}$ ({\sl Gaia}) & 10.689 & 0.3620 & 0.223 & sdL \\
$G_{RP}$ ({\sl Gaia}) & 8.627 & 0.4512 & 0.259 & esdL/usdL \\
$M_i$ (PS1)  & 11.094 & 0.4394 & 0.274 & sdL \\
$M_i$ (PS1)  & 8.385 & 0.5546 & 0.388 & esdL/usdL \\
$M_z$ (PS1)  & 10.345 & 0.3627 & 0.241 & sdL \\
$M_z$ (PS1)  & 8.569 & 0.4197 & 0.291 & esdL/usdL \\
$M_y$ (PS1)  & 9.888 & 0.3248 & 0.184 & sdL \\
$M_y$ (PS1)  & 8.890 & 0.3424 & 0.201 & esdL/usdL \\
$M_J$ (MKO) & 9.279 & 0.2242 & 0.200 & sdL \\
$M_J$ (MKO) & 7.883 & 0.3016 & 0.110 & esdL/usdL \\
$M_H$ (MKO) & 8.972 & 0.2157 & 0.167 & sdL \\
$M_H$ (MKO) & 7.339 & 0.3299 & 0.106 & esdL/usdL \\
$M_K$ (MKO) & 8.647 & 0.2144 & 0.187 & sdL \\
$M_K$ (MKO) & 6.760 & 0.3685 & 0.124 & esdL/usdL \\
\hline
\end{tabular}
\end{table}

\section{{\sl Gaia} observations of L subdwarfs}
\label{sgaia} 
The ESA's {\sl Gaia} \citep{gaia16} astrometric survey have measured precise parallaxes and proper motions \citep{lind18} for $\sim$ 1.332 billion stars in its second data release \citep[DR2;][]{gaia18}. The {\sl Gaia} survey has three optical pass-bands ($G$, $G_{\rm BP}$ and $G_{\rm RP}$) that are not very sensitive for ultra-cool dwarfs. Ultra-cool subdwarfs is slightly easier to be detected by {\sl Gaia} than dwarfs with the same spectral types. As ultra-cool subdwarfs have brighter magnitude in the optical \citep[see fig. 3 in][]{zha13}.  
 
Twenty of 66 known L subdwarfs in Table \ref{tsdlm} were observed by  {\sl Gaia}. Table \ref{tgaia} shows the {\sl Gaia} astrometry and photometry of these 20 L subdwarfs. As our UKIDSS-SDSS sruvey is relatively deep, only two of our L subdwarfs are in {\sl Gaia} DR2; UL0753+20 is from this work and ULAS J134749.79+333601.7 was presented in Paper I. The rest are previous known L subdwarfs. 

Fig. \ref{fhrd} shows the HRD of 20 L subdwarfs in comparison to field stars and L dwarfs in {\sl Gaia} DR2.  L subdwarfs have slightly bluer $G-G_{\rm RP}$ colour than L dwarfs; however, L subdwarfs do not appear as `sub' dwarfs in the HRD of $G$ band absolute magnitude ($M_G$) versus $G-G_{\rm RP}$. Because L subdwarfs not only have bluer $G-G_{\rm RP}$ but also brighter $G$ band absolute magnitude than L dwarfs therefore are shifted to upper left towards to M dwarfs. Early- to mid-type L subdwarfs could be as bright as M dwarfs at $G$ band. L subdwarfs appear as `sub' dwarfs in the HRD of $z_{\rm P1} - y_{\rm P1}$, $z_{\rm P1} - J$, and $J-K$ colours at which they are much bluer than L dwarfs (see Fig. \ref{fsptc}).   

To better understand the properties of L subdwarfs and estimate the distance of L subdwarfs that are not in {\sl Gaia} DR2, we studied the correlation between spectral type and absolute magnitude of L0--7 subdwarfs. Fig. \ref{fsptmag} shows the first-order polynomial fits of relationships between their spectral types and {\sl Gaia}, PS1, and MKO absolute magnitudes. Table \ref{tmabs} shows the coefficients of these fittings (see Table \ref{tcolour} for L0--9 dwarfs). The $G_{\rm BP}$ band magnitudes of known L subdwarfs/dwarfs are likely close to {\sl Gaia}'s detection limit; therefore, these $M_{\rm GBP}$ in Fig. \ref{fsptmag} (b) might not be reliable. In general, the $G$ to $J$ absolute magnitudes get brighter from dL to sdL and esdL/usdL subclasses. The esdL/usdL subclasses have similar $M_H$ to L dwarfs but fainter $M_K$. These are similar to what were shown in fig. 3 of \citet{zha13}. However, the $M_H$ of sdL subclass seams fainter at L0 but brighter at L7 than dL, esdL/usdL subclasses. The $M_K$ of sdL subclass is also fainter at L0 but brighter at L7 than esdL/usdL subclasses.  

Fig. \ref{fvtan} shows the tangential velocities of 20 L subdwarfs observed by {\sl Gaia}. These field stars in Fig. \ref{fvtan} have an median $V_{tan}$ of $\sim$ 36 km s$^{-1}$. Five of these seven sdL subdwarfs have $V_{tan} > 75$ km s$^{-1}$. All the esdL/usdL subdwarfs have $V_{tan} > 120$ km s$^{-1}$, and most of them are between 200 and 400 km s$^{-1}$. The esdL and usdL subclasses generally have halo kinematics, which is consisted to the esdM/usdM subclasses \citep[on the classification of][]{lep07}. The sdL subclass mostly have thick disc kinematics. There are a few relatively more metal-poor ([Fe/H] $\approx -1$) sdL subdwarfs have halo kinematics: UL0212+06, SD1333 (Paper I) and VVV J12564163$-$6202039 \citep{smit18}.

\section{Conclusions and future directions}
\label{ssum}

We present the discovery of 27 L subdwarfs including 5 esdL and 22 sdL subdwarfs. These new objects were classified according to the L subdwarf classification scheme presented in Paper I. Six of these L subdwarfs have spectral types between L3 and L8 and are likely substellar objects, while the other 21 L0--L1 subdwarfs are likely VLMS. We measured their proper motions and estimated their spectroscopic distances. We also measured the RV of three that have X-shooter spectroscopy. Our SDSS-UKIDSS programme has confirmed/classified 35 L subdwarfs in total (amongst a full known population of 66), including 11 probable BDs and 24 VLMS.

We also interpret one of our candidates (UL2332+12) as a mildly metal-poor unresolved binary consisting of a blue $\sim$L6p primary and a $\sim$T4p secondary. UL2332+12 has a high probability of thick disc membership by its kinematics. Metal-poor BD binaries are rare, but their properties and binary fraction may be very useful for our understanding of substellar formation in the early Galaxy \citep[e.g.][]{bate14,stam09}.

We have assessed optical to mid-infrared colours of the L subdwarf population, using colour--spectral type and 2-colour diagrams, comparing with both L dwarfs and main-sequence stars. We found that L subdwarfs of different metallicity subclasses can be well separated from L dwarfs and main-sequence stars using a range of optical/infrared colours. Colour spaces have been identified in which preferential selection can be made of the full range of L subdwarf subclasses, as well as separating stellar and substellar subdwarfs. This analysis shows that the photometric systems employed by the PS1, VST and VISTA surveys provide strong future potential for expansion of the known L subdwarf population out to its metallicity extremes, for which (based on the current sample) around a third will be substellar objects with the remaining two-thirds VLMS.

We also note that the PS1 $i_{\rm P1}, z_{\rm P1}$, and $y_{\rm P1}$ filters have similar wavelength coverage and transmission profiles to those planned for the next decade's Large Synoptic Survey Telescope \citep[LSST;][]{lsst17} and the {\sl Chinese Space Station Optical Survey} ({\sl CSS-OS}). The LSST will observe the southern hemisphere in six pass-bands to single-visit depths of 23.4, 22.2, and 21.6 mag and co-added depths of 26.4, 25.2, and 24.4 mag in the $i, z$, and $y$ bands, respectively, from 2022. A large area in the northern sky missed by the LSST will be covered by the {\sl CSS-OS}. The {\sl CSS-OS} will observe $\sim$ 17 500 deg$^2$ ($|b| > 15$ deg and $|Dec.| > 20$ deg) of the sky in seven bands to depths of 25.9, 25.2, and 24.4 mags in $i, z$, and $y$, respectively. The LSST could provide relative parallax distances for objects well brighter than its single-visit depth using the {\sl Gaia}'s reference frame. Meanwhile, the {\sl CSS-OS} have much deeper single-visit depth than the LSST, therefore, is better in detecting higher proper motion cool objects beyond LSST's single-visit depth. 

Furthermore, ESA's {\sl Euclid} \citep{laur11} space survey telescope is scheduled to launch in 2021, and aiming to observe half of the sky in four pass-bands to depths of 25 mag in VIS band and 24 mag in $Y, J$ and $H$ bands. Its slit-less spectroscopy will observe the 0.92--1.85 $\mu$m wavelength to a depth of $H \approx 19.5$ mag and could be used to identify T subdwarfs by their $Y/J$ index \citep{bur06c,mace13a}. The NASA's {\sl Wide-Field InfraRed Survey Telescope} \citep[{\sl WFIRST};][]{sper15} is planning to observe 2000 deg$^2$ of the sky in its high-latitude survey to depths of $Y=26.7, J=26.9, H=26.7$, and F184 = 26.2 from mid-2020s. 

These future optical and NIR sky surveys will provide great opportunity to the study large numbers of extremely metal-poor L subdwarfs and halo degenerate BDs (esdT/Y and usdT/Y types) in the near future. In particular, the $z$ to $H$ bands (covered by these facilities) will probe very large volumes for L subdwarfs, with the most extreme examples remaining reasonably bright in these bands.

The $K$ band flux of ultra-cool objects are the most sensitive wavelength to metallicity disparity. The $J-K$ colour is very useful in the identification and characterization of ultra-cool subdwarfs, particularly for T subdwarfs that have similar $J$- and $H$-band spectral profile to T dwarfs but stronger suppressed flux in $K$ band \citep[e.g.][]{burn14}. However, the $K$ band filter is not included in current survey strategies of both {\sl Euclid} and {\sl WFIRST}.
The  {\sl Euclid} and {\sl WFIRST} surveys will gain a lot more impacts on the science of ultra-cool objects, if it could include a $K$ band filter and extend the red cut-off wavelength of their NIR spectroscopy from $\sim$ 1.9 to 2.2 $\mu$m. The {\sl WFIRST} has a larger aperture size than the {\sl Euclid}, and would have a better capability in the $K$ band detection of T subdwarfs, which become very faint. 

We further discussed the BD transition-zone and properties of transitional BDs following Papers II and III. Degenerate BDs have an essentially different evolution from VLMS but their observational distinction are blurred by transitional BDs. Firstly, because the existence of the BD transition-zone was not widely realised. Secondly, least-massive stars, transitional and degenerate BDs are mixed in the L dwarf domain. Although, in the L dwarf domain, young or massive degenerate BDs are crossing, field transitional BDs are making a long stay or slowly crossing, meanwhile older least-massive stars are more permanent (at early-type L). The BD transition-zone worth in-depth studies by modelling and observation, and that would help us to better understand observations of substellar populations. For example, there is a lack of objects at the L/T transition \citep{bur07b}. This is firstly because the rapid evolution of BD atmospheres at $\sim$1200 K stretched the spectral subtype sampling. Secondly, the L/T transition is at the bottom of the BD transition-zone and next to the abundant degenerate BDs that crossed the BD transition-zone. 

Twenty of these 66 known L subdwarfs were observed by the {\sl Gaia} with precise astrometry. L subdwarfs do appear as `sub' dwarfs on the HRD with some specific colours (e.g. $z_{\rm P1} - y_{\rm P1}$, $z_{\rm P1}-J$ and $J-K$). Their absolute magnitudes are brighter in optical to $J$ band and fainter in $K$ band than L dwarfs. The esdL and usdL subclasses generally have halo kinematics and the sdL subclass mostly have thick disc kinematics, which is consisted to M subdwarfs. Five of these 20 L subdwarf in {\sl Gaia} DR2 are transitional BDs. Degenerate BDs or T subdwarfs are too faint for the {\sl Gaia} survey. However, we could get the precise astrometry of a T subdwarf if it has a bright companion observed by the {\sl Gaia}. For example, WISE J200520.38+542433.9 \citep{mace13b} is an sdT8 subdwarf companion (separated by 188.5 arcsec) to Wolf 1130 that has a {\sl Gaia} DR2 distance of 16.5587$\pm$0.0094 pc. Note that the uncertainty of the {\sl Gaia} distance of Wolf 1130 is about two-thirds of the projected separation from its cool companion. 

More wide binary systems contain both metal-poor BDs and FGKM stars are expected to be discovered in the near future \citep[e.g.][]{maro17}. Such systems can be used as benchmarks to characterize metal-poor BDs and test new ultra-cool atmosphere models and substellar evolutionary models. As we could have precise measurements of distances (from {\sl Gaia}) and abundances (from high resolution spectroscopy) of these bright FGKM primaries, which could be applied to their metal-poor BD companions.

\section*{Acknowledgements}
Based on observations made with the Gran Telescopio Canarias (GTC), installed in the Spanish Observatorio del Roque de los Muchachos of the Instituto de Astrof{\'i}sica de Canarias, in the island of La Palma. Based on observations collected at the European Organisation for Astronomical Research in the Southern Hemisphere under ESO programmes 094.C-0202, 095.C-0878, 096.C-0130, 096.C-0974, and 0101.C-0626. 

This work is based in part on data obtained as part of the UKIRT Infrared Deep Sky Survey. The UKIDSS project is defined in \citet{law07}. UKIDSS uses the UKIRT Wide Field Camera \citep[WFCAM;][]{casa07}. The photometric system is described in \citet{hew06}, and the calibration is described in \citet{hodg09}. The pipeline processing and science archive are described in \citet{irwi04} and \citet{hamb08}. 

Funding for the Sloan Digital Sky Survey (SDSS) has been provided by the Alfred P. Sloan Foundation, the Participating Institutions, the National Aeronautics and Space Administration, the National Science Foundation, the U.S. Department of Energy, the Japanese Monbukagakusho, and the Max Planck Society. The SDSS Web site is \url{http://www.sdss.org/}. The SDSS is managed by the Astrophysical Research Consortium (ARC) for the Participating Institutions. The Participating Institutions are The University of Chicago, Fermilab, the Institute for Advanced Study, the Japan Participation Group, The Johns Hopkins University, Los Alamos National Laboratory, the Max-Planck-Institute for Astronomy (MPIA), the Max-Planck-Institute for Astrophysics (MPA), New Mexico State University, University of Pittsburgh, Princeton University, the United States Naval Observatory, and the University of Washington.

The Pan-STARRS1 Surveys (PS1) and the PS1 public science archive have been made possible through contributions by the Institute for Astronomy, the University of Hawaii, the Pan-STARRS Project Office, the Max-Planck Society and its participating institutes, the Max Planck Institute for Astronomy, Heidelberg and the Max Planck Institute for Extraterrestrial Physics, Garching, The Johns Hopkins University, Durham University, the University of Edinburgh, the Queen's University Belfast, the Harvard-Smithsonian Center for Astrophysics, the Las Cumbres Observatory Global Telescope Network Incorporated, the National Central University of Taiwan, the Space Telescope Science Institute, the National Aeronautics and Space Administration under Grant No. NNX08AR22G issued through the Planetary Science Division of the NASA Science Mission Directorate, the National Science Foundation Grant No. AST-1238877, the University of Maryland, Eotvos Lorand University (ELTE), the Los Alamos National Laboratory, and the Gordon and Betty Moore Foundation.

This publication makes use of data products from the Two Micron All Sky Survey \citep{skr06} and the VLT Survey Telescope ATLAS survey \citep{shan15}. This publication has made use of data  from the VISTA Hemisphere Survey, ESO Progamme, 179.A-2010 (PI: McMahon) and the VIKING survey from VISTA at the ESO Paranal Observatory, programme ID 179.A-2004. Data processing has been contributed by the VISTA Data Flow System at CASU, Cambridge and WFAU, Edinburgh. This publication makes use of data products from the Wide-field Infrared Survey Explorer, which is a joint project of the University of California, Los Angeles, and the Jet Propulsion Laboratory/California Institute of Technology, funded by the National Aeronautics and Space Administration.

This work presents results from the European Space Agency (ESA) space mission {\sl Gaia}. {\sl Gaia} data is being processed by the {\sl Gaia} Data Processing and Analysis Consortium (DPAC). Funding for the DPAC is provided by national institutions, in particular the institutions participating in the {\sl Gaia} MultiLateral Agreement (MLA). The {\sl Gaia} mission website is \url{https://www.cosmos.esa.int/gaia}. The {\sl Gaia} archive website is \url{https://archives.esac.esa.int/gaia}.

This research has benefitted from the M, L, T, and Y dwarf compendium housed at \url{http://DwarfArchives.org}. This research has benefited from the SpeX Prism Spectral Libraries, maintained by Adam Burgasser at \url{http://www.browndwarfs.org/spexprism}. This publication makes use of VOSA, developed under the Spanish Virtual Observatory project supported from the Spanish MICINN through grant AyA2008-02156. 

ZHZ is supported by the PSL fellowship. DJP and HRAJ acknowledge support from the UK's Science and Technology Facilities Council, grant numbers ST/M001008/1, ST/N001818/1, and ST/J001333/1. MCGO acknowledges the financial support of  the Spanish Virtual Observatory project supported from the Spanish ministry of Education and Competitiveness (MINECO) through grant AYA2014-55216. NL is funded by the project number AYA2015-69350-C3-2-P from Spanish Ministry of Economy and Competitiveness (MINECO). EM is funded by the MINECO under grant AYA2015-69350-C3-1-P. DH is supported by Sonderforschungsbereich SFB 881 `The Milky Way System' (subproject A4) of the German Research Foundation (DFG). FA received funding from the French `Programme National de Physique Stellaire' (PNPS) and the `Programme National de Plan\'etologie' of CNRS (INSU). The computations of atmosphere models were performed  in part on the Milky Way supercomputer, which is funded by the Deutsche Forschungsgemeinschaft (DFG) through the Collaborative Research Centre (SFB 881) `The Milky Way System' (subproject Z2) and hosted at the University of Heidelberg Computing Centre, and at the {\sl P\^ole Scientifique de Mod\'elisation Num\'erique} (PSMN) at the {\sl \'Ecole Normale Sup\'erieure} (ENS) in Lyon, and at the {\sl Gesellschaft f{\"u}r Wissenschaftliche Datenverarbeitung G{\"o}ttingen} in collaboration with the Institut f{\"u}r Astrophysik G{\"o}ttingen. The authors thank the referee, J. Davy Kirkpatrick for the useful and constructive comments.

\end{document}